\documentclass[fleqn,usenatbib]{mnras}

\usepackage[subrefformat=parens]{subcaption}
\usepackage[whole]{bxcjkjatype} 

\usepackage[T1]{fontenc}

\DeclareRobustCommand{\VAN}[3]{#2}
\let\VANthebibliography\thebibliography
\def\thebibliography{\DeclareRobustCommand{\VAN}[3]{##3}\VANthebibliography}

\usepackage{graphicx}	
\usepackage{amsmath}	
\usepackage{amssymb}	
\usepackage{multicol}        
\usepackage{bm}		
\usepackage{pdflscape}	
\usepackage{physics} 
\usepackage{cases}

\usepackage{newtxtext,newtxmath} 

\title[Magnetic field transport in thick discs]{Magnetic field transport in geometrically thick discs:\\
multi-dimensional effects on the field strength and inclination angle}

\author[R. Yamamoto and S. Takasao]{
Ryoya Yamamoto,$^{1}$\thanks{E-mail: ryoya@astro-osaka.jp}
Shinsuke Takasao,$^{1}$
\\
$^{1}$Department of Earth and Space Science, Graduate School of Science, Osaka University, Toyonaka, Osaka 560-0043, Japan
}

\date{Accepted XXX. Received YYY; in original form ZZZ}

\begin{document}
\label{firstpage}
\pagerange{\pageref{firstpage}--\pageref{lastpage}}
\maketitle

\begin{abstract}
We theoretically investigate the magnetic flux transport in geometrically thick accretion discs which may form around black holes. We utilize a two-dimensional (2D) kinematic mean-field model for poloidal field transport which is governed by both inward advection and outward diffusion of the field. 
Assuming a steady state, we analytically show that the multi-dimensional effects prevent the field accumulation toward the centre and reduce the field inclination angle.
We also numerically investigate the radial profile of the field strength and the inclination angle for two geometrically thick discs for which (quasi-)analytic solutions exist: radiatively inefficient accretion flows (RIAFs) and super-Eddington accretion flows.
We develop a 2D kinematic mean-field code and perform simulations of flux transport to study the multi-dimensional effects. The numerical simulations are consistent with our analytical prediction. We also discuss a condition for the external field strength that RIAF can be a magnetically arrested disc. 
This study could be important for understanding the origin of a large-scale magnetic field that drives jets and disc winds around black holes.

\end{abstract}




\section{Introduction}
To understand the origin of magnetically driven outflows, it is crucial to elucidate the distribution of large-scale poloidal magnetic fields, including their strength and inclination \citep[e.g.,][]{Contopoulos1994, Kudoh1997, Fukumura2010, JacqueminIde2019, Dihingia2021}. This is because the large-scale magnetic fields neither dissipate through local magnetic diffusion nor vanish when accreted on to black holes.
The disc acquires magnetic fields from the accretion flow from outside the disc during its growth \citep[e.g.,][]{Lovelace1976, Hawley2015, Takasao2022}, and the imported magnetic fields distribute within the disc according to the disc's advection and effective magnetic diffusion \citep[e.g.,][]{Lubow1994, Beckwith2009}. 
In addition to the magnetic flux transport, disc dynamo could also be important for the generation of the disc poloidal field \citep[e.g.,][]{Stepanovs2014}. However, the effectiveness of dynamos seems to depend on the initial magnetic field strength and disc thickness and the importance remains elusive \citep[e.g.,][]{Hogg2018, Liska2020}.
Recent AGN observations suggest that Radio Loud AGNs with jets have stronger magnetic fields compared to Radio Quiet AGNs without jets \citep{LopezRodriguez2023} at a scale of $\gtrsim 5$~pc, emphasizing the importance of magnetic flux transport.

The amount of magnetic flux brought into the central black hole is crucial for the driving of relativistic jets \citep[e.g.,][]{Blandford1977, Tchekhovskoy2011}.
It is believed that when a rotating black hole accumulates a significant amount of poloidal magnetic fields, it forms a magnetically arrested disc (MAD) that can drive powerful jets \citep[e.g.,][]{Narayan2003, Tchekhovskoy2011, McKinney2012}.
Supermassive black holes like Sgr A* exhibit recurrent flares \citep{GRAVITY2018}, and recent simulation studies suggest that flares can be driven by magnetic reconnection in the inner disc if it is in the MAD state \citep{Dexter2020, Porth2021, Ripperda2022} (a similar flare mechanism is also discussed in the context of accretion of protostars \citep{Takasao2019}). Recent observations of X-ray binary (XRB) systems suggest the existence of MAD \citep{Zdziarski2022}, and future X-ray spectroscopic observations will find evidence of MAD \citep[e.g.,][] {Inoue2023}.
Understanding the conditions for MAD manifestation is crucial for revealing the origin of jets.

However, the mechanism determining the disc magnetic field distribution remains poorly understood, making it challenging to discuss the magnetization of black holes.
The transport of global poloidal magnetic fields is one of the challenging problems in astrophysics, primarily due to the need to simultaneously track a wide range of scales. 
The typical size of the X-ray binary disc is considered to be $\sim10^5-10^6 r_{\rm g}$ ($r_{\rm g} = GM/c^2$ is the gravitational radius) 
\citep[e.g.,][]{Alfonso-Garzon2018, Hynes2019}.
\citet{Ressler2020a, Ressler2020b} performed three-dimensional magnetohydrodynamic (3D MHD) simulations covering up to seven orders of magnitude in scale using a zoom-in technique.
Such an approach is powerful when the backreaction from smaller scales to larger scales is negligible.
As the smaller scales can affect the larger scales during the system's evolution, there is still a need to solve a wide range of spacetime simultaneously.

3D MHD simulations that cover a wide range of scales are numerically challenging. For this reason, a one-dimensional kinematic mean-field model (1D model), which integrates the disc structure in the vertical direction, has been widely used \citep{Lubow1994, Okuzumi2014, Guilet2014}. Such a model is useful for modelling geometrically thin discs such as a standard disc \citep{Shakura1973}. \citet{Lubow1994} demonstrated for geometrically thin discs that the following dimensionless parameter $D$ is key for determining the efficiency of inward magnetic flux transport and the inclination of the magnetic field: 
\begin{align}
D \approx \frac{1}{P_{\rm m}h},\label{eq:D_Pm}
\end{align}
where $P_{\rm m}$ represents the effective magnetic Prandtl number due to turbulence, and $h$ is the aspect ratio of the disc.

It is believed that jet-related discs are thick discs in many cases \citep[e.g.,][]{Cao2011, Begelman2014, Dhang2023}. 
Although 1D models are a powerful tool for geometrically thin discs, a simple application to geometrically thick discs could be problematic because the model ignores a detailed vertical structure.
There have been theoretical studies examining the multi-dimensional effects of discs. \citet{Guilet2012} studied flux transport by considering the vertical structure of the discs, but their models adopt the thin disc approximation \citep[see also][]{Lovelace2009}. In addition, the radial field distribution (or the radial gradient of the field strength), which is intrinsically a result of magnetic flux transport, is treated as an input parameter. \citet{Li2021} investigated flux transport in vertically structured discs using their 2D axisymmetric model. However, the impact of the geometrical thickness on the disc field structure remains unclear.

In this study, we analytically and numerically investigate the distribution of poloidal magnetic fields achieved in geometrically thick discs around black holes.
We examine the magnetic field distribution in two types of geometrically thick discs: radiatively inefficient accretion flows (RIAF; e.g., \citealt{Narayan1994, Yuan2014}) and slim discs \citep[e.g.,][]{Abramowicz1988}.
The structure of this paper is as follows.
Section~2 provides an overview of past 1D models regarding magnetic flux transport. We consider the vertical structure of the disc's internal magnetic field, which was partly ignored in the 1D approximation. We analytically derive the parameter $D_{\rm eff}$ applicable to thick discs. We then introduce 1D and 2D kinematic mean field models used for the numerical validation of $D_{\rm eff}$. We also present the disc models used for validation.
Section~3 presents detailed results of the magnetic field distribution, focusing on the steady state.
In Section~4, our results are compared with previous studies with a particular focus on the vertical magnetic field distribution. We also discuss the onset condition of MAD in a RIAF.
Section~5 provides a summary of our results.

\section{Methods}
\subsection{Analytic estimation}
\label{sec:Basic_eq}

\subsubsection{Overview of kinematic 1D mean-field model}
\label{sec:1Dmodel}

Here we introduce a 1D model formulated by \citet{Lubow1994}.
The evolution of the axisymmetric poloidal magnetic field in the disc ($-H<z<H$, where $H \approx c_{\rm S}/\Omega$, $c_{\rm S}$ is sound velocity, $\Omega$ is rotational angular velocity) is calculated in cylindrical coordinates ($R,z,\phi$). The aspect ratio of the disc is denoted as
\begin{align}
h &\equiv \frac{H}{R}.
\end{align}
The disc is assumed to be geometrically thin ($h\ll1$),
and the magnetic field within the disc is approximated as follows:
\begin{align}
    B_R(R,z)&\simeq B_{R, \rm surf}\times\left(\frac{z}{H}\right), \label{eq:1DBR} \\
    B_z(R,z)&\simeq B_{z, \rm mid}, \label{eq:1DBz} 
\end{align}
where $B_{R, \rm surf}=B_R(R, z=H)$ and $B_{z, \rm mid} = B_z (R, z=0)$.

The induction equation governs the time evolution of the poloidal magnetic field.
In an axisymmetric configuration, it is helpful to utilize a flux function $\psi(R,z)$ defined by $\bm{B}=B_R\bm{e}_R + B_z \bm{e}_z = \nabla\times(\psi \bm{e}_\phi/R)$. The following relations hold:
\begin{align}
    B_R &= -\frac{1}{R}\frac{\partial\psi}{\partial z}, \\
    B_z &=  \frac{1}{R}\frac{\partial\psi}{\partial R}. \label{eq:Bz_cyl}
\end{align}
The induction equation for $\psi(R,z)$ is given by equation (10) of \citet{Lubow1994}:
\begin{align} \label{eq:1Dinduction}
\frac{\partial\psi}{\partial t} =
-v_R\frac{\partial\psi}{\partial R}
- R\eta J_\phi,
\end{align}
where $v_R$ is the radial advection velocity and $\eta$ is the magnetic diffusion coefficient,
which depend on a disc model (see Section \ref{sec:thick_disc}).
The azimuthal current density $J_\phi$ is given by
\begin{align} \label{eq:current}
    J_\phi = (\nabla\times \bm{B})_\phi
           = \left(\frac{\partial B_R}{\partial z} - \frac{\partial B_z}{\partial R}\right).
\end{align}
The 1D model averages equation (\ref{eq:1Dinduction}) in the vertical direction with weighting by $1/\eta$ over the range of $-H < z < H$ \citep[e.g.,][]{Ogilvie2001, Okuzumi2014}, which yields
\begin{align} \label{eq:avaraged_inducition_eq}
    \frac{\partial\psi_0}{\partial t}
    = - v^*_R\frac{\partial\psi_0}{\partial R} - \frac{R \eta^*}{2H}K_\phi,
\end{align}
where
\begin{align}
    \frac{1}{\eta^*} &\equiv \frac{1}{2H} \int^{H}_{-H} \frac{1}{\eta} dz, \\
    v^*_R  &\equiv \frac{\eta^*}{2H} \int^H_{-H} \frac{v_R}{\eta} dz.
\end{align}
Here, $\psi_0=\psi(R,z=0)$ is the flux function on the equatorial plane of the disc, and $\psi$ is assumed to be approximately constant in the vertical direction. $K_\phi$ is the azimuthal component of the surface current defined as
\begin{align} \label{eq:integ_current1}
    K_\phi = \int^{H}_{-H} J_\phi dz.
\end{align}
We determine $K_\phi$ from Biot-Savart's law using $\psi_0$ \citep[see e.g.,][]{Lubow1994}. Using equation (\ref{eq:current}), we also get
\begin{align} \label{eq:integ_current2}
K_\phi = 2B_{R, \rm surf} - \int^{H}_{-H} \frac{\partial B_{z}}{\partial R} dz .
\end{align}
This is the equation to calculate $B_{R,\rm surf}$ in the 1D models.

The second term in the right-hand side of the equation~(\ref{eq:integ_current2}) includes multi-dimensional effects as it depends on the vertical structure and the disc thickness \citep[e.g.,][]{Okuzumi2014}. 
Using equation (\ref{eq:1DBz}), we find
\begin{align} \label{eq:1DBRs_hDBz}
    B_{R, \rm surf} \approx \frac{1}{2} K_\phi + h D_{B_z0} B_{z, \rm mid},
\end{align}
where
\begin{align}
B_{z, \rm mid} &= \frac{1}{R} \frac{\partial \psi_0}{\partial R}, \\
D_{B_z0} &\equiv \frac{d {\rm ln} B_{z, \rm mid} }{d {\rm ln} R}.
\end{align}
We will examine the validity of equation~(\ref{eq:1DBRs_hDBz}) for geometrically thick discs by performing two-dimensional numerical calculations (see Section \ref{sec:Results}).
When $|B_{R, \rm surf}/B_{z, \rm mid}|\gg H/R$, we can further simplify the equation~(\ref{eq:integ_current2}) as follows:
\begin{align} \label{eq:integ_current_approx}
    K_\phi\approx 2B_{R, \rm surf}
\end{align}
\citep[e.g.,][]{Lubow1994, Guilet2014}. Under the assumption, the radial component of the magnetic field on the disc surface can be expressed as
\begin{align} \label{eq:1DBRs}
    B_{R, \rm surf} \approx \frac{1}{2} K_\phi.
\end{align}
We note that $B_{z, \rm mid}$ is unaffected by the choice of the calculation methods of $B_{R, \rm surf}$ as long as the accretion velocity and the resistivity do not depend on $B_{R, \rm surf}$ (see the induction equation~\ref{eq:avaraged_inducition_eq}). $K_{\phi}$ is calculated exactly from Biot-Savart's law.

In the steady state, the following equation holds true based on equations (\ref{eq:avaraged_inducition_eq}):
\begin{align}
    B_{z, \rm{mid}} &= \frac{D}{2} K_\phi. \label{eq:Bz_Kphi}
\end{align}
Substituting equation~(\ref{eq:1DBRs}) into equation~(\ref{eq:Bz_Kphi}) yields the following equation:
\begin{align}
    \frac{B_{R, \rm{surf}}}{B_{z, \rm{mid}}} &\approx D^{-1}. \label{eq:BRs_Bz}
\end{align}
$D$ is a dimensionless parameter defined as
\begin{align} 
    D\equiv \frac{\eta^*}{\left|v^*_R\right| H}. \label{eq:D}
\end{align}
Equation~(\ref{eq:BRs_Bz}) indicates that $D^{-1}$ is a measure of the inclination of the magnetic field at the disc surface.
The situation of $D<1$ corresponds to the case where advection is dominant. In such a case, the magnetic flux is efficiently transported to the centre, and the inclination of the magnetic field at the disc surface is also large.
In contrast, $D>1$ corresponds to the case where diffusion is dominant. The magnetic flux transport toward the centre is inefficient, and the magnetic field is more vertical at the disc surface.
Thus, $D$ serves as an indicator of the efficiency of magnetic flux transport in the 1D model (see Section 4 of \citealt{Lubow1994}).

\subsubsection{Evaluation of the multi-dimensional effect}
\label{subsec:high_order_ana}

The 1D model ignores a detailed vertical magnetic field structure inside the disc by assuming that the disc is thin. The assumption breaks down for geometrically thick discs. As a result, the vertical distribution of the magnetic field can deviate from a linear distribution, as represented by equations~(\ref{eq:1DBR}) and (\ref{eq:1DBz}). Here, we analytically evaluate the effect of the thickness of the disc on the vertical distribution of the magnetic field inside the disc.

We expand the magnetic field in the disc as follows:
\begin{align}
    B_R(R,z) &= \zeta B_R^{(1)}(R) + \zeta^3 B_R^{(3)}(R) + \mathcal{O}\left(\zeta^5\right), \label{eq:B_R_expanded}\\
    B_z(R,z) &= B_z^{(0)}(R) + \zeta^2 B_z^{(2)}(R) + \mathcal{O}\left(\zeta^4\right), \label{eq:B_z_expanded}
\end{align}
where
\begin{align}
    \zeta \equiv \frac{z}{H}.
\end{align}
Substituting equations~(\ref{eq:B_R_expanded}) and (\ref{eq:B_z_expanded}) into the induction equation for a steady state, we obtain the following:
\begin{align}
    0 = - v_R(R,z) B_z(R,z) + \eta(R,z) \left( \pdv{B_R(R,z)}{z} - \pdv{B_z(R,z)}{R}\right). 
\end{align}
After some calculations, we get the following equation:
\begin{align} \label{eq:high_order_induction}
    &\left[(D_*^{-1} + D_{B_z0})B_z^{(0)} - h^{-1} B_R^{(1)}\right] \nonumber \\ 
    & + \left[(D_{B_z2}-2D_H + D_*^{-1})B_z^{(2)} - h^{-1} 3B_R^{(3)}\right]\zeta^2
    + \mathcal{O} \left(\zeta^4\right)= 0,
\end{align}
where
\begin{align}
    D_* &\equiv \frac{\eta}{|v_R|R}, \\
    D_{B_z0} &\equiv \frac{d {\rm ln} B_z^{(0)}}{d {\rm ln} R}, \\ 
    D_{B_z2} &\equiv \frac{d {\rm ln} B_z^{(2)}}{d {\rm ln} R}, \\
    D_{H} & \equiv \frac{d {\rm ln}H}{d {\rm ln} R} .
\end{align}
When the vertical dependence on $\eta$ and $v_R$ is sufficiently small such that $\eta = \mathcal{O}(\zeta^0)$ and $v_R = \mathcal{O}(\zeta^0)$, equation (\ref{eq:high_order_induction}) requires that the following two relations must hold true regardless of $\zeta$:
\begin{align}
    \frac{B_R^{(1)}}{B_z^{(0)}} &= h (D_*^{-1} + D_{B_z0}) = D^{-1} + h D_{B_z0}, \label{eq:BR1_Bz0}\\
    \frac{B_R^{(3)}}{B_z^{(2)}} &= \frac{1}{3}h (D_*^{-1} + D_{B_z2} -2D_H). \label{eq:BR3_Bz2}
\end{align}

Equation~(\ref{eq:BR1_Bz0}) is an extension of equation~(\ref{eq:BRs_Bz}). By comparing the two equations, we can rewrite equation~(\ref{eq:BR1_Bz0}) as 
\begin{align} \label{eq:BR_Bz_mod}
    \frac{B_R^{(1)}}{B_z^{(0)}} = D_{\rm eff}^{-1},
\end{align}
where
\begin{align} \label{eq:D_eff}
    D_{\rm eff} \equiv \frac{1}{D^{-1} + h D_{B_z0}}.
\end{align}
$D_{\rm eff}$ is similar to $D$ but includes multi-dimensional effects due to the thickness of the disc and magnetic field gradient. One can easily find that $D_{\rm eff} \approx D$ in the limit of a thin disc ($h \ll 1$). When the magnetic field strength decreases monotonically outward from the centre, $D_{B_z0} < 0$, and thus $D_{\rm eff} > D$. The difference between them becomes more significant as $h$ increases. For a thick disc ($h \sim 1$), $D_{\rm eff} > D$, which indicates that the multi-dimensional effects effectively increase magnetic diffusivity.
Jet-related discs are commonly thick discs, but the multi-dimensional effects suppress the magnetic flux accumulation. Considering these results, careful investigations of flux transport are necessary for thick discs.

The assumption of the linear distribution (equations~\ref{eq:1DBR} and \ref{eq:1DBz}) will break down when multi-dimensional effects are important.
To show this, we evaluate the magnitude of higher-order components of the magnetic field.
By substituting equations~(\ref{eq:B_R_expanded}) and (\ref{eq:B_z_expanded}) into the divergence-free condition $\nabla \cdot \bm{B} = 0$, we obtain
\begin{align} \label{eq:gauss}
    \left[\left(1+D_{B_R1} - D_H\right)\frac{B_R^{(1)}}{R} + \frac{2B_z^{(2)}}{H}\right]\zeta
     + \mathcal{O}\left(\zeta^3\right)= 0,
\end{align}
where
\begin{align}
    D_{B_R1} &\equiv \frac{d {\rm ln} B_R^{(1)}}{d {\rm ln} R}.
\end{align}
Equation~(\ref{eq:gauss}) requires that
\begin{align} \label{eq:Bz2_Bz0}
    \frac{B_z^{(2)}}{B_z^{(0)}}
    = -\frac{1}{2} h^2 (D_{C} + D_{B_z0})(D_*^{-1} + D_{B_z0}),
\end{align}
where we have used equation~(\ref{eq:BR1_Bz0}) to eliminate $B_R^{(1)}$ and defined
\begin{align}
    D_{C} \equiv \frac{d {\rm ln} (D_*^{-1}+D_{B_z0})}{d {\rm ln} R}.
\end{align}
Equation~(\ref{eq:Bz2_Bz0}) is a direct comparison between the 0th and 2nd order components of $B_z$. If $B_z^{(2)}/B_z^{(0)}\approx 1$, the linear distribution approximation given by equation~(\ref{eq:1DBz}) is invalid. Since $B_z^{(2)}/B_z^{(0)}\propto h^2$, higher-order components should be more significant for thicker discs.

\subsection{Geometrically thick disc models}
\label{sec:thick_disc}
To highlight the impact of disc thickness on magnetic flux transport, we perform numerical calculations for geometrically thick discs. We compare the results of the 1D and axisymmetric 2D models to examine the limitation of the assumptions in the 1D model.

Our disc models are the following two: the self-similar solution of the radiatively inefficient accretion flow \citep[RIAF;][]{Narayan1994} and the quasi-analytical solution of the super-Eddington accretion flow \citep[SEAF, slim disc and standard disc;][]{Watarai2006}. They are representative models of geometrically thick discs around black holes. Both are advection-dominated accretion flows (ADAF), but RIAF and SEAF correspond to optically thin and thick discs, respectively.
The following scaling laws hold for both models:
\begin{align}
    \Sigma &\propto f^{-1/2}R^{-1/2}, \\
    \Bar{v}_R &\propto -\alpha f R^{-1/2}, \\
    H &\propto f^{1/2}R,
\end{align}
where $\alpha$ is the $\alpha$-parameter \citep[][]{Shakura1973}. $\Sigma$ and $\Bar{v}_R$ are the surface density and density-weighted average advection velocity defined as
\begin{align}
    \Sigma    & \equiv \int^\infty_{-\infty} \rho dz,\\
    \Bar{v}_R & \equiv \frac{1}{\Sigma} \int^\infty_{-\infty} \rho v_R dz, 
\end{align}
respectively.
$f$ is the advection parameter, defined as the ratio of the advective cooling rate per unit area to the viscous heating rate per unit area, i.e.,
\begin{align}
    f\equiv \frac{Q_{\rm adv}^-}{Q_{\rm vis}^+}.   
\end{align}
We have $f=1$ in cases of no radiative cooling.

The RIAF solution is a self-similar solution obtained when $f$ is a constant value throughout the disc. The aspect ratio $h$ is an increasing function of $f$.
Assuming isothermal in the vertical direction, the density is written as
\begin{align}
    \rho(R,z)=\rho_{\rm mid}(R)\exp[-\frac{z^2}{2H(R)^2}],
\end{align}
where $H = \sqrt{\Pi/\Sigma}/\Omega$. This density structure is used only for visualization.
A different semi-analytical solution has been proposed recently \citep[][]{Xu2023}. The application of such other disc models will be our future work.

\citet{Watarai2006} constructed the SEAF solution by assuming that the disc radiation is blackbody radiation and the radiation pressure dominates over the gas pressure ($p_{\rm rad} \gg p_{\rm gas}$). The radius dependence of $f$ is expressed as
\begin{align} \label{eq:adv_param_watarai}
    f(x) = \frac{1}{2} \left(C^2x^2 + 2 - C x\sqrt{C^2x^2+4}\right),
\end{align}
where $C$ is a dimensionless constant of order unity
(see equation~24 of \citealt{Watarai2006}) \footnote{
In \citet{Watarai2006}, the dimensionless constant is denoted as $D$.
However, as we have already used $D$ (equation \ref{eq:D}), we instead use $C$ to denote the constant to avoid confusion.
}.
Also, $x \equiv (R/r_S) (\Dot{M}/\Dot{M}_{\rm Edd})^{-1}$
($r_{\rm S}=2GM/c^2$ and $\Dot{M}_{\rm Edd} = L_{\rm Edd}/c^2$ are the Schwarzschild radius and the Eddington accretion rate, respectively).
$f$ takes a value in the range $0 \leq f \leq 1$.

SEAF has the characteristic radius that determines the slim region. Photons travel diffusively in the disc because of their high gas density. When the accretion time becomes shorter than the diffusion time of photons, the photons are trapped inside the disc. The photon trapping radius can be expressed as $R_{\rm trap} \approx H/(c/\tau) v_R$ ($\tau$ is the optical depth). Within the photon trapping radius, the cooling effect of radiation becomes inefficient, causing the disc to expand and form a slim disc \citep[e.g.,][]{Abramowicz1988, Watarai2006}. In terms of the accretion rate, the photon trapping radius can be written as
\begin{align}
    R_{\rm trap} \approx 85r_{\rm S} \left(\frac{\dot{M}}{100\dot{M}_{\rm Edd}}\right). 
\end{align}
For $R \lesssim R_{\rm trap}$, $f\rightarrow 1$ ($f\propto R^0$) and the advection cooling is dominant. Such a geometrically and optically thick disc is commonly referred to as a "slim disc". For $R \gtrsim R_{\rm trap}$, $f\propto R^{-2}$ and the radiative cooling dominates the advection cooling. Therefore, the outer disc can be described as a standard disc.
The quasi-analytical solution by \citet{Watarai2006} has also been reproduced in recent 2D radiation hydrodynamic simulations (\citealt{Kitaki2021, Yoshioka2022}).

The density structure of SEAF is introduced only for better visualization of the results.
The temperature of SEAF has a non-isothermal distribution given by $T(R, z) =T_{\rm mid}(R) (1-z^2/H^2)$, and it is assumed that the polytropic relation $P\propto \rho^\Gamma$ holds. Therefore, the density distribution is given by
\begin{align}
    \rho(R, z) = \rho_{\rm mid} (R) \left(1-\frac{z^2}{H(R)^2}\right)^{N},
\end{align}
where $N = 1/(\Gamma - 1)$ (polytropic index) and $H = (2N+3)\sqrt{\Pi/\Sigma}/\Omega$.
SEAF has a photon trapping radius (approximately $85 r_{\rm S}$) and a radius where the "cold" (gas-pressure-dominated) standard disc appears (approximately $500 r_{\rm S}$), resulting in a significantly varying aspect ratio in the radial direction.

We describe the fiducial sets of parameters in this study. Both models assume a black hole mass of $10M_\odot$, $\alpha=0.01$, and a specific heat ratio of $\Gamma=4/3$. Regarding $\alpha$, we refer to the results of past 3D MHD simulations on MRI turbulent discs, which show $\alpha = \mathcal{O} (0.01) - \mathcal{O} (0.1)$ \citep[e.g.,][]{Hawley2013, Suzuki2014, Takasao2018}. 
We adopt the mass accretion rates of $10^{-3} \dot{M}_{\rm Edd}$ and $100 \dot{M}_{\rm Edd}$ for RIAF and SEAF, respectively.
The disc structure of RIAF remains unchanged regardless of the mass accretion rate, while the mass accretion rate affects the photon trapping radius in SEAF.
For the RIAF model, we adopt a fiducial value of $f=1$. The aspect ratios $h$ for these parameter values are $h\approx 0.53$ and $\approx 3$ for the RIAF and slim disc in SEAF, respectively. We also have $C\approx1.77$ for the SEAF solution (equation \ref{eq:adv_param_watarai}).

We characterise the density-weighted magnetic diffusivity using the magnetic Prandtl number
\begin{align} \label{eq:Pm_def}
    P_{\rm m} \equiv \Bar{\nu}/\Bar{\eta},
\end{align}
where the density-weighted viscosity coefficient $\Bar{\nu}$ is given by
\begin{align} \label{eq:nu_weighted}
    \Bar{\nu} \equiv \frac{1}{\Sigma}\int^\infty_{-\infty}\rho \nu dz = \frac{2}{3} \alpha \Omega^{-1} \frac{\Pi}{\Sigma}.
\end{align}
Here, $\Omega=v_\phi/R$ and $\Pi \equiv \int^\infty_{-\infty} p dz$ represent the disc rotation angular velocity and surface pressure, respectively, both of which are provided by the disc model.
As the relation $v_R\approx \bar{\nu}/R$ holds in a viscous disc, we obtain equation~(\ref{eq:D_Pm}).

Once the magnetic diffusivity $\bar{\eta}$ is given, $D$ can be obtained as follows:
\begin{align}
D =
    \left\{
    \begin{array}{ll}
    X(f, \alpha)  P_{\rm m}^{-1} & \textrm{for~RIAF} \\ 
    0.22f^{-1/2}P_{\rm m}^{-1} & \textrm{for~SEAF} 
    \end{array}
    \right. \label{eq:D_thick_disc}
\end{align}
Assuming $\Gamma = 4/3$, $X(f, \alpha)$ is described as 
\begin{align}
    X(f, \alpha) = \frac{2\alpha^2f^{-1/2}}{ \left[\left(5+2/f\right)^2 + 18\alpha^2\right]^{1/2} - (5+2/f) }.
\end{align}
When $f=1$ and $\alpha=0.01$, $X \approx 1.56$.

\subsection{Numerical methods}
\label{sec:numerical_methods}
To directly investigate the multi-dimensional effects discussed in Section~\ref{subsec:high_order_ana}, we perform magnetic flux transport calculations using a two-dimensional axisymmetric model in spherical coordinates $(r, \theta, \phi)$ (hereinafter referred to as the 2D model), and compare the results with those obtained using a 1D model. We describe the computational methods for the 1D and 2D models below.

\subsubsection{1D model}
We solve equation~(\ref{eq:avaraged_inducition_eq}) and use the equation~(\ref{eq:1DBRs}) to obtain $B_{R, \rm surf}$.
The surface current density $K_\phi$ in the diffusion term (the second term of equation~\ref{eq:avaraged_inducition_eq}) is calculated in the same way as in \citet{Lubow1994} (see Section~3 of \citealt{Lubow1994}), but we set $\lambda=10^{-3}$ in our calculations, whereas \citet{Lubow1994} used their equation~(27).
We calculate the advection term (the first term of equation~\ref{eq:avaraged_inducition_eq}) using second-order upwind differencing with the Monotonized Central (MC) interpolation and the MUSCL method \citep{vanLeer1997} (first-order upwind differencing was used in \citealt{Lubow1994}).  We use second-order strong stability preserving Runge–Kutta methods (SSPRK, \citealt{Gottlieb2009}) for time integration. The time step is determined by the CFL condition:
\begin{align}
    \Delta t = C_{\rm CFL} \min_{i} \left(\frac{\Delta R_{i}}{|v^*_{R, i}|}, \frac{H_i \Delta R_i}{\eta^*_i}\right),
\end{align}
where $\Delta R_{i}=R_{i+1/2}-R_{i-1/2}$ ($R_i$ denotes the position of the $i$-th cell center), and the CFL number is $C_{\rm CFL}=0.4$. The logarithmically spaced cells are used, and the cell number $N_R$ is set to match the spatial resolution of the corresponding 2D models.

The discs are initially threaded by a uniform magnetic field, $\psi (R, t=0)= (1/2) R^2 B_0$ ($B_0$ is the uniform imposed field strength). The field evolves in response to the advection and effective magnetic diffusion. $v^{*}_R$ and $\eta^*$ are given according to the disc models and are assumed to be constants with time. We ignore the back reaction of the Lorentz force on the disc flows. We also assume that $v_R$ and $\eta$ are uniform in the vertical direction of the disc and set $v^*_R=\Bar{v}_R$ and $\eta^*=\Bar{\eta} = P_{\rm m} \Bar{\nu}$ for simplicity.
The velocity generally varies with height in response to the coupling between the magnetic field and the plasma, as suggested by 3D MHD simulations \citep[e.g.,][]{BaiStone2013,Lesur2013,Suzuki2014,Takasao2018}. However, there are no established analytic methods that self-consistently describe both the radial and vertical velocity structures \citep[for the analytic study of the local disc, see, e.g., ][]{Guilet2012}. For this reason, we adopt the vertically uniform velocity in this study.

As the inner boundary condition, we impose that $B_z = \rm const$. This condition approximates the boundary condition of the 2D models. We note that a simple outflow boundary ($\psi = \rm const$ or $B_z = 0$) behaves as a sink of the field (see also Appendix \ref{app:2D_Inner_Boundary}).

For comparisons with 2D models, we define the radius of an effective disc outer edge inside the numerical domain, $R_{\rm trunc}$. 
We set $v^*_R=\eta^*=0$ for $R>R_{\rm trunc}$.

\subsubsection{2D model}\label{sec:2Dmodel}

The time evolution equation for the poloidal magnetic field is solved using the flux function $\psi$ in the 2D spherical coordinate system $(r, \theta)$. The relationship between $\psi$ and the components of the magnetic field is given as follows:
\begin{align}
    B_r &= \frac{1}{r^2 \sin\theta} \frac{\partial \psi}{\partial \theta}, \label{eq:2DBr} \\
    B_\theta &= - \frac{1}{r\sin\theta} \frac{\partial \psi}{\partial r}. \label{eq:2DBtheta}
\end{align}
Here, $\psi$ satisfies
$\bm{B}=\nabla\times(\psi \bm{e}_\phi/r\sin\theta)$.
Therefore, the induction equation is given by
\begin{align} \label{eq:2Dinduction}
    \frac{\partial \psi}{\partial t} 
    = -v_r \frac{\partial \psi}{\partial r} - \frac{v_\theta}{r} \frac{\partial \psi}{\partial \theta}
    + \eta\left[\frac{\partial^2 \psi}{\partial r^2} 
    + \frac{\sin\theta}{r^2}\frac{\partial}{\partial \theta}\left(\frac{1}{\sin\theta}\frac{\partial \psi}{\partial \theta}\right)\right].
\end{align}
The first and second terms of equation~(\ref{eq:2Dinduction}) are the advection terms, and the third and fourth terms are the diffusion terms. Note that our code can handle the time evolution of the magnetic field, unlike the 2D model of \citet{Li2021}.

The velocity field $(v_r, v_\theta)$ is given by a disc model within the disc and is set to 0 outside the disc. Assuming that the disc surface is located at a height of scale height $H$ from the equatorial plane, we set
\begin{align} \label{eq:smoothing_vel}
    v_R(r, \theta) = \Bar{v}_R(R)\frac{1}{2}\left[\tanh\left(\frac{H-|z|}{\Delta H}\right) + 1\right],
\end{align}
where $(R, z) = (r\cos \theta, r\sin \theta)$ and $\Delta H = 0.1 H$. We then set $v_r$ and $v\theta$ as
\begin{align}
    v_r (r, \theta)& = v_R \sin \theta, \\
    v_\theta (r, \theta) &= v_R \cos \theta.
\end{align}

The magnetic diffusion coefficient $\eta$ is assumed to be constant in the vertical direction, including above the disc (i.e., $\eta(R, z) = \Bar{\eta}(R) = P_{\rm m} \Bar{\nu}(R)$), and its value at each radius $R$ is determined by the disc model. 
Although this study focuses only on steady-state solutions, the steady-state is independent of the method used to determine the magnetic diffusion coefficient outside the disc.
To prevent numerical instability near the poles, we set $\eta(r, \theta) = 0$ for the physical domain mesh that borders the $\theta$ direction boundaries. We also set $v_R$ and $\eta$ to zero for $R > R_{\rm trunc}$, respectively, to prevent magnetic field inflow/outflow from the external boundary for $R < R_{\rm trunc}$.
Setting the velocity field to zero and the magnetic diffusion coefficient to a finite value above the disc ensures that the magnetic field distribution above the disc in the steady state is a potential field (i.e., $J_\phi=0$). This is consistent with the requirements of a 1D model. The magnetic field distribution within and above the disc in the steady state does not depend on the magnitude of the magnetic diffusion coefficient above the disc.

We solve the advection terms with the second-order upwind scheme using the MC limiter and the diffusion terms with second-order central differencing, achieving second-order spatial accuracy. We also integrate the time evolution with second-order SSPRK same as the 1D model. The time step is determined by the CFL condition ($C_{\rm CFL}=0.4$), given by
\begin{align}
    dt = \min_{i, j} 
    \left(C_{\rm CFL}\frac{\Delta r_{i}}{|v_{r_{i,j}}|}, C_{\rm CFL}\frac{r_i \Delta\theta_{j}}{|v_{\theta_{i,j}}|},
    \frac{\Delta r_{i}^2}{2\eta_{i,j}}, \frac{(r_i \Delta\theta_{j})^2}{2\eta_{i,j}}\right),
\end{align}
where the subscripts $i$ and $j$ denote the labels of the radial and latitudinal locations of the cell centre.
The domain is descretized such that $\Delta r_i = r_{i+1/2} - r_{i-1/2}$ and $\Delta \theta_j = \theta_{j+1/2} - \theta_{j-1/2}$.

We set a uniform vertical magnetic field for the initial condition, similar to the 1D model (i.e. $\psi(r, \theta, t=0) = (1/2) B_0 (r \sin\theta)^2$).

We set the boundary conditions for $\psi$ at the inner and outer boundaries in the $r$ direction such that $rB_r=\text{const}$ (see Appendix \ref{app:2D_Inner_Boundary} for details). Additionally, we set reflective boundary conditions for $\psi$ in the $\theta$ direction.

We adopt a logarithmic mesh structure in the $r$ direction with the same number of meshes as the $R$ direction of the 1D model. For the $\theta$ direction, we use a compressed non-uniform mesh so that the grid spacing at the equator is 2.5 times smaller than that at the poles. Table \ref{tab:simulation_setup} summarises the information about the domain size and resolution.
As shown in Appendix \ref{app:convergence_check}, our numerical resolution is sufficient to obtain the converged results.
Our code for the 2D models utilizes the framework of the publicly available MHD code, Athena++ \citep{Stone2020}.

\begin{table*}
 \caption{
    $r_{\text{min}}$ and $r_{\text{max}}$ are the minimum and maximum spherical radii of the calculation domain of 2D models, respectively. $N_r$ is the number of cells in the $r$ direction of 2D models, while $N_R$ is that in the $R$ direction of 1D models. We set $N_r = N_R$. $N_\theta$ is the number of cells in the $\theta$ direction of 2D models. $(H/\Delta r\theta)_{\text{min}}$ and $(H/r\Delta \theta)_{\text{max}}$ are the minimum and maximum resolutions within the disc, respectively. 
    The aspect ratios of cells, $\Delta r/ r\Delta \theta$, at the pole and equatorial plane are also listed.
    }\label{tab:simulation_setup}
	\centering
	\begin{tabular}{llllllllll} 
		\hline
		  model & $r_{\rm min}$ & $r_{\rm max}$ & $R_{\rm trunc}$ & $N_r~(N_R)$ & $N_{\theta}$ & $(H/r\Delta \theta)_{\rm min}$ & $(H/r\Delta \theta)_{\rm max}$ & $(\Delta r/ r\Delta \theta)_{\theta=\pi/2}$ & $(\Delta r/ r\Delta \theta)_{\theta=0}$\\
		\hline
		  RIAF  & $r_{\rm S}$ & $200r_{\rm S}$  & $100r_{\rm S}$ & $128$ & $80$ & $18$ & $18$ & $1.58$ & $0.64$\\
    
		  SEAF & $r_{\rm S}$ & $2000r_{\rm S}$ & $750r_{\rm S}$ & $180$ & $80$ & $13$ & $36$ & $1.61$ & $0.65$\\
    
		\hline
	\end{tabular}

\end{table*}

\section{Results}
\label{sec:Results}

\subsection{Overview}
We first review the results for $P_{\rm m}=1$. The dependence on $P_{\rm m}$ is investigated in Section~\ref{subsec:Pm_dependences}.

\subsubsection{Overview of steady-state solutions}
\begin{figure} 
    \centering
    \includegraphics[width=0.495\columnwidth]{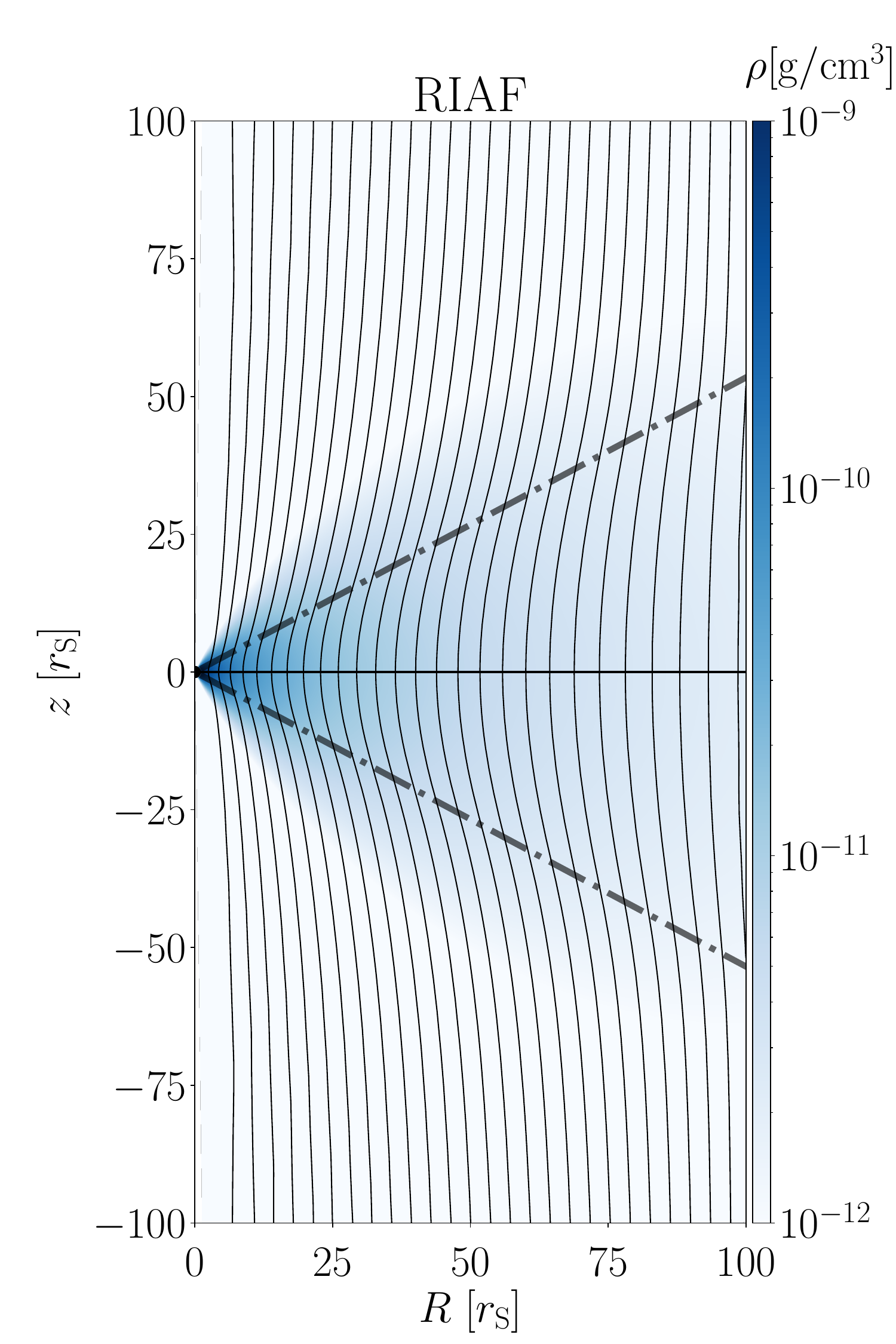}
    \includegraphics[width=0.495\columnwidth]{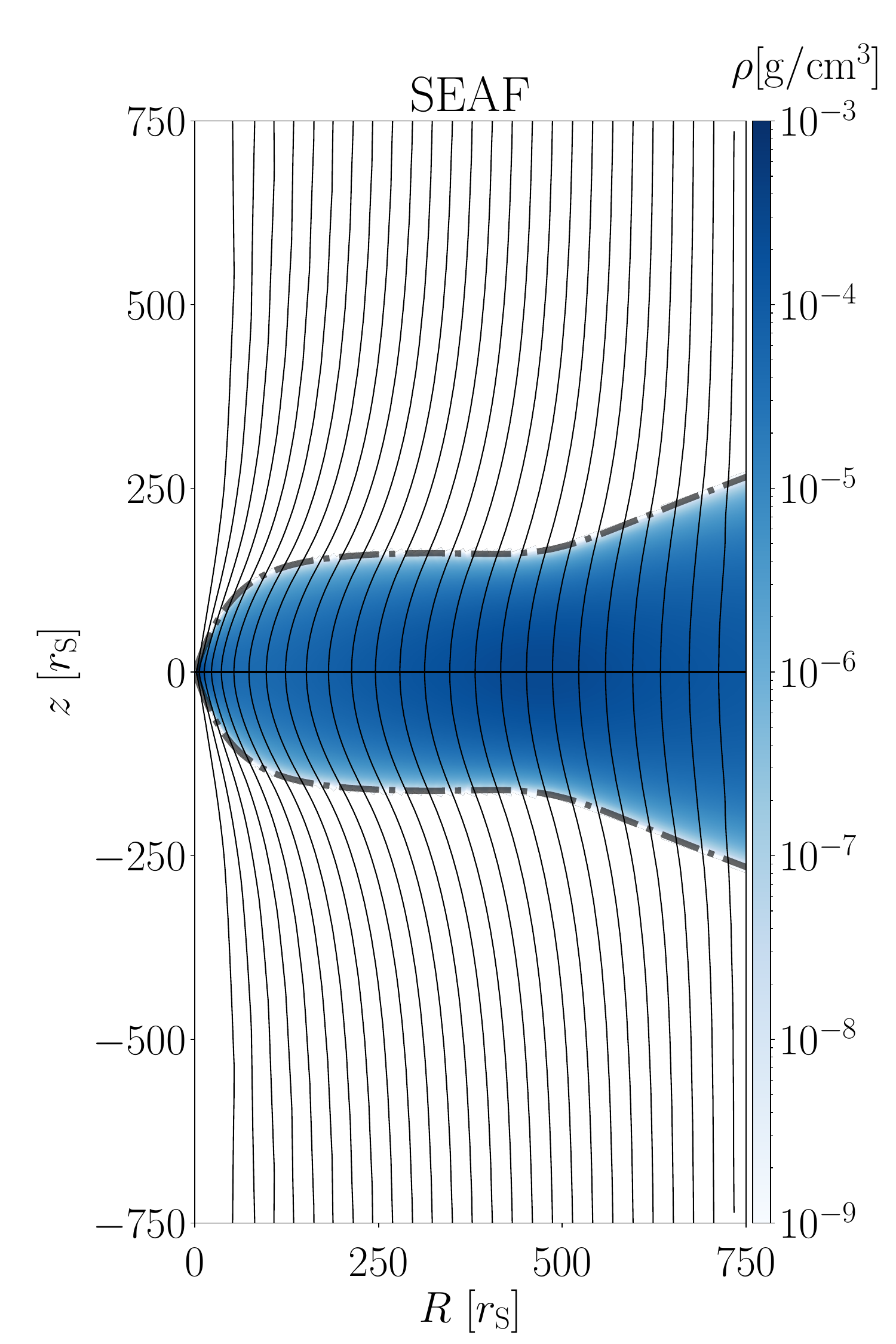}
    \includegraphics[width=0.495\columnwidth]{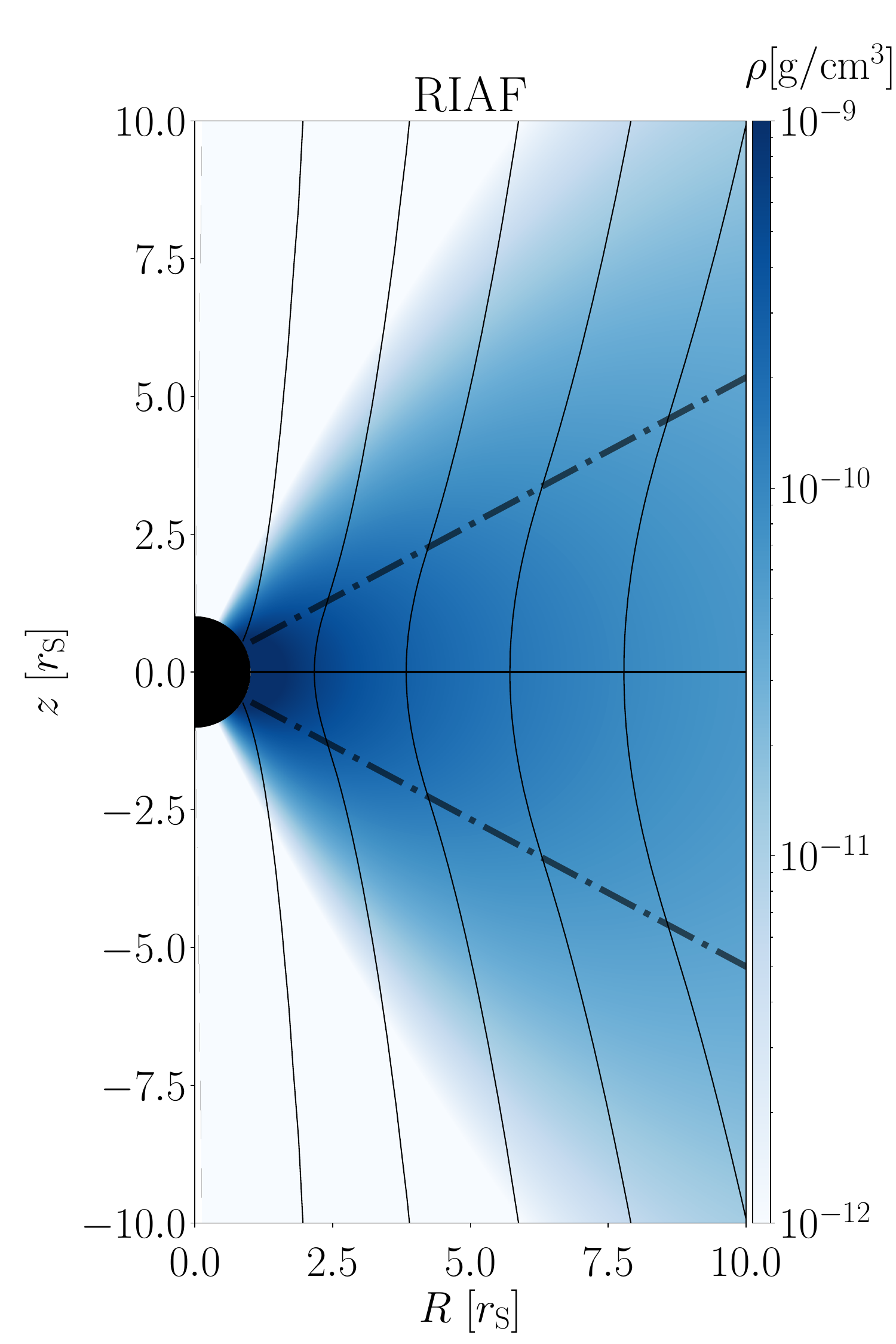}
    \includegraphics[width=0.495\columnwidth]{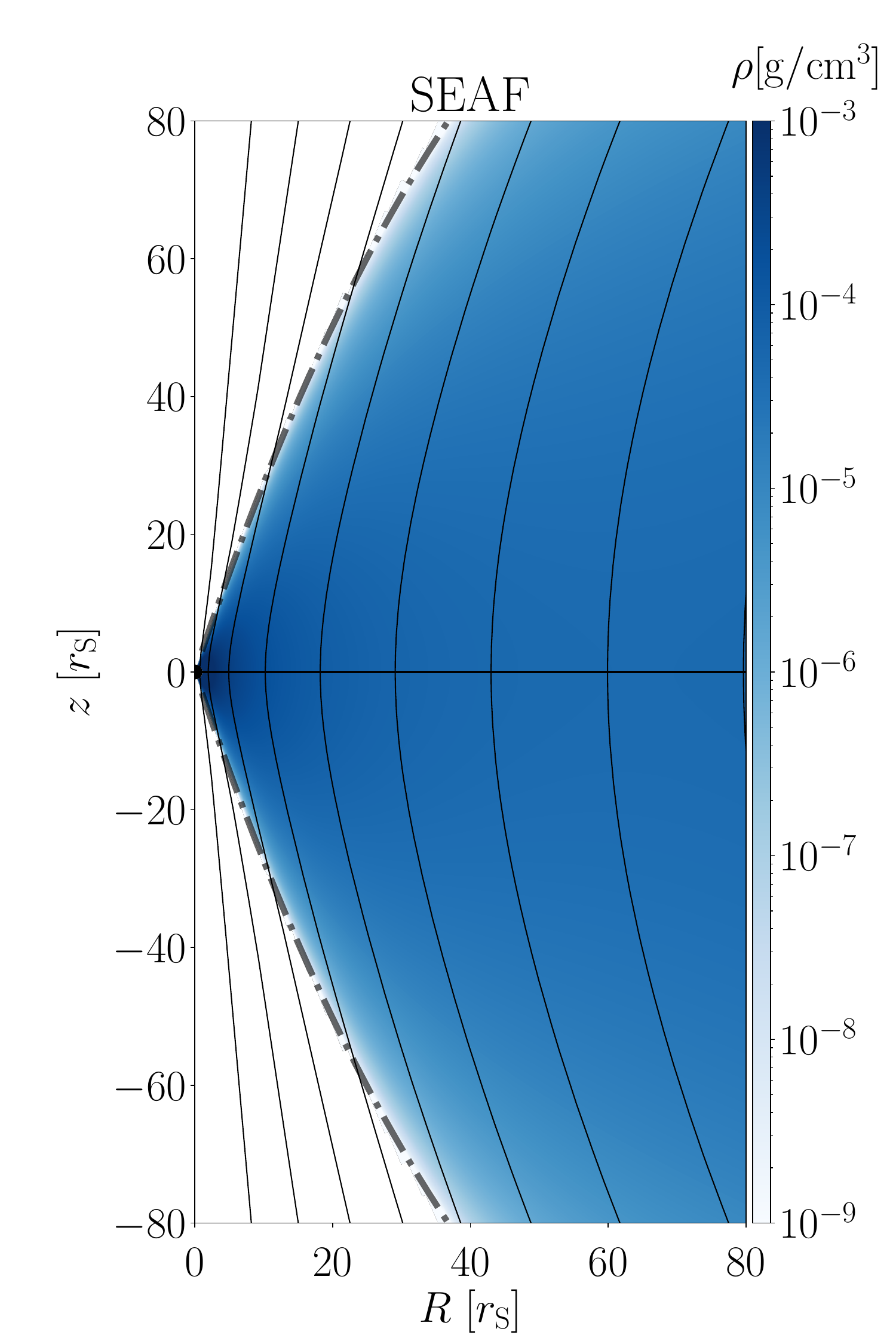}
  \caption{
  The poloidal field structures in steady states (solid lines). The left and right panels show the results of RIAF and SEAF, respectively. The bottom panels are zoom-in images of the top panels. In the case of SEAF, the zoomed-in region approximately corresponds to the slim disc region. The figure shows the density distribution by colour to indicate the disc structure. The pressure scale height is shown as dashed-dotted lines. The results are for $P_{\rm m}=1$.
  }
  \label{fig:field_line}
\end{figure}

In this study, we focus only on the steady state. For a better comparison between 1D and 2D models, the results of the 2D model are displayed in cylindrical coordinates. The solid lines in Fig. \ref{fig:field_line} represent the poloidal magnetic field structures of the RIAF and SEAF obtained from the 2D model.
The disc density distribution is also indicated by colour for better visualization of the disc structure.
The disc aspect ratios $h$ are approximately $0.5$ for RIAF and $3$ for the slim disc region in SEAF.
The upper panel represents the global magnetic field distribution inside $R < R_{\rm trunc}$, while the lower panel represents the magnetic field distribution focused on the inner region (approximately the slim region in SEAF). Furthermore, in this section, we assume a fiducial magnetic Prandtl number of $P_{\rm m} = 1$.

The timescale until the magnetic flux transport reaches a steady state can be estimated by the relaxation timescale defined as $t_{\rm relax}(R) \equiv HR/\eta^{*}$ \citep[e.g.,][]{Lubow1994, Lovelace2009}. The relaxation timescale at $R=R_{\rm trunc}$ is approximately $t_{\rm relax} \approx 1.5\times 10^5 r_{\rm g}/c$ for RIAF and $t_{\rm relax} \approx 3.5\times10^8 r_{\rm g}/c$ for SEAF. The state shown in Fig.~\ref{fig:field_line} corresponds to a later time compared to $t_{\rm relax}$.
The magnetic field above the disc has already become a potential field.

\begin{figure*}
    \centering
    \includegraphics[keepaspectratio, width=0.8\columnwidth]{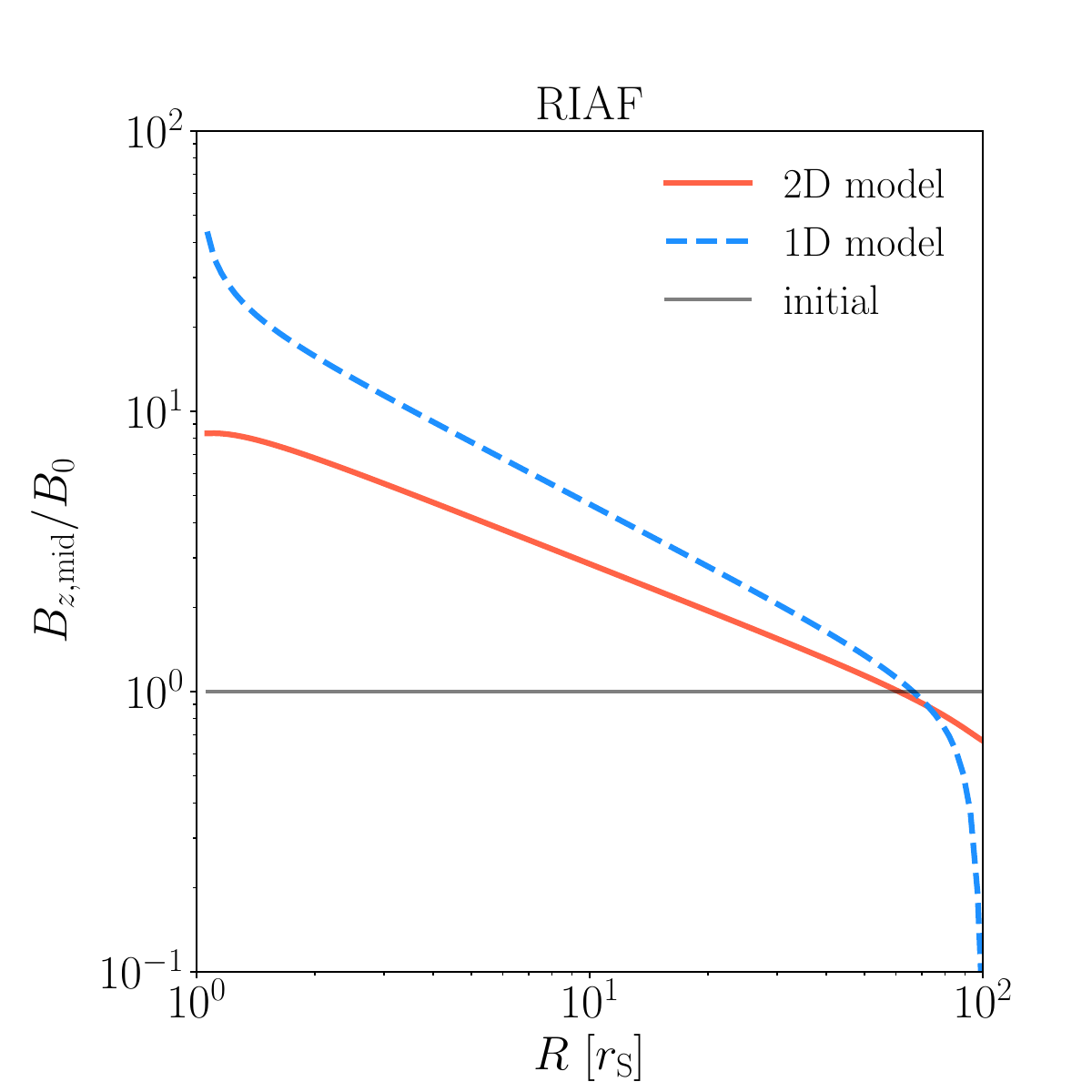}
    \includegraphics[keepaspectratio, width=0.8\columnwidth]{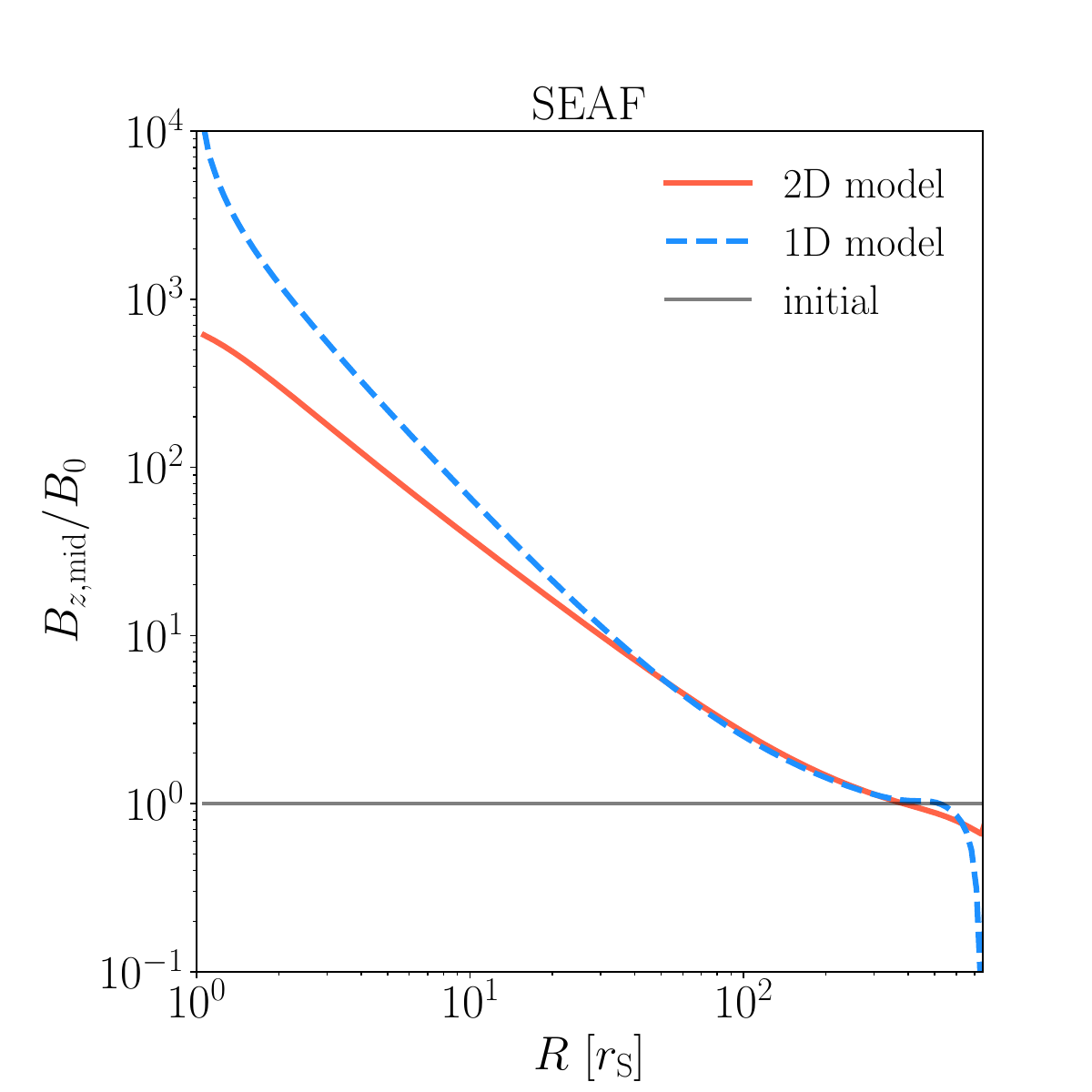}
    \includegraphics[keepaspectratio, width=0.8\columnwidth]{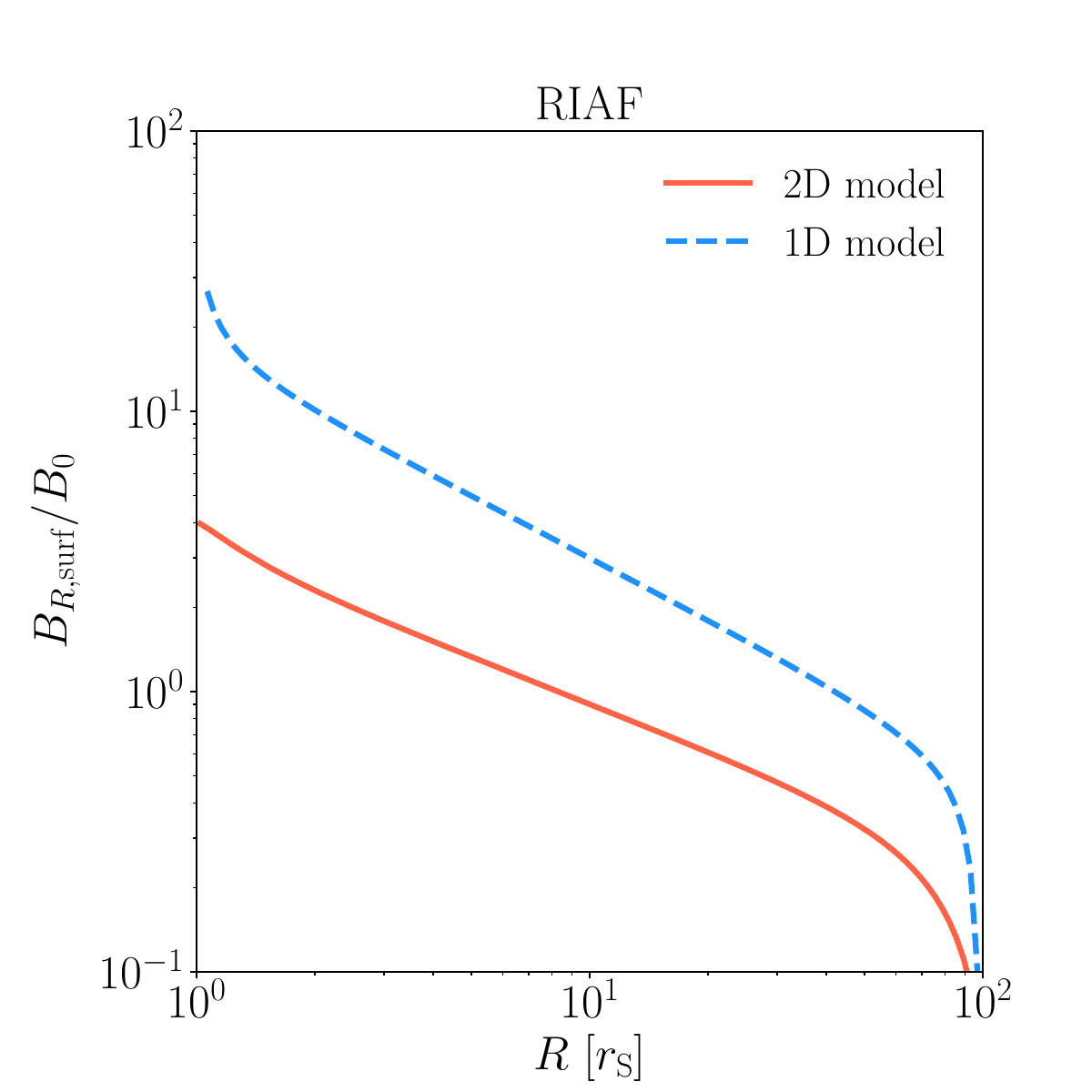}
    \includegraphics[keepaspectratio, width=0.8\columnwidth]{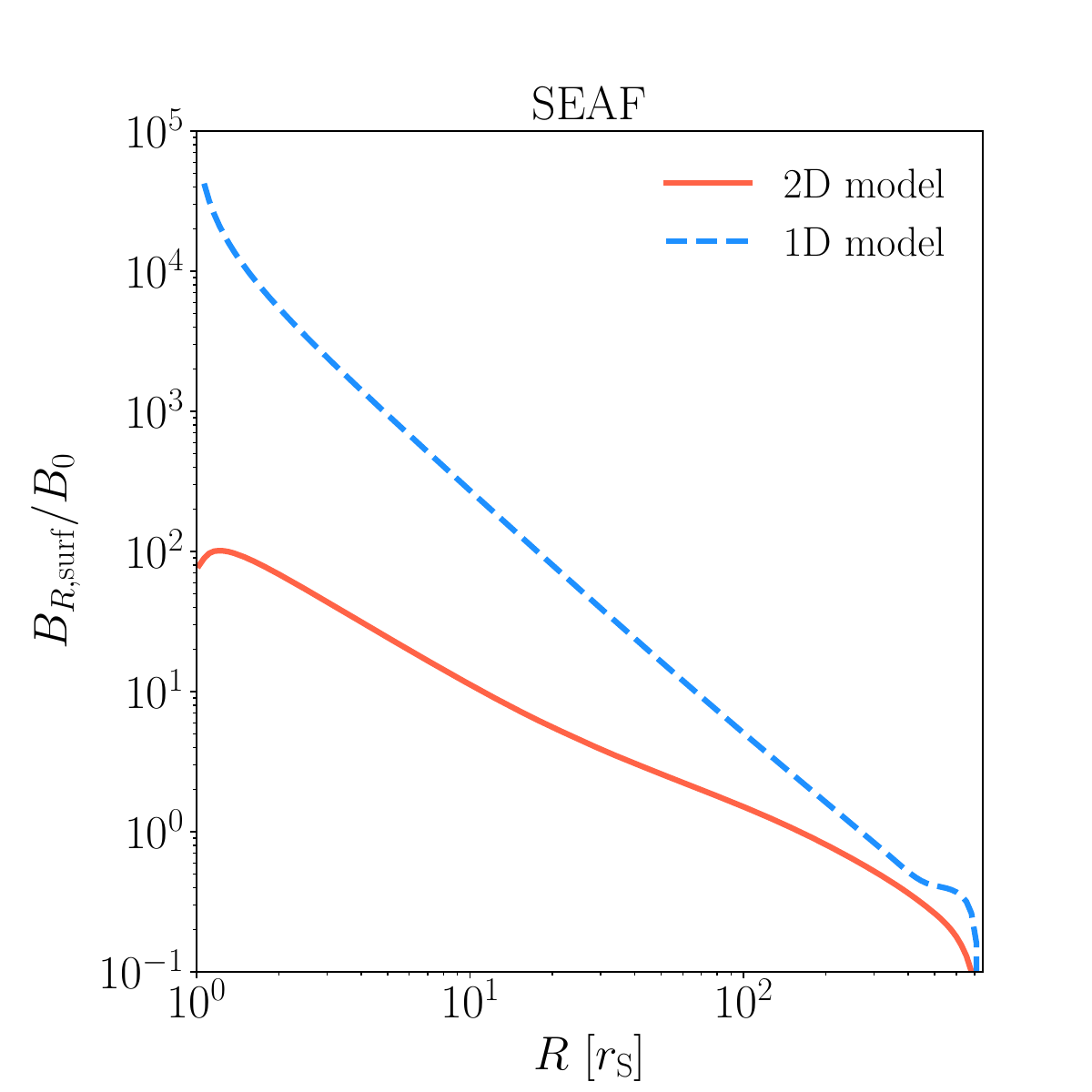}
  \caption{
  Comparison between the 1D and 2D models. The red solid lines denote the 2D results, while the blue dashed lines indicate the 1D results. The top and bottom panels show the radial distribution of $B_{z,\rm mid}$ and $B_{R, \rm surf}$, respectively. The horizontal grey lines in the top panels show the initial profile of $B_{z,\rm mid}$.
  }
  \label{fig:1D2DBz_BR}
\end{figure*}

Fig.~\ref{fig:1D2DBz_BR} compares the field distributions between the 1D and 2D models.
The top panels show the vertical component at the disc equator, $B_{z, \rm mid} = B_z(R, 0)$, and the bottom panels display the radial component at the disc surfaces, $B_{R, \rm surf} = B_R(R, H)$.
For the 1D model, we use the relation $B_{R, \rm surf} = K_\phi/2$ (equation \ref{eq:1DBRs}).
The figure demonstrates significant differences between the 1D and 2D models.
Both $B_{z, \rm mid}$ and $B_{R, \rm surf}$ are smaller in the 2D model. Furthermore, we find that the distribution of $B_{z, \rm mid}$ is more diffusive in the 2D model.

\begin{figure}
    \centering
    \includegraphics[keepaspectratio, width=0.6\columnwidth]{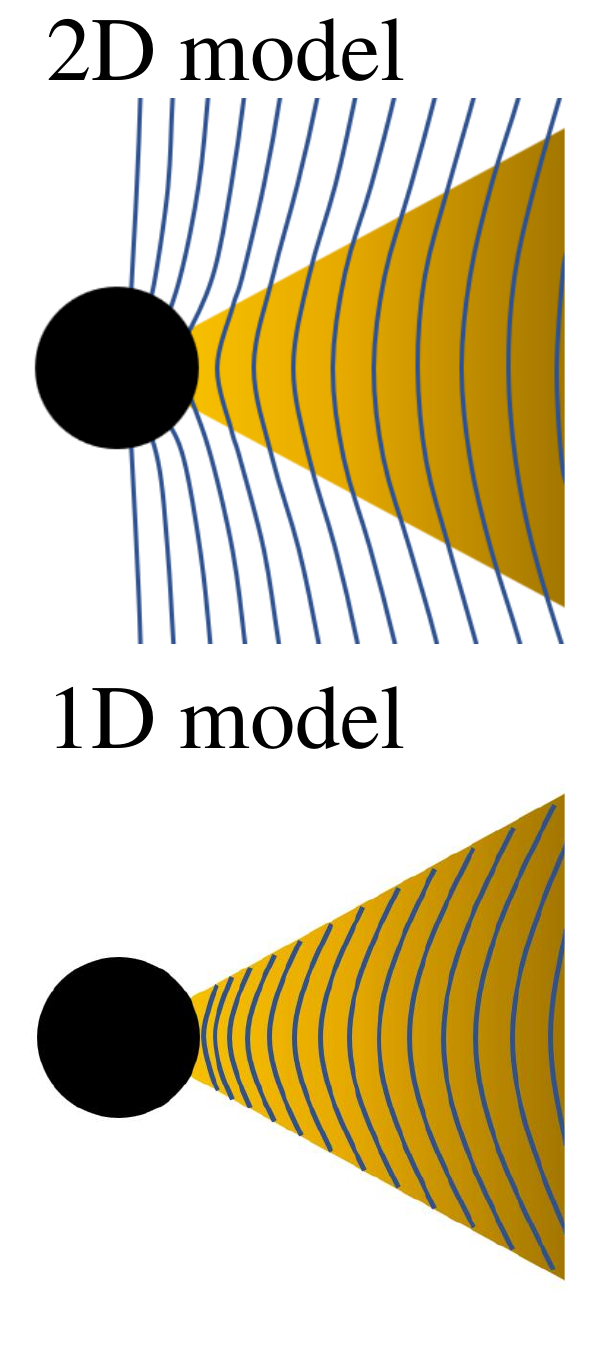}
  \caption{
Schematic pictures of the magnetic field structures just around the inner boundary in the 2D (top) and 1D (bottom) models. Note that in the 1D model, the poloidal field accumulates more significantly at the inner edge of the disc and has a larger inclination than in the 2D model.
    }
  \label{fig:1D2DInnerSchm}
\end{figure}

The 1D model exhibits a steeper $B_{z, \rm mid}$ distribution near the inner boundary than the 2D model. 
This steep structure is seen in previous studies \citep[e.g.,][]{Lubow1994, Okuzumi2014, Guilet2014}. The 1D model does not allow the magnetic field to go inside the inner boundary, which results in the accumulation of the field there. On the other hand, in the 2D model, the magnetic field can spread around the polar regions (Fig.~\ref{fig:field_line}), which leads to a smoother distribution.
Fig.~\ref{fig:1D2DInnerSchm} illustrates the differences in the magnetic field distribution between the 1D and 2D models near the inner boundary.
Fig.~\ref{fig:1D2DBz_BR} indicates that the 1D model can significantly overestimate the magnetic field strength near the central object. This result is important when discussing the injection of a magnetic field into the central object.

\subsubsection{Importance of multi-dimensional effects}\label{subsec:multi-dim-effects}

\begin{figure}
    \centering
    \includegraphics[keepaspectratio, width=0.8\columnwidth]{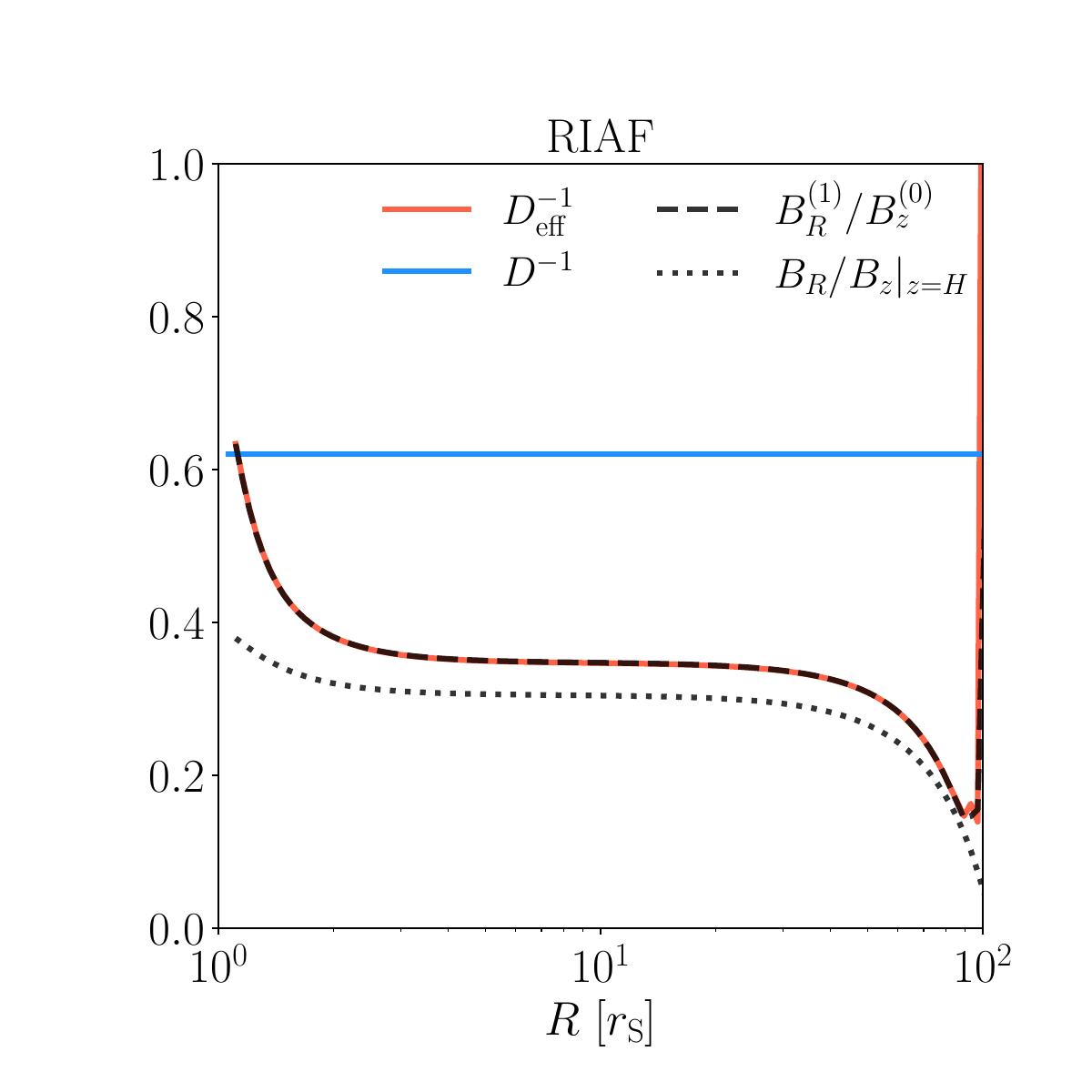}
    \includegraphics[keepaspectratio, width=0.8\columnwidth]{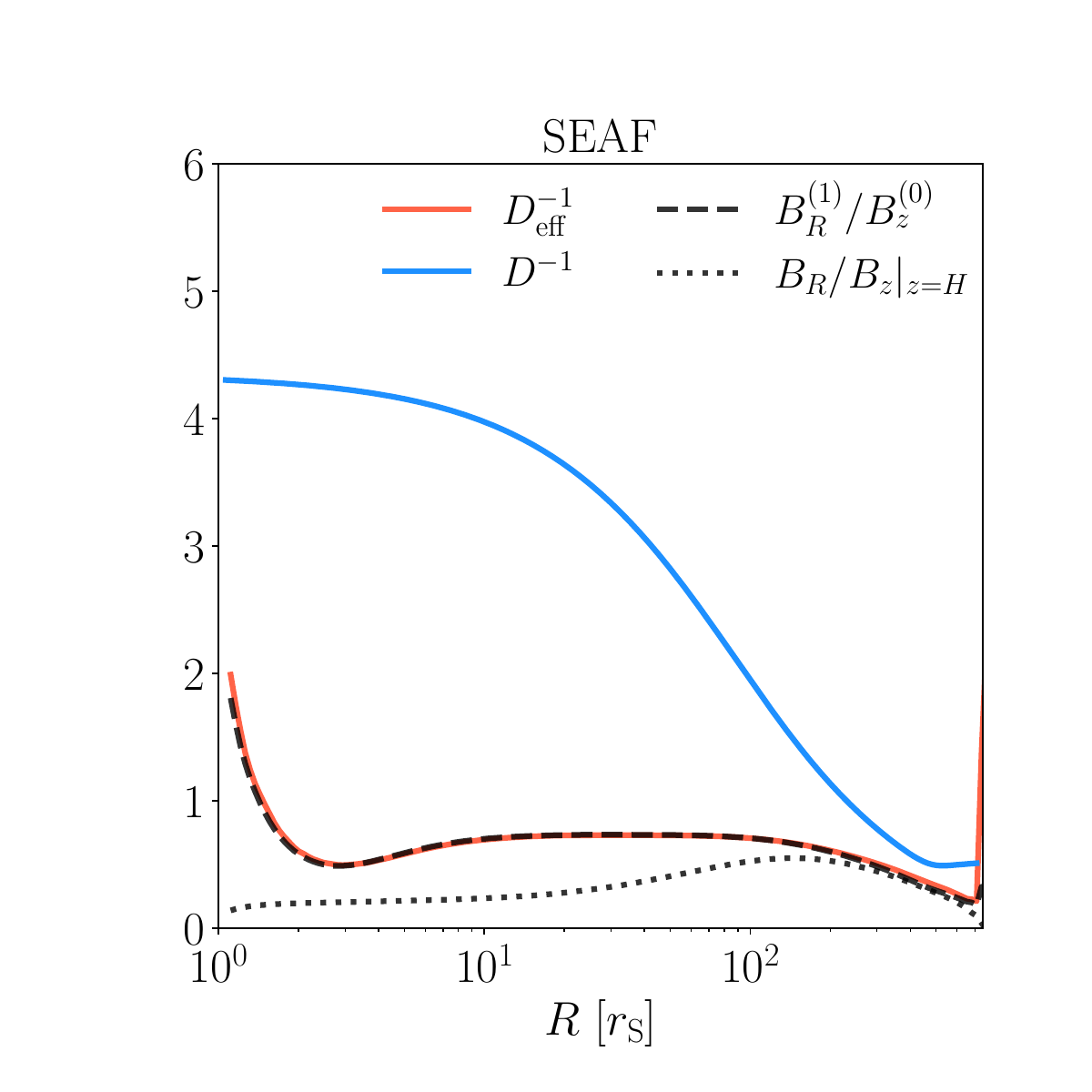}
  \caption{
  Comparison of the field inclination with the parameters $D^{-1}$ and $D_{\rm eff}^{-1}$. The dashed lines denote the field inclination defined by the linear components of the field, $B_R^{(1)}/B_z^{(0)}$. The dotted lines indicate the field inclination $B_R/B_z |_{z=H}$. The red lines denote $D_{\rm eff}^{-1}$, while the blue lines show $D^{-1}$. The results are for 2D models with $P_{\rm m} = 1$.
  } \label{fig:D_eff}
\end{figure}

We analyse the 2D results to study the multi-dimensional effects.
Fig.~\ref{fig:D_eff} show the radial distributions of $D$ and $D_{\rm eff}$ and compare them with the inclination of the field.
The figure confirms the theoretical prediction of $B_R^{(1)}/B_z^{(0)}=D_{\rm eff}^{-1}$ (equation \ref{eq:BR_Bz_mod}), where the lowest-order terms $B_R^{(1)}$ and $B_z^{(0)}$ are obtained by fitting the magnetic field distribution with a 15th-order function in terms of $\zeta(=z/H)$.
In the case of RIAF, we find $B_R/B_z|_{z=H}\approx D_{\rm eff}^{-1}$ except near the inner boundary (see black dotted and red lines). In the case of SEAF, we find a greater mismatch between $B_R/B_z|_{z=H}$ and $D_{\rm eff}^{-1}$. As we will see later, the mismatch is caused by higher-order components in the magnetic fields. Nevertheless, $D_{\rm eff}^{-1}$ provides a much better approximation for $B_R/B_z|_{z=H}$ (red line) than the classical estimate $D^{-1}$ (blue line) over a wide range of radius.
We note that $D^{-1}$ significantly overestimates the inclination in both disc models, which emphasizes the impact of the disc thickness. 

The result of $D_{\rm eff}>D$ indicates that the effective magnetic diffusivity is enhanced by multi-dimensional effects. 
The multi-dimensional effects are prominent only when $P_{\rm m}\lesssim 1$. We will see this in Section~\ref{subsec:Pm_dependences}.

\begin{figure}
    \centering
    \includegraphics[keepaspectratio, width=0.8\columnwidth]{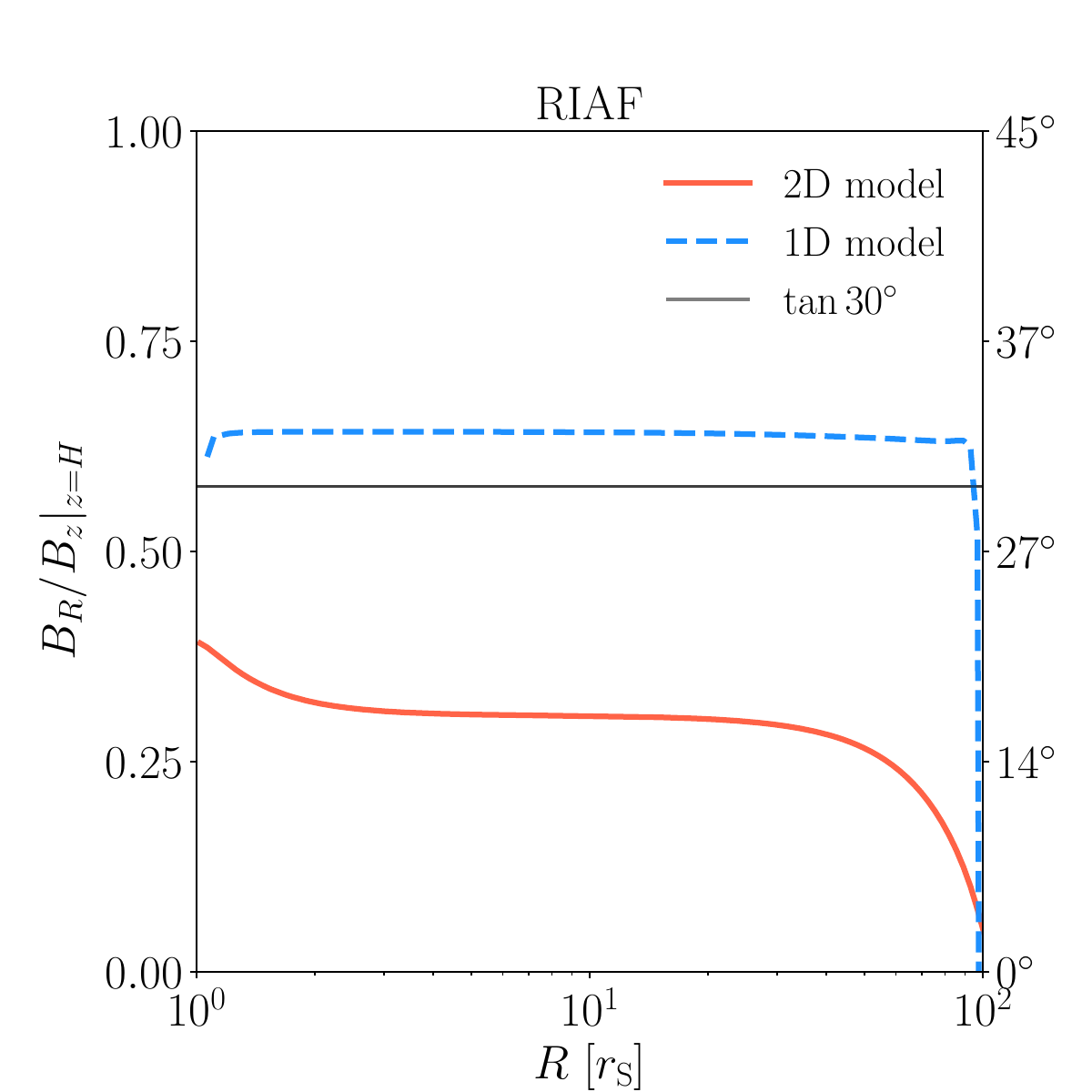}
    \includegraphics[keepaspectratio, width=0.8\columnwidth]{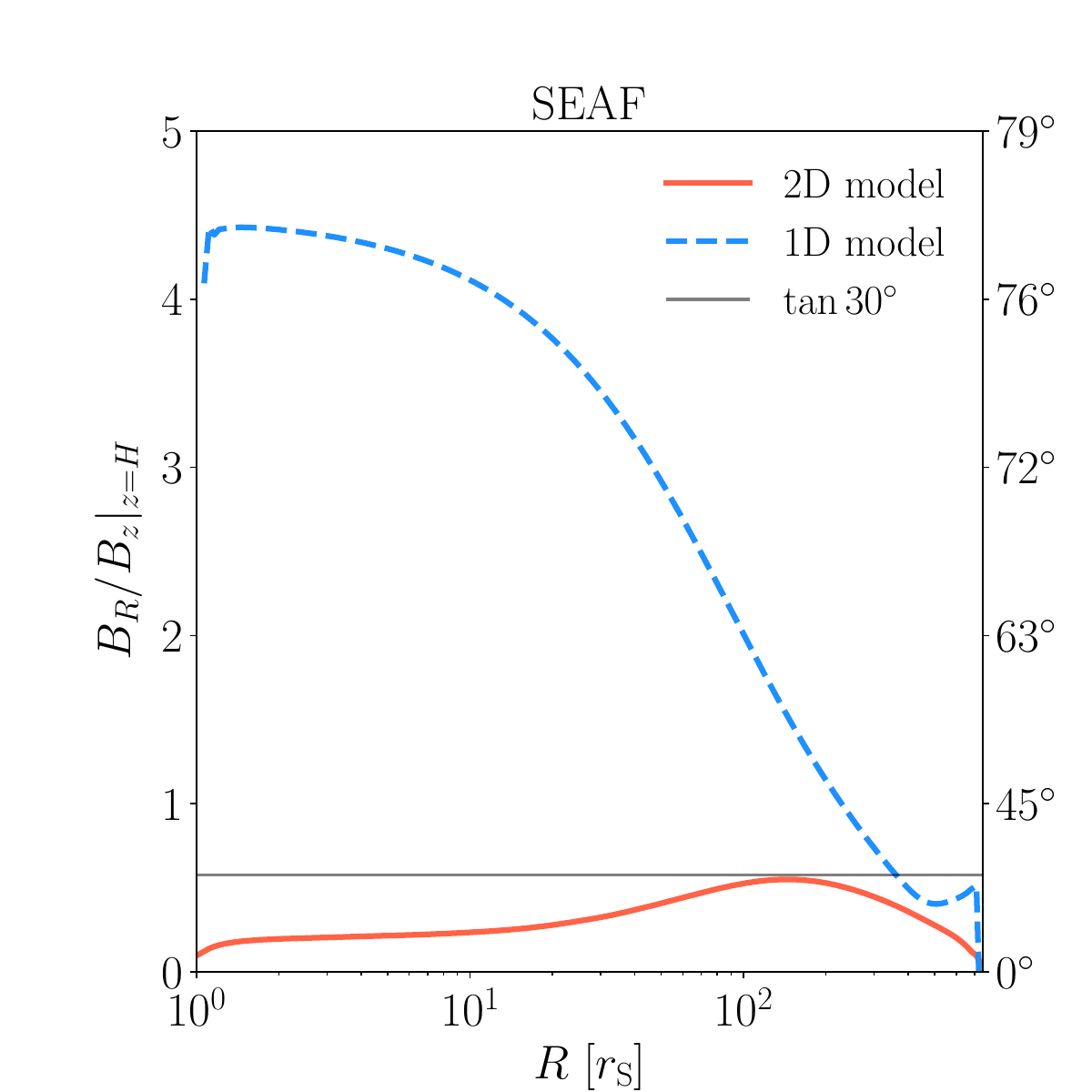}
  \caption{
  Comparison of $B_R/B_z |_{z=H}$ between the 1D and 2D models. The red solid lines denote the results for the 2D models, while the blue dashed lines show the results for the 1D models. The red lines in this figure correspond to the dotted lines in Fig.~\ref{fig:D_eff}.
  The horizontal grey lines denote the inclination angle of 30$^{\circ}$, above which the BP wind can blow. The results are for $P_{\rm m}=1$.
  }
  \label{fig:1D2DBRBz}
\end{figure}

The field inclination is a key factor for driving the magnetocentrifugal wind \citep[][hereafter BP wind]{Blandford1982}. 
If the field inclination is larger than $30^\circ$, the disc may drive the BP wind.
Fig.~\ref{fig:1D2DBRBz} compares the field inclination in the 1D and 2D models. As expected from Fig.~\ref{fig:D_eff}, the inclination is smaller in the 2D models. The 1D model predicts that the BP wind can appear in both RIAF and SEAF. However, in the 2D model, the condition for the BP wind is not satisfied for $P_{\rm m}=1$.

\begin{figure}
    \centering
    \includegraphics[keepaspectratio, width=0.9\columnwidth]{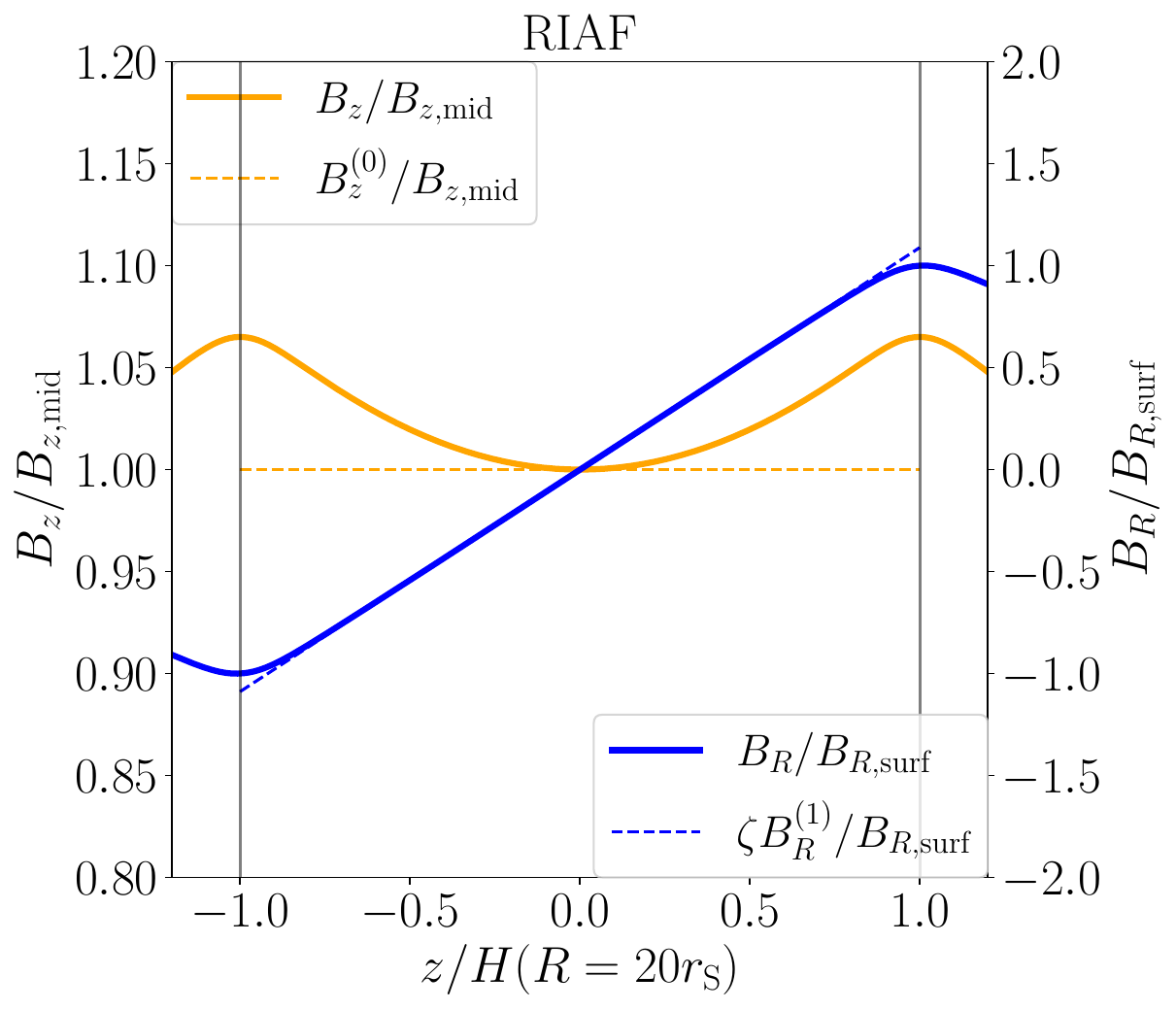}
    \includegraphics[keepaspectratio, width=0.9\columnwidth]{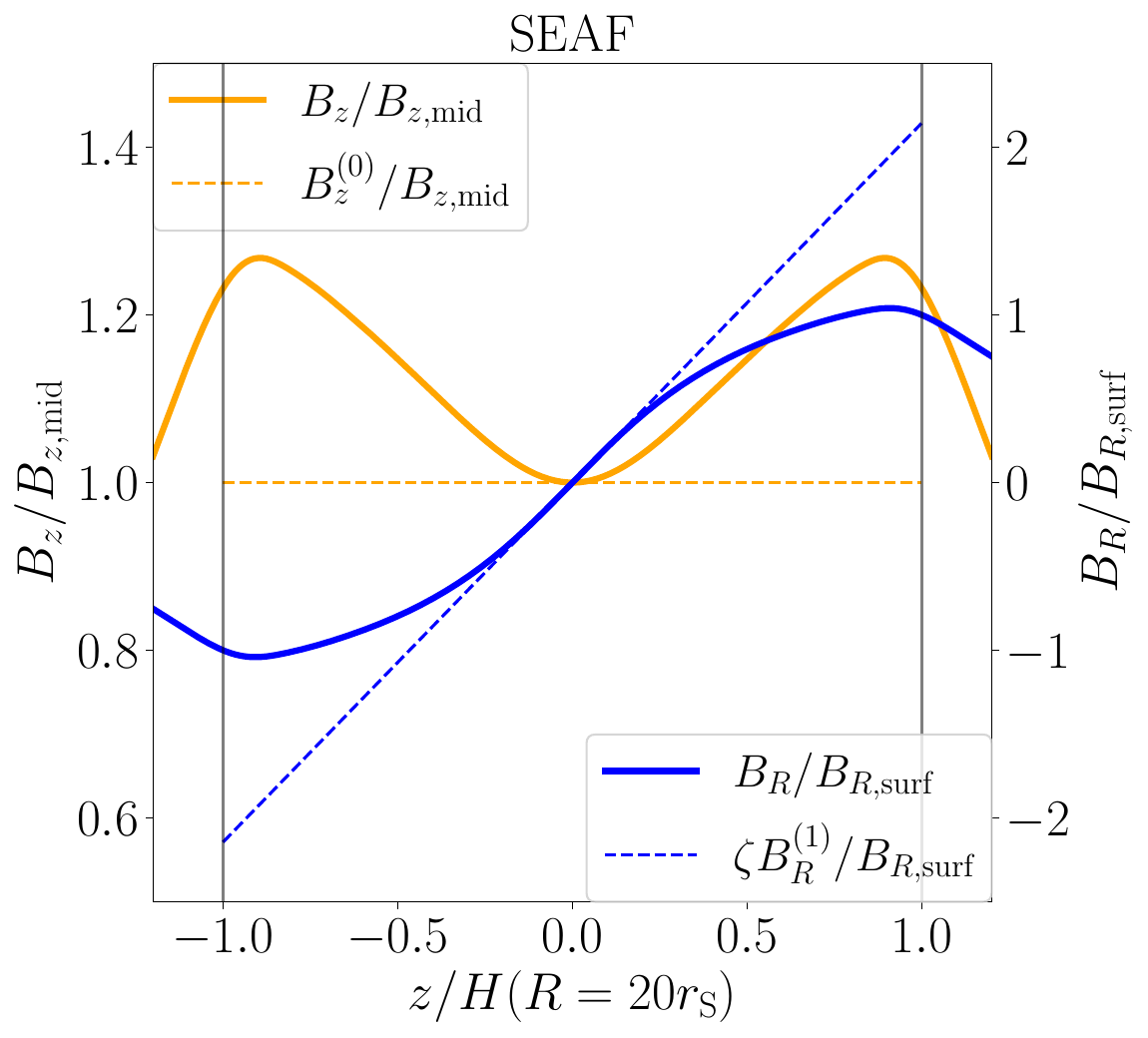}
  \caption{
  The vertical profiles of $B_R$ and $B_z$ at $R=20r_S$ in the 2D models. The orange and blue solid lines denote $B_z$ and $B_R$, respectively. The dashed lines indicate the linear terms of the field. The vertical black lines denote the height of the scale height, $z=H$ ($\zeta=1$). For SEAF, the radius of $R=20r_S$ is inside the slim region.
  } 
  \label{fig:2DBRBzVer}
\end{figure}

In Section~\ref{subsec:high_order_ana}, we show that the multi-dimensional effects of the disc thickness lead to nonlinear field profiles in the disc vertical direction. Fig.~\ref{fig:2DBRBzVer} examines the distributions of $B_R$ and $B_z$ at $R=20r_{\rm S}$ in the 2D model. For SEAF, the profile in the slim region is shown.
The solid lines represent the numerical results, and the dashed lines represent the lowest-order terms obtained by fitting the magnetic field distribution with a 15th-order function in terms of $\zeta(=z/H)$ (corresponding to $B_R^{(1)}$ and $B_z^{(0)}$ in equations~(\ref{eq:B_R_expanded}) and (\ref{eq:B_z_expanded}), respectively).
In the case of RIAF, the 1D approximation almost holds around the equatorial plane, but $B_z$ deviates from the dashed line near the disc surfaces. The breakdown of the 1D approximation is more pronounced in SEAF. The higher-order components of the magnetic fields are significant even around the equatorial plane. SEAF shows larger deviations than RIAF because it has a larger aspect ratio $h$ (RIAF has $h \approx 0.5$, while the slim region of SEAF has $h\approx3$).

\begin{figure}
    \centering
    \includegraphics[keepaspectratio, width=0.8\columnwidth]{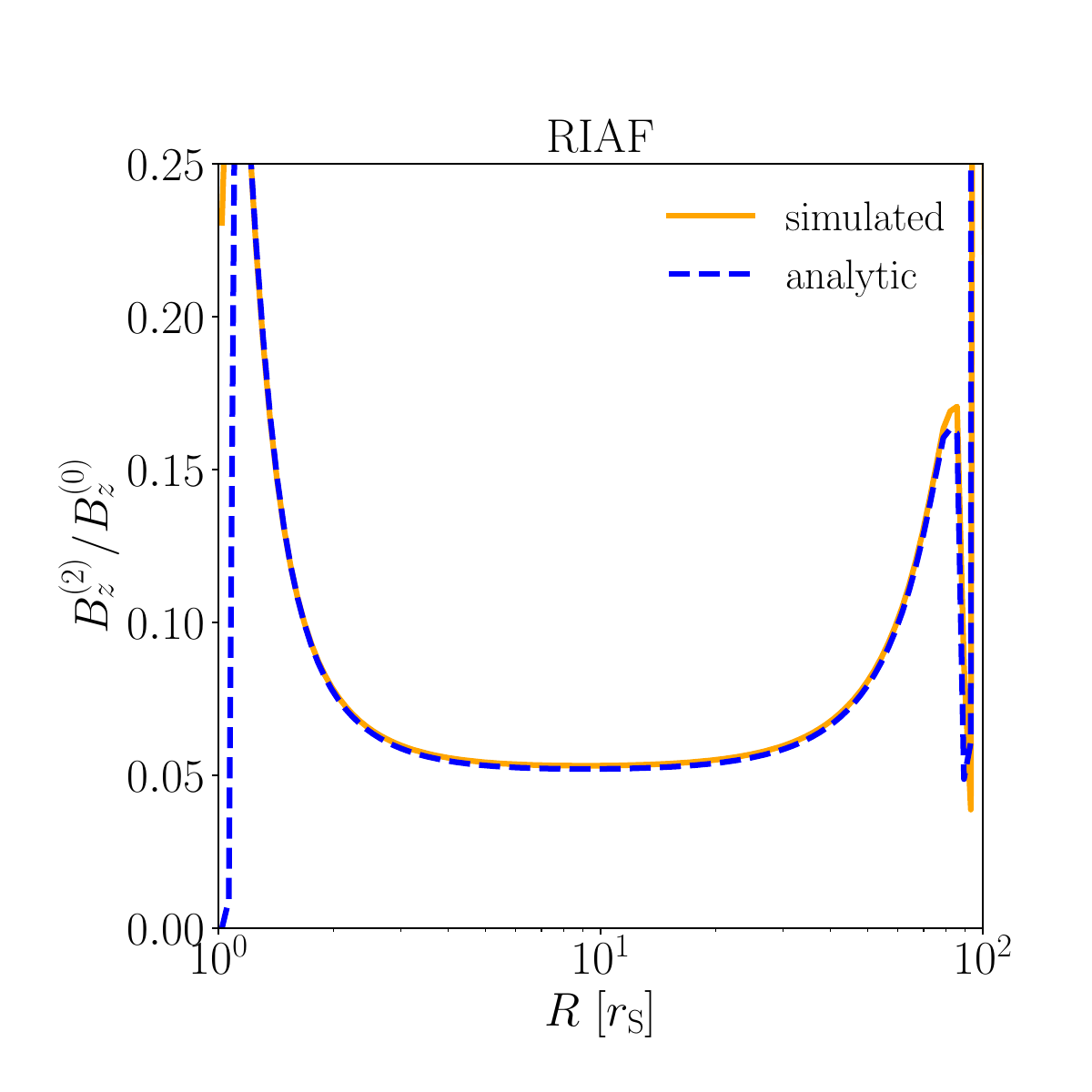}
    \includegraphics[keepaspectratio, width=0.8\columnwidth]{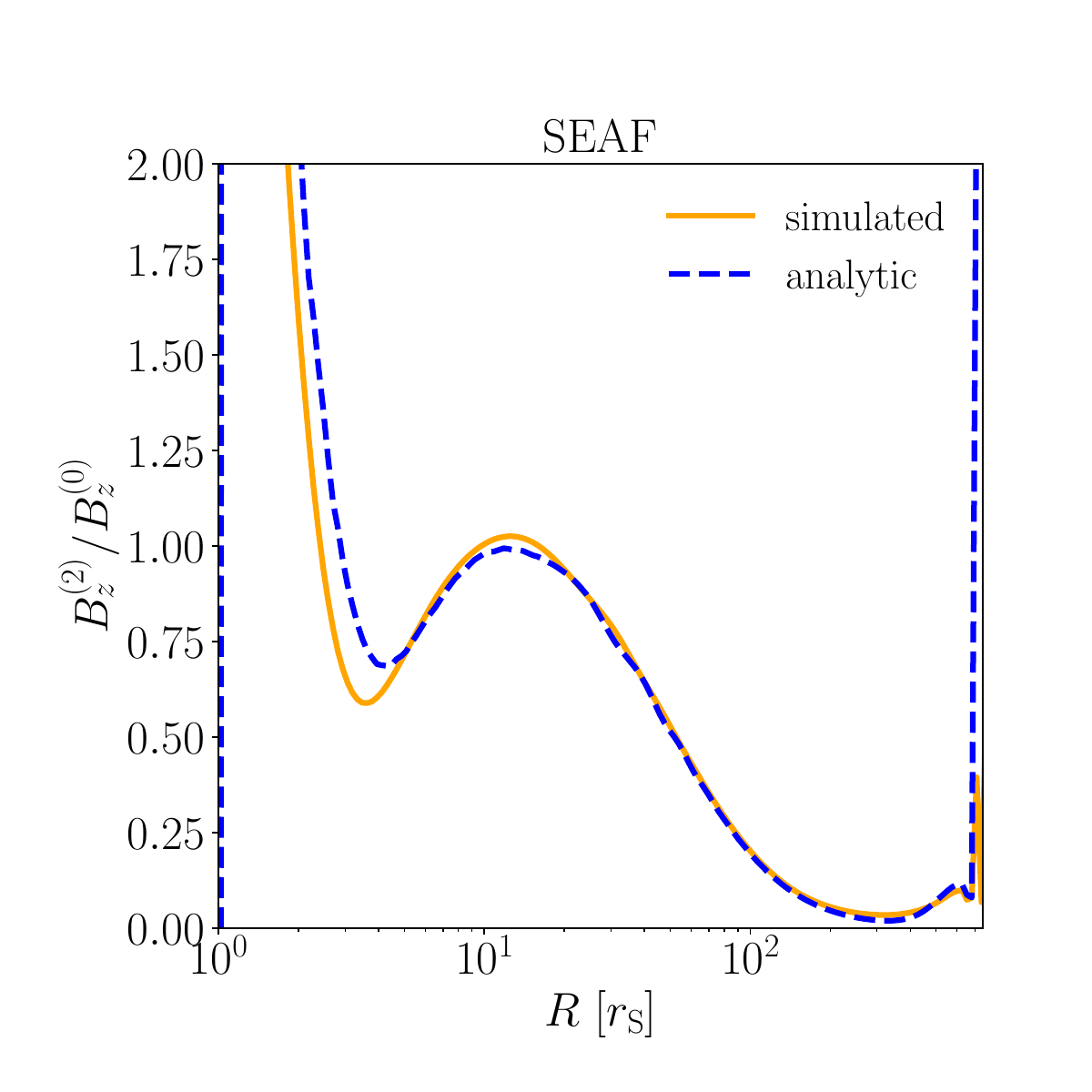}
  \caption{
The radial profile of $B_z^{(2)}/B_z^{(0)}$, which denotes the relative importance of the second-order term. The orange solid lines show the numerical results, and the blue dashed lines denote the analytic prediction (equation~\ref{eq:Bz2_Bz0}). The results are for $P_{\rm m}=1$.
  } 
  \label{fig:2DBz2_Bz0}
\end{figure}

The nonlinearity is examined by looking at $B_z^{(2)}/B_z^{(0)}$. Fig.~\ref{fig:2DBz2_Bz0} shows the plot of $B_z^{(2)}/B_z^{(0)}$ and the prediction of equation~(\ref{eq:Bz2_Bz0}). One can see that equation~(\ref{eq:Bz2_Bz0}) holds well at all radii except in the very vicinity of the inner boundary. This result demonstrates the validity of the discussion in Section~\ref{subsec:high_order_ana}, particularly the prediction of equation~(\ref{eq:Bz2_Bz0}) that $B_z^{(2)}/B_z^{(0)}\propto h^2$.

\begin{figure}
    \centering
    \includegraphics[keepaspectratio, width=0.8\columnwidth]{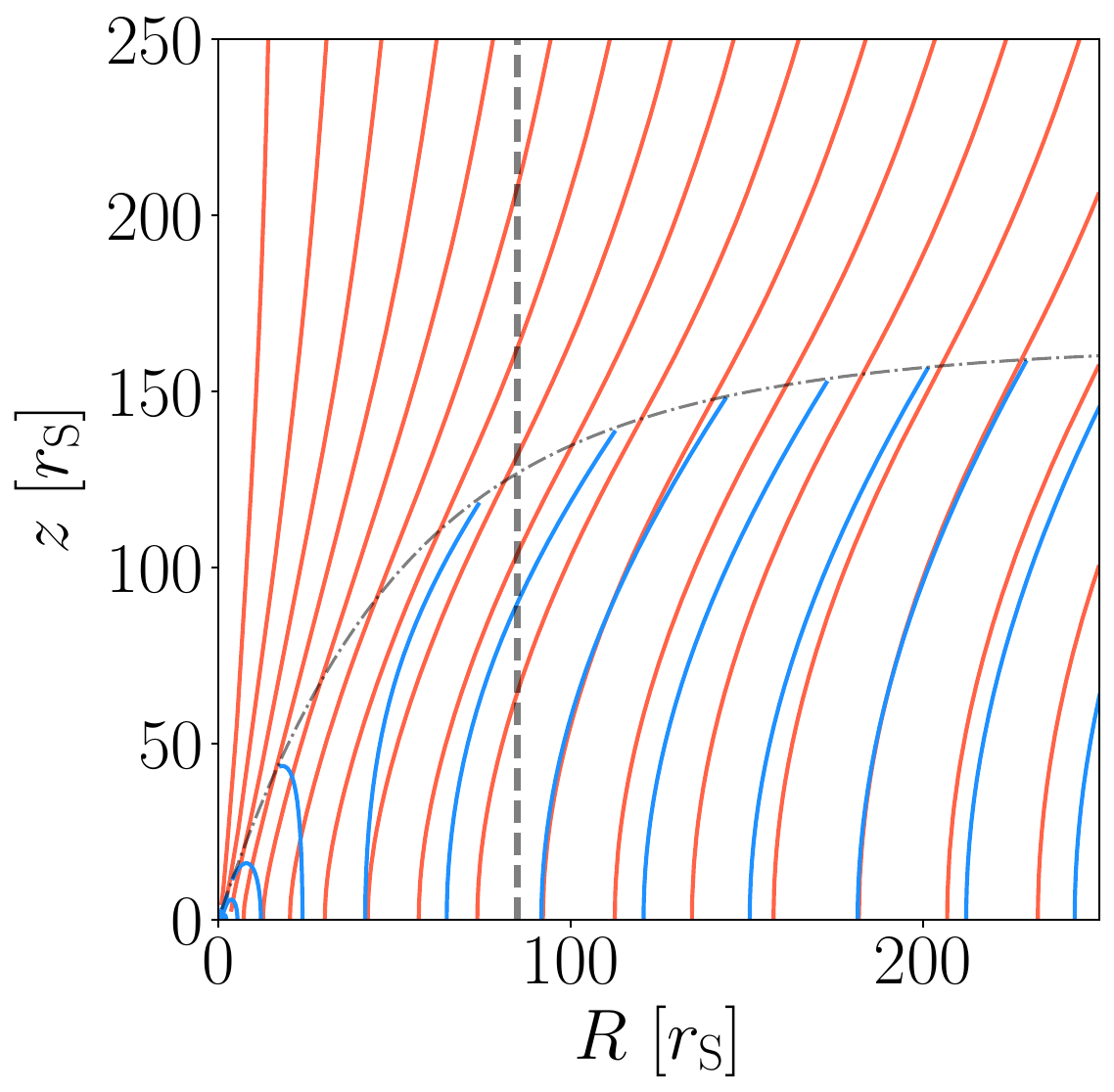}
  \caption{
  The poloidal field structures of SEAF obtained by the 1D (blue lines) and 2D (red lines) models.
  The dashed line shows the photon trapping radius ($R_{\rm trap}\approx85r_{\rm S}$).
  } \label{fig:field_line_slim_1D2D}
\end{figure}

\begin{figure}
    \centering
    \includegraphics[keepaspectratio, width=0.8\columnwidth]{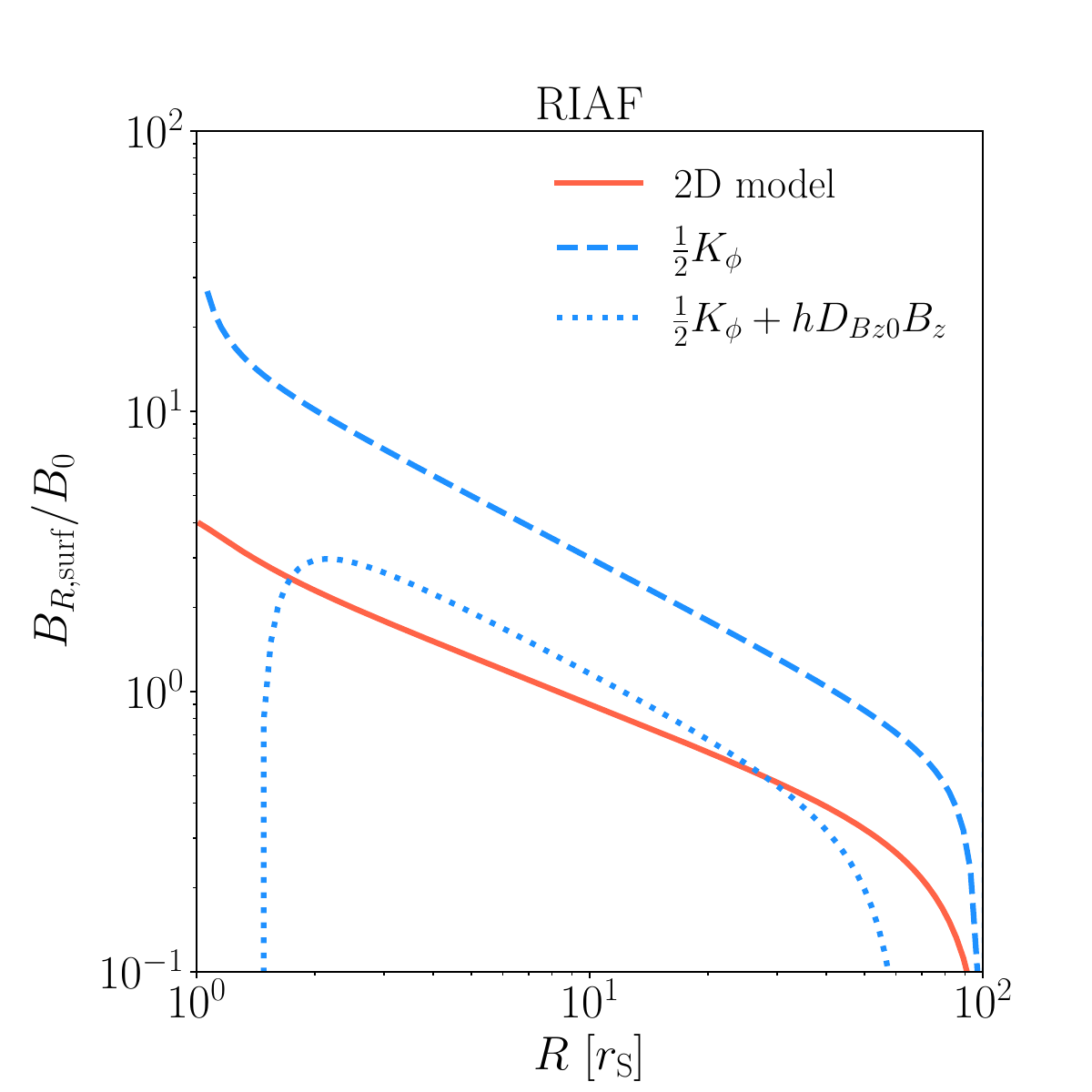}
    \includegraphics[keepaspectratio, width=0.8\columnwidth]{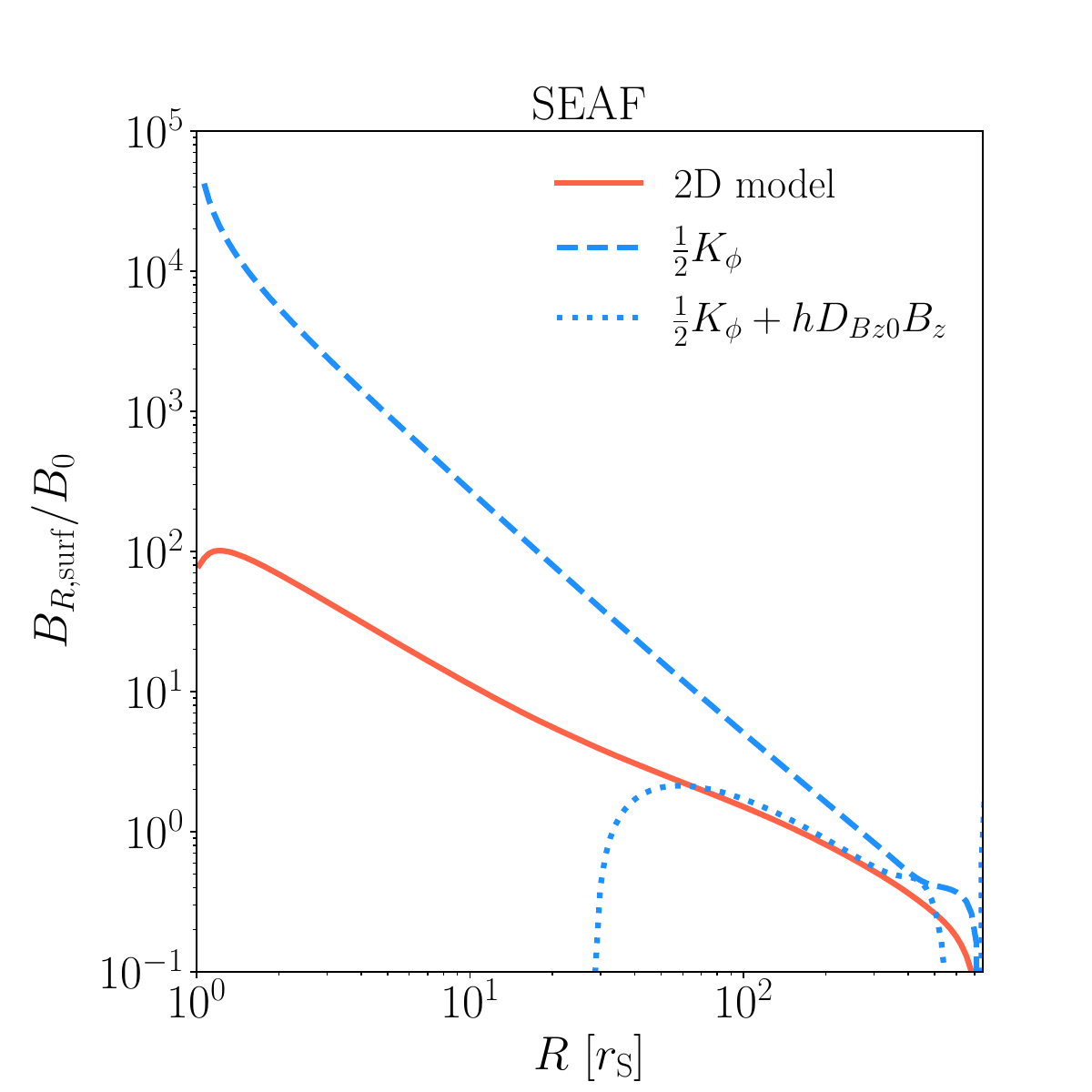}
  \caption{
  Comparison of the 2D models to the 1D models with different approximation levels. The red lines denote the field profiles obtained by the 2D models. The blue lines are for the 1D models. The dotted blue lines show the results based on equation~(\ref{eq:1DBRs_hDBz}), while the dashed blue lines display the results based on equation~(\ref{eq:1DBRs}). The latter ignores the effect of the radial gradient of the vertical field. The results are for $P_{\rm m}=1$.
  } \label{fig:1DBR_with_hDBz}
\end{figure}

As mentioned in Section~\ref{sec:1Dmodel}, the 1D model can partially take into account the multi-dimensional effects using equation~(\ref{eq:1DBRs_hDBz}).
We examine the applicability of this 1D approach through the comparison with the 2D model. Fig.~\ref{fig:1DBR_with_hDBz} compares the different predictions of the surface field, $B_{R,\rm surf}$. The blue dashed lines denote the results of the 1D model using the relation $B_{R, \rm surf} = K_\phi/2$ (equation~\ref{eq:1DBRs}). 
The blue dotted lines indicate the results of the 1D model using the relation $B_{R, \rm surf} = K_\phi/2 + hD_{B_z0}B_{z, \rm mid}$ (equation~\ref{eq:1DBRs_hDBz}), which takes into account the field gradient. The red solid lines show the results of the 2D models.
In the case of RIAF, the 1D prediction of equation~(\ref{eq:1DBRs_hDBz}) agrees well with the 2D result except near the boundaries. In the case of SEAF, the 1D prediction matches the 2D result only outside the slim region. 
We note that $B_{R,\rm surf}$ based on equation~(\ref{eq:1DBRs_hDBz}) is negative in the inner region. 
Fig. \ref{fig:field_line_slim_1D2D} compares the poloidal field structures of SEAF obtained by the 1D and 2D models. It is shown that the field near the centre is inclined inward, which is unrealistic.
In Appendix~\ref{app:multidimensional_effects}, we demonstrate for the SEAF model that the unphysical structure emerges because the 1D model cannot handle the multidimensional effects correctly even if the radial gradient of $B_{z,\rm mid}$ is considered in the calculation of $B_{R,\rm surf}$.

Fig. \ref{fig:field_line_slim_1D2D} shows that in the slim region (i.e. inside the photon trapping radius), the field structure of the 1D model significantly deviates from that of the 2D model, although they match in the standard disk region (outside $R_{\rm trap}$). This is because higher-order components in the magnetic fields are critically important to determine $B_{R, \rm surf}$ in the slim region.
For example, the fitting to our 2D result shows $B_{R}^{(i)}/B_{R}^{(1)}\approx -1.72, 4.30, -7.93, 8.45$ for $i=3, 5, 7, 9$, respectively, at $R=20r_S$. 

The following analysis demonstrates an example of the interplay among high-order terms.
Using equations (\ref{eq:BR1_Bz0}), (\ref{eq:BR3_Bz2}), and (\ref{eq:Bz2_Bz0}), we can express $B_R^{(3)}$ as follows:
\begin{align}
    \frac{B_R^{(3)}}{B_R^{(1)}} &= \frac{B_R^{(3)}}{B_z^{(2)}}\cdot \frac{B_z^{(2)}}{B_z^{(0)}}\cdot \left(\frac{B_R^{(1)}}{B_z^{(0)}}\right)^{-1}\\ \nonumber
    &= - \frac{1}{6}h^2\left(D_*^{-1}+D_{B_z2} -2D_H \right)\left(D_C+D_{B_z0}\right),
\end{align}
At $R=20r_S$ (inside the slim region), $h\approx 2.5$, $D_H\approx 1$, and $D_*^{-1} \approx 1.5$. $D_*^{-1} + D_{B0}\approx 0.3$.
In addition, the gradient of the high-order term $D_{B_z2}$ is approximately $-1.4$ and plays a role in increasing the absolute magnitude of $B_R^{(3)}/B_R^{(1)}$. As a result, $|B_R^{(3)}/B_R^{(1)}|\approx 1.7$, which shows the significance of the third-order component and explains the mismatch of the field structures of the two models (Fig. \ref{fig:field_line_slim_1D2D}). 
In the standard disc region, however, the high-order term is unimportant mainly owing to its thickness and weak radial dependence of $H$ ($D_H \ll 1$): $|B_R^{(3)}/B_R^{(1)}|\approx 3\times 10^{-2}$ at $R=200 r_{\rm S}$. This explains the good agreement between the field structures of the two models in that region.

\subsection{Parameter search}\label{subsec:parameter_search}

\subsubsection{Dependence on aspect ratio} \label{subsec:aspect_dependences}
As we are interested in the flux transport in thick discs, we study the dependence of the solution on the disc aspect ratio $h$, with a particular focus on RIAF.
In the RIAF solutions, when we set $f=1, 0.01, 0.001$, the disc aspect ratios are approximately $h\approx0.5, 0.1, 0.03$, respectively.
We use different mesh numbers for discs with different thicknesses to resolve the disc structure. Namely, $(N_r, N_{\theta}) = (128, 80),~(360, 200),~(800, 460)$ for the cases $f = 1, 0.01, 0.001$, respectively.
Again, we align the number of mesh points in the $R$ direction of the 1D model with the number of mesh points in the $r$ direction of the 2D model.

\begin{figure}
    \centering
    \includegraphics[keepaspectratio, width=0.8\columnwidth]{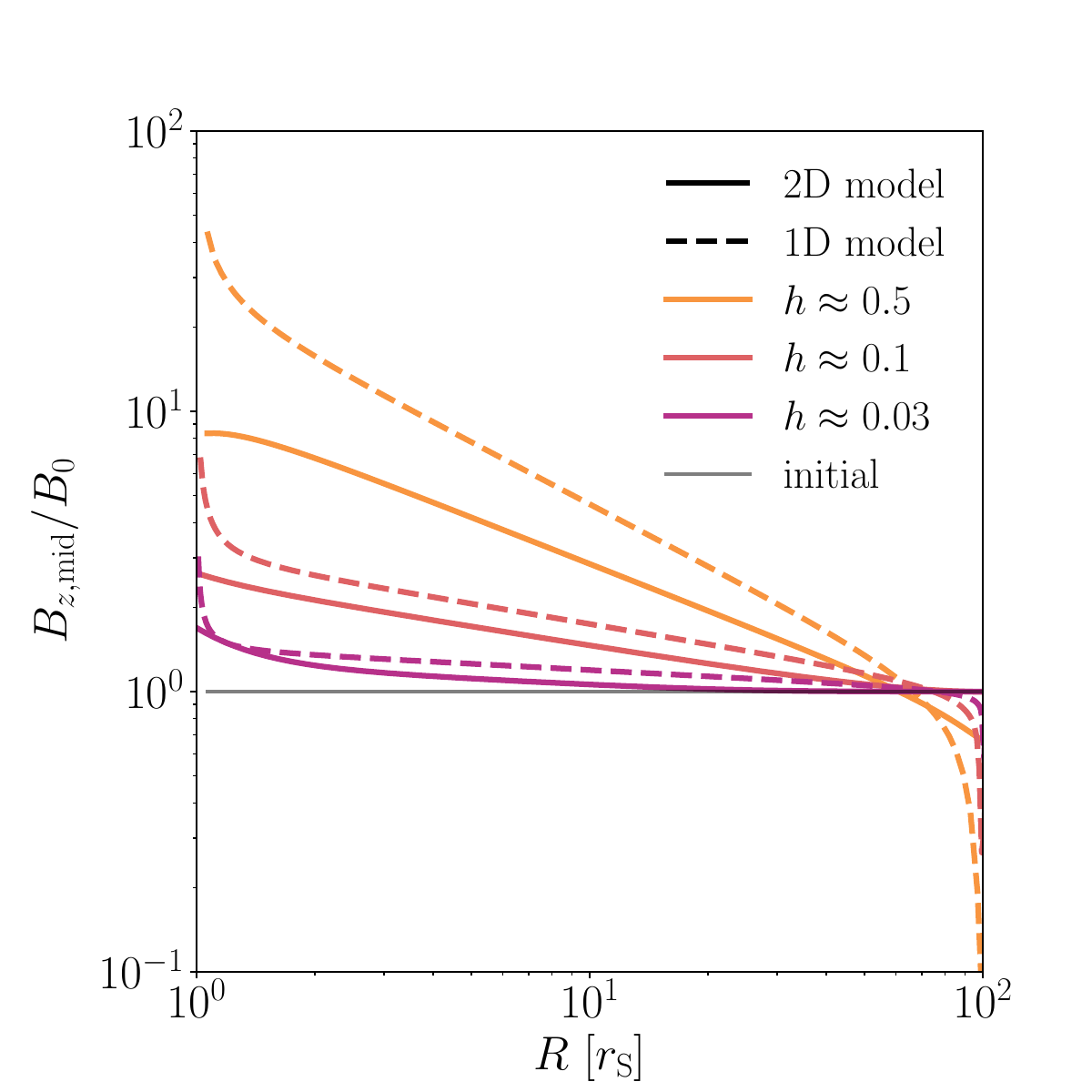}
    \includegraphics[keepaspectratio, width=0.8\columnwidth]{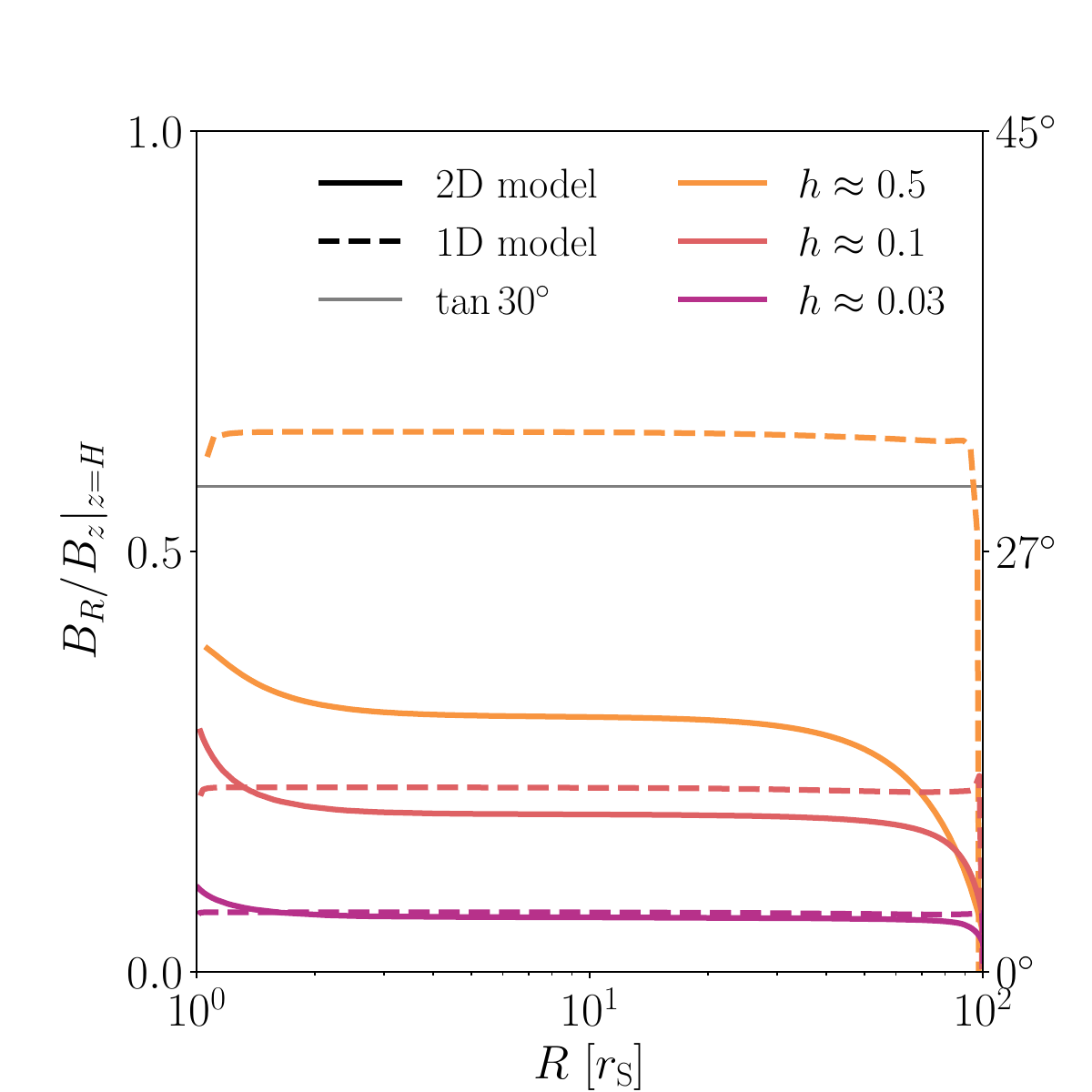}
    \caption{
    Dependence of the field structure on the disc aspect ratio $h$ for the RIAF case. The coloured solid lines show the results for 2D models, and the dashed lines display the 1D results. The grey line in the upper panel denotes the initial profile, while the grey line in the lower panel denotes the field inclination angle corresponding to 30$^{\circ}$. We set $P_{\rm m}=1$. The inclination for the 1D results is calculated using equation~(\ref{eq:1DBRs}).
    }
    \label{fig:RIAF_f}
\end{figure}

Fig.~\ref{fig:RIAF_f} presents the results with different $h$. The upper panel displays $B_{z, \rm mid}$.
For the thickest model ($h\approx 0.5$), the results differ significantly between the two models. However, the 2D model shows results close to the 1D model for $h\lesssim 0.1$ except near the inner boundary, which suggests that the assumption of the linear field distribution is valid for such thin discs.
The result that a thinner disc has a weaker magnetic field is also expected because a thinner disc shows a larger $D$ for a given $P_{\rm m}$.

The lower panel of Fig.~\ref{fig:RIAF_f} shows the inclination of the field at the disc surface. Again, the discrepancy between the 1D and 2D models is significant for the thick disc but is insignificant for thinner discs ($h \lesssim 0.1$). Therefore, the higher-order components of the magnetic field are unimportant for thin discs, and $D_{\rm eff}^{-1}\approx D^{-1}$ is a good measure of the field inclination (equation~\ref{eq:BRs_Bz}).

\subsubsection{Dependence on effective magnetic Prandtl number} \label{subsec:Pm_dependences}
3D MHD shearing box simulations suggest that the magnetic Prandtl number $P_{\rm m}$ is of order unity \citep{Guan2009, Lesur2009, Fromang2009, Kapyla2020}. In geometrically thick discs, the magnetic Prandtl number and $D$ are related as $D\sim (h P_{\rm m})^{-1} \sim P_{\rm m}^{-1}$ (see equation~\ref{eq:D_Pm}), which implies that the efficiency of magnetic flux transport strongly depends on the local quantity $P_{\rm m}$. Considering this, we investigate the $P_{\rm m}$ dependence of the solutions by performing a parameter search.

\begin{figure} 
    \centering
    \includegraphics[width=0.495\columnwidth]{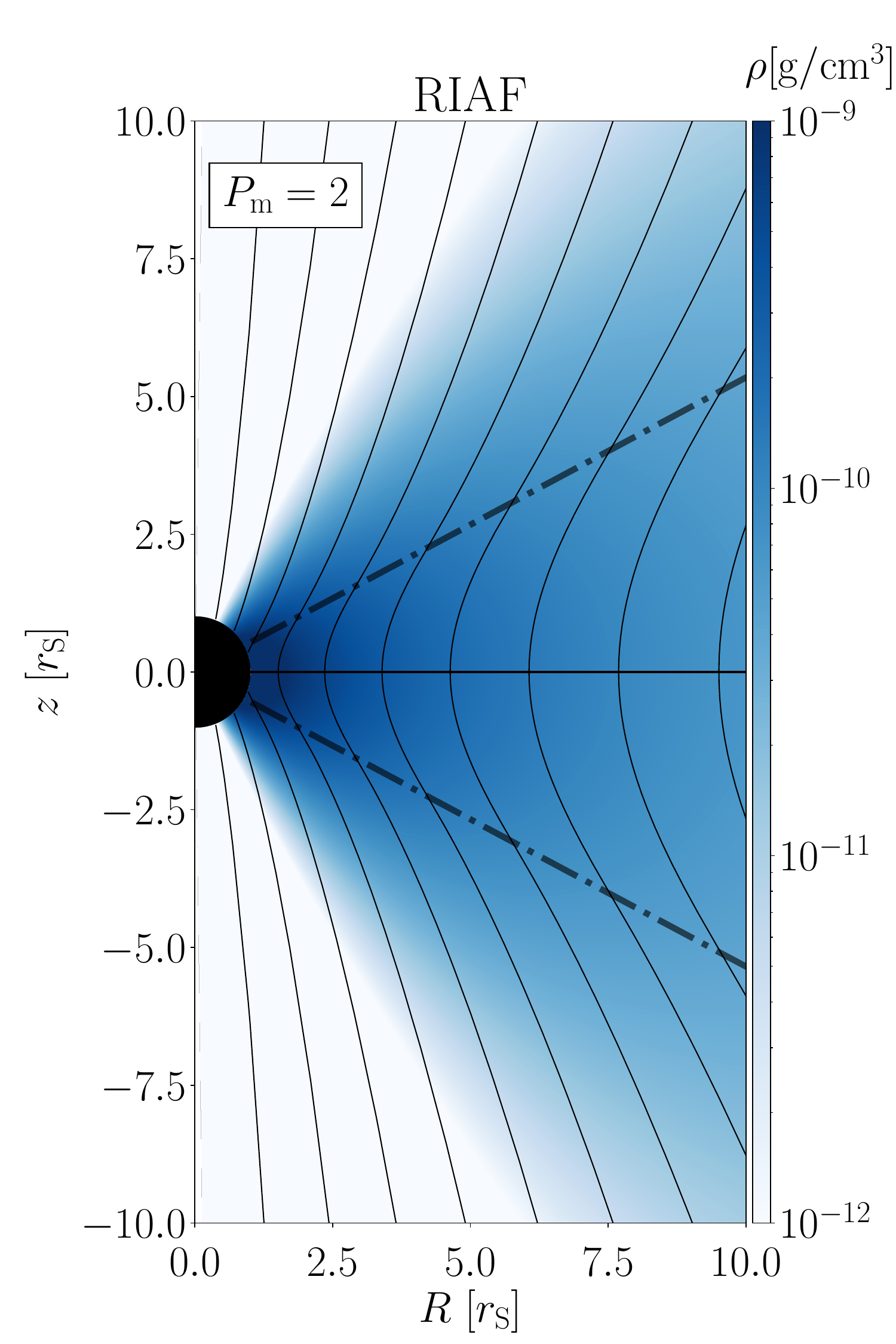}
    \includegraphics[width=0.495\columnwidth]{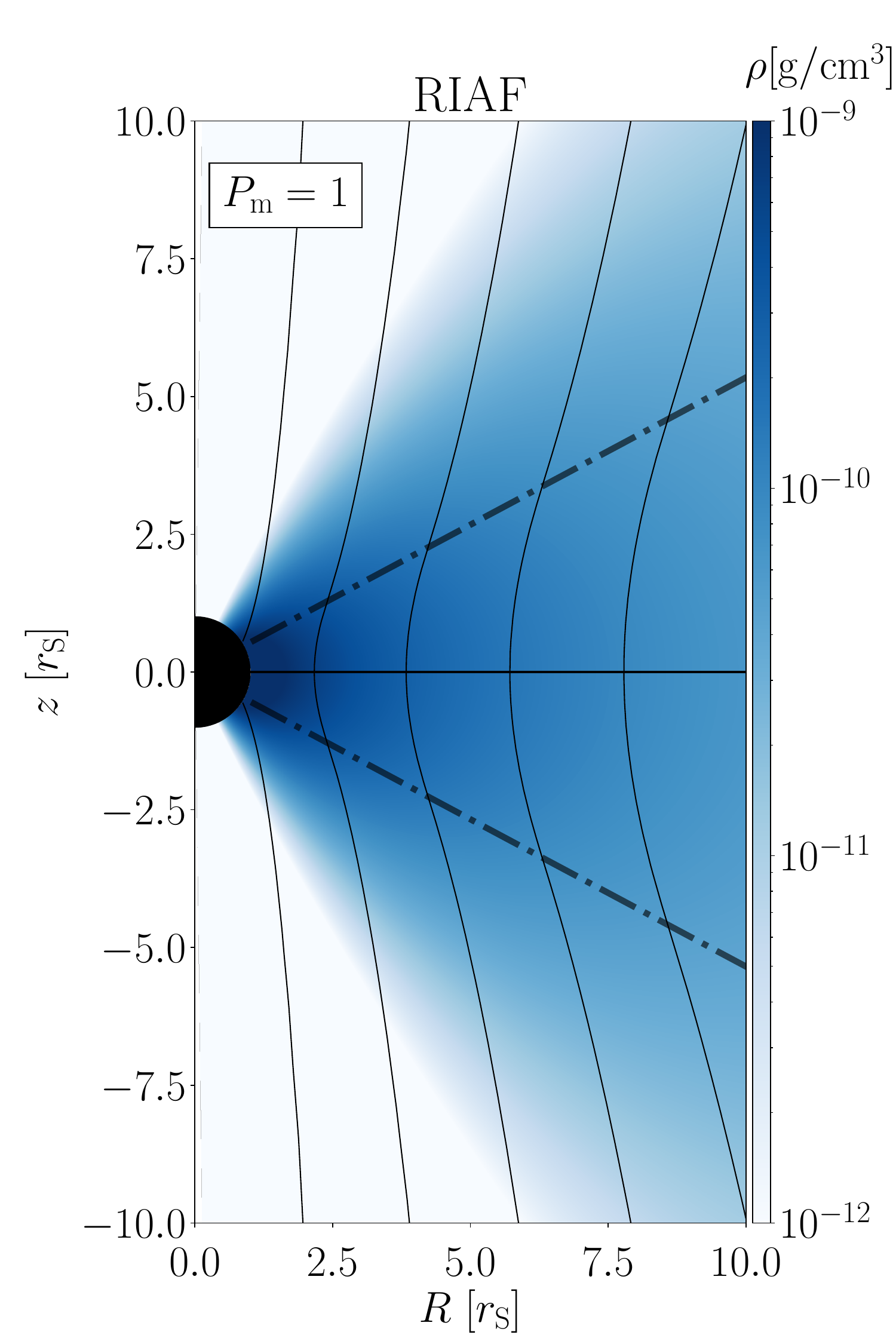}
    \includegraphics[width=0.495\columnwidth]{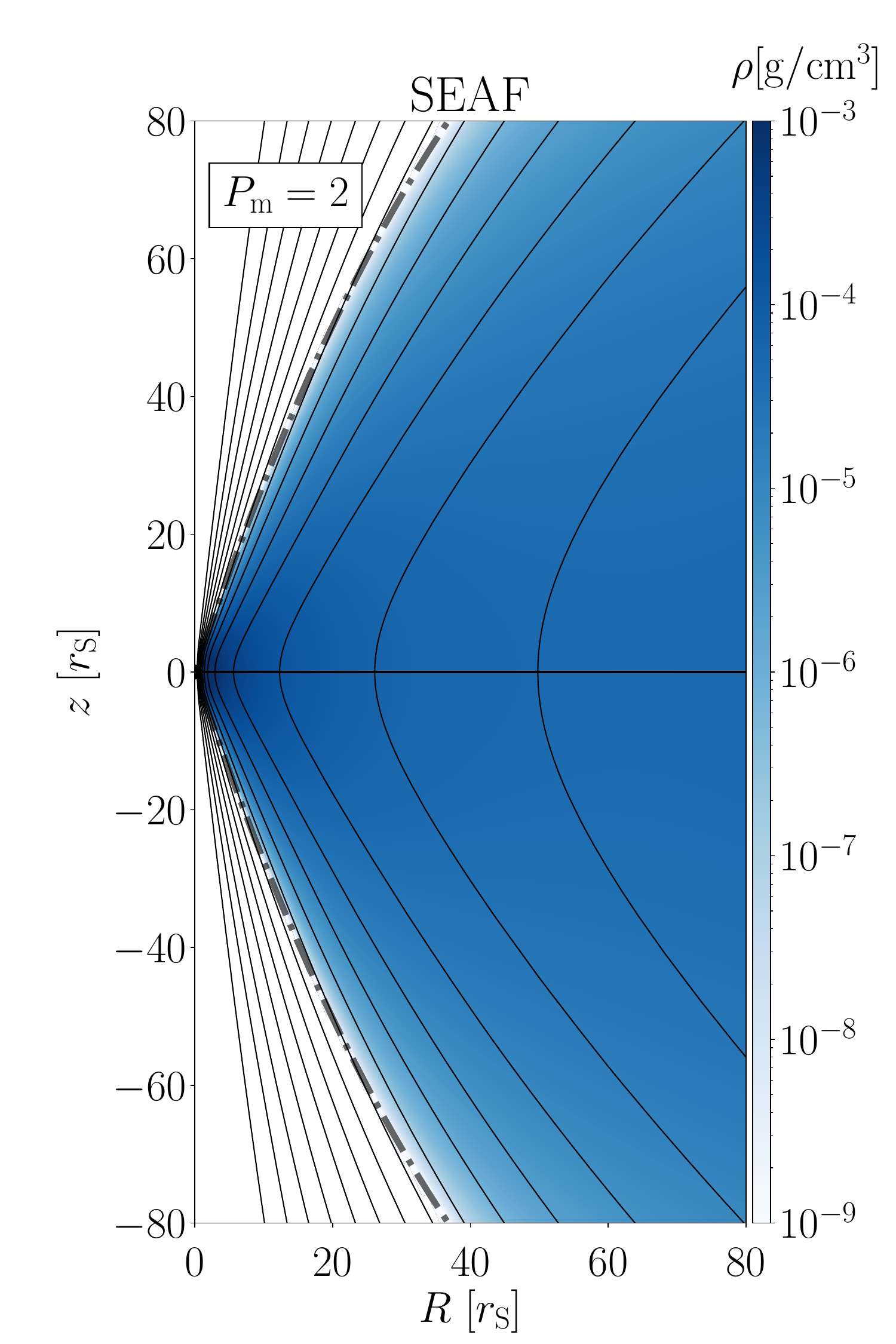}
    \includegraphics[width=0.495\columnwidth]{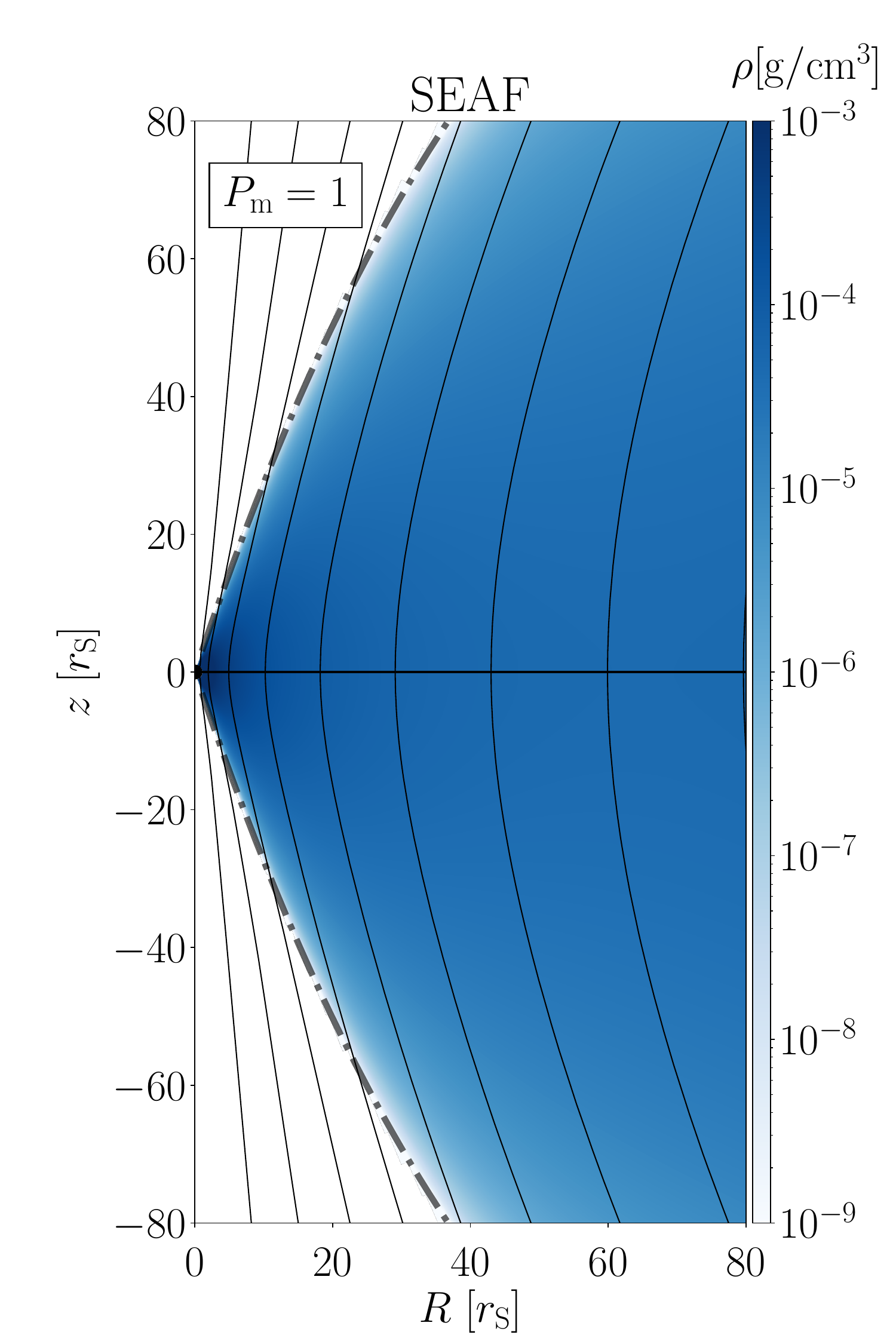}
  \caption{
The poloidal field structures for different $P_{\rm m}$. The left and right panels show the results for $P_{\rm m}=2$ and 1, respectively. The colour shows the density of the disc models, and the dashed lines denote the disc surface.
  }
  \label{fig:field_line_highPm}
\end{figure}

\begin{figure}
    \centering
    \includegraphics[keepaspectratio, width=0.8\columnwidth]{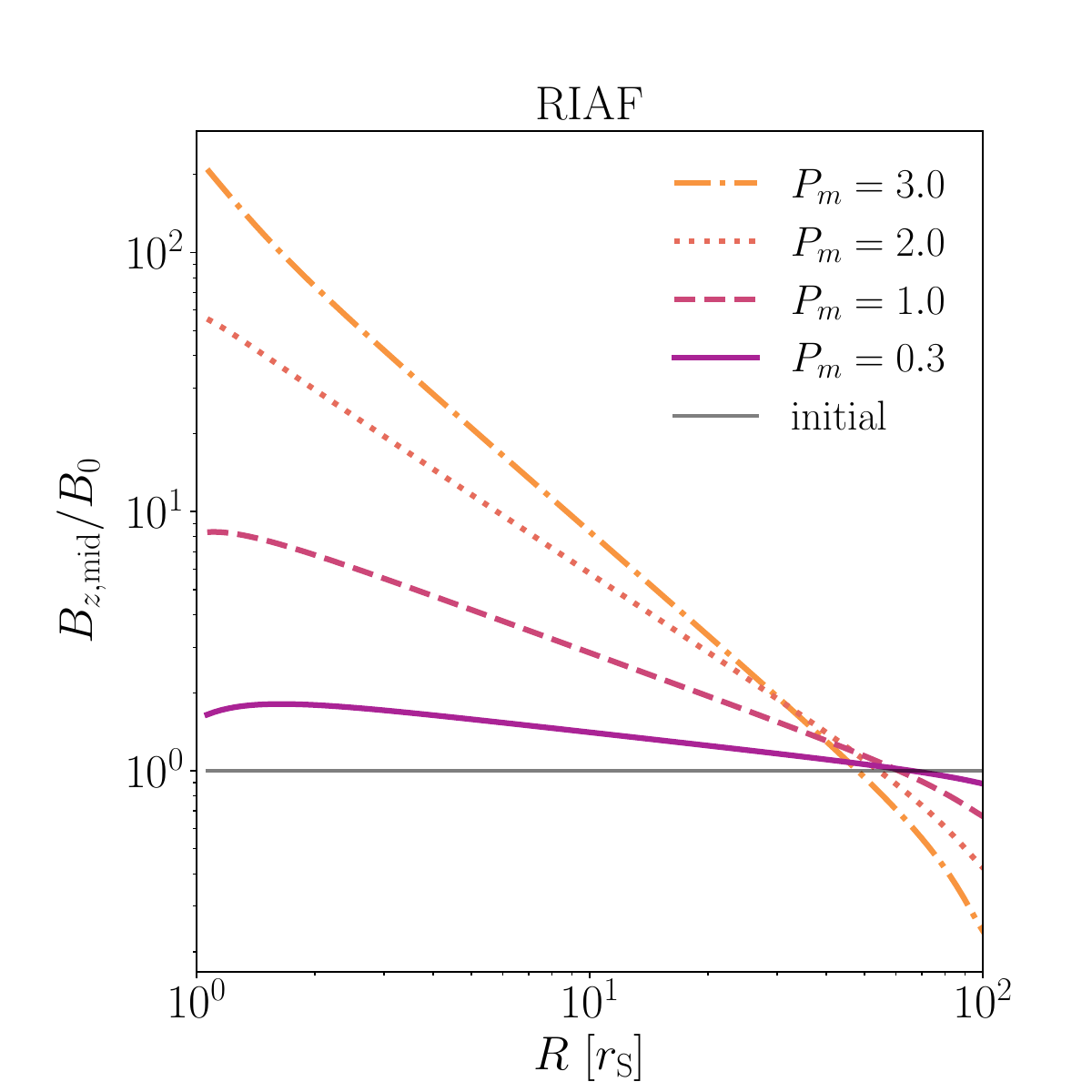}
    \includegraphics[keepaspectratio, width=0.8\columnwidth]{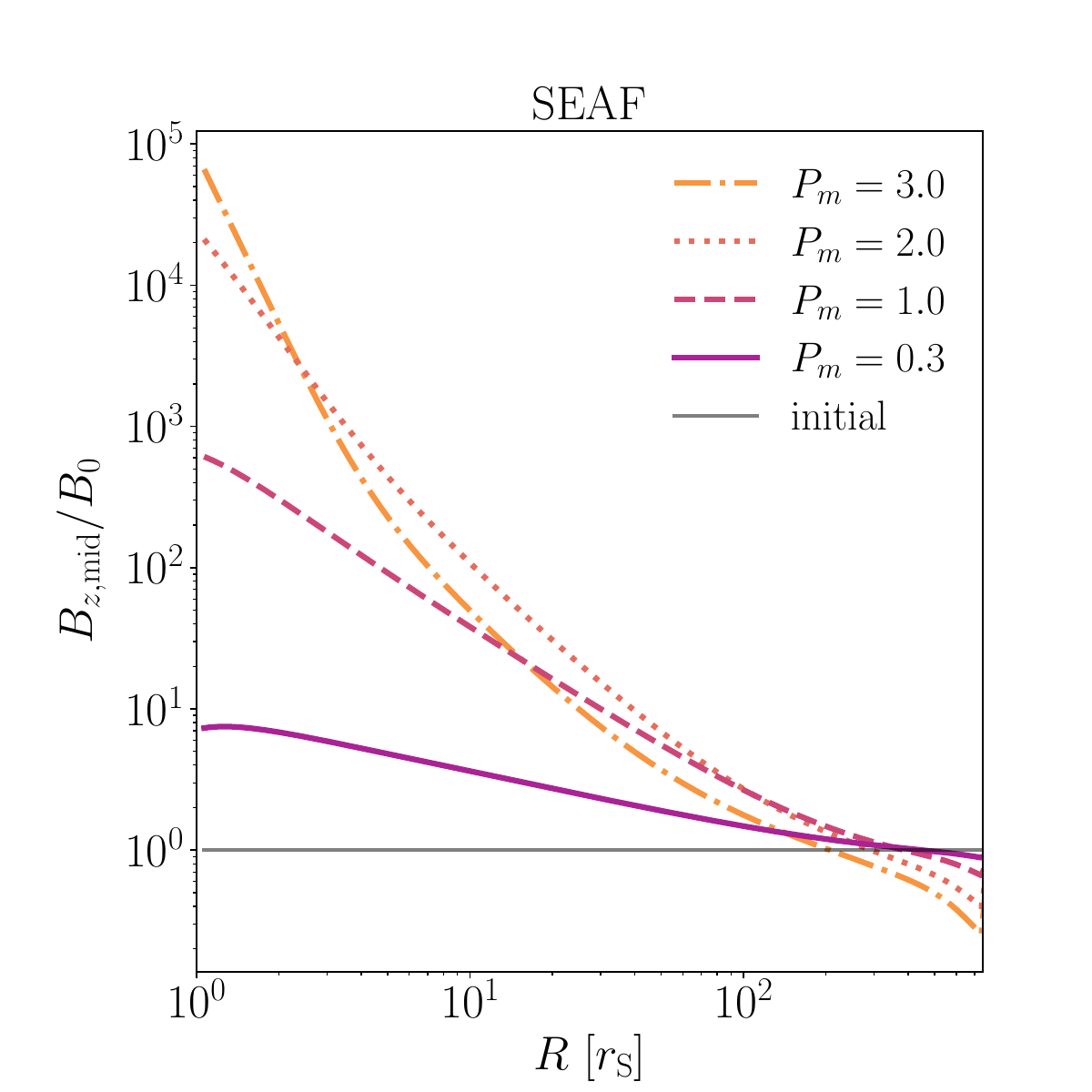}
  \caption{
  The $P_{\rm m}$ dependence of the $B_{z,\rm mid}$ profile. The grey lines denote the initial profile.
  } \label{fig:Bz_mid_Pm}
\end{figure}

Figs.~\ref{fig:field_line_highPm} and \ref{fig:Bz_mid_Pm} compare the results with different $P_{\rm m}$. Fig.~\ref{fig:field_line_highPm} shows the difference in the field structure between the models with $P_{\rm m}=2$ (left) and 1 (right). The figure shows that the discs with larger $P_{\rm m}$ accumulate magnetic fields more efficiently. It is found that even a twofold increase in the magnetic Prandtl number leads to significant changes in the magnetic field.
Fig. \ref{fig:Bz_mid_Pm} quantitatively compares the distributions of $B_{z, \rm mid}$ for different $P_{\rm m}$. 
When comparing the cases of $P_{\rm m} = 0.3$ and $P_{\rm m} = 3.0$, there is an order of magnitude difference in the magnetic field strength near the inner boundary for RIAF and about four orders of magnitude difference for SEAF. 
Therefore, the distribution of the disc field strongly depends on the local properties of effective viscosity and magnetic diffusion arising from turbulence and other factors.

\begin{figure}
    \centering
    \includegraphics[keepaspectratio, width=0.8\columnwidth]{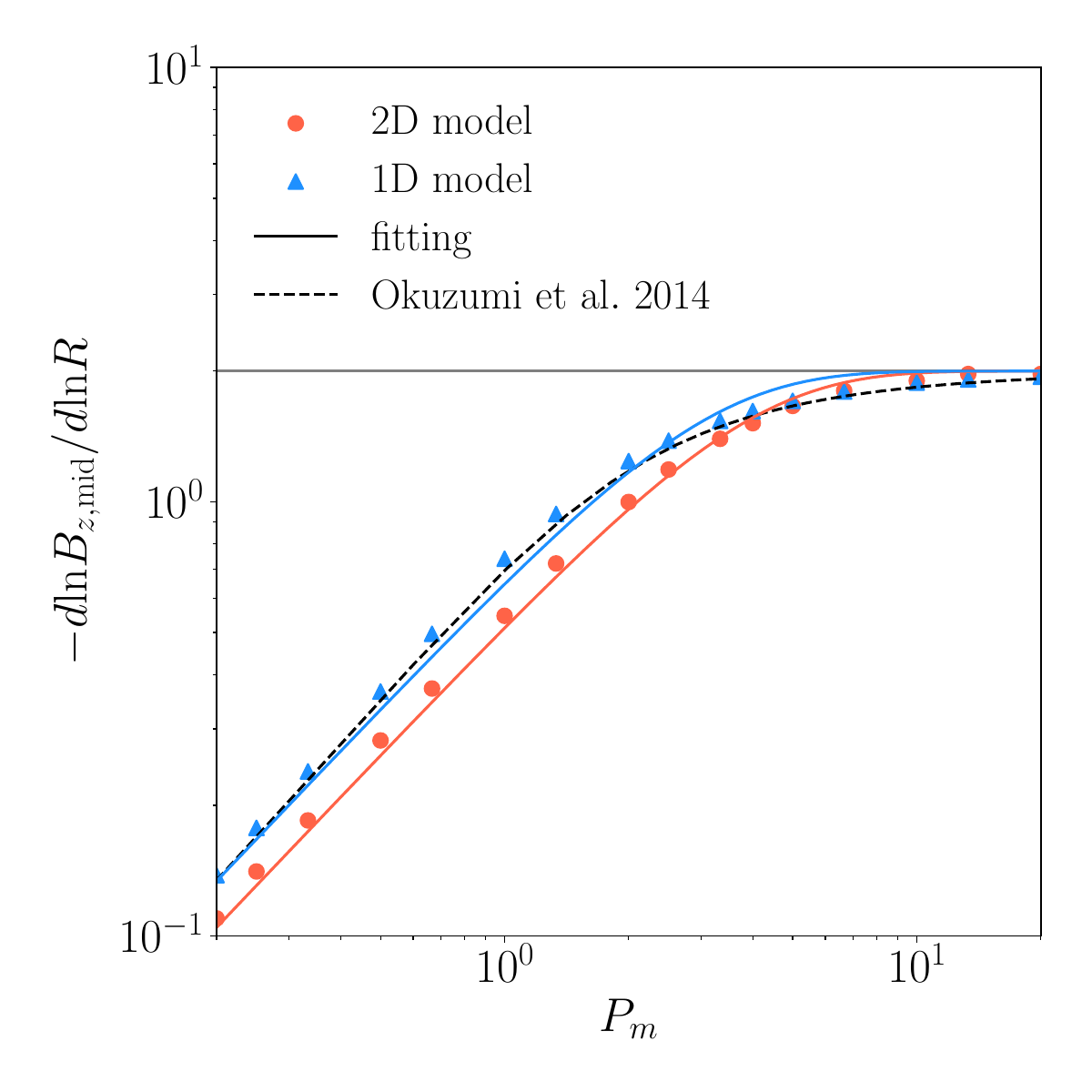}
    \caption{
    The relationship between $P_{\rm m}$ and the power law index of $B_{z,\rm mid}$ ($-d{\rm ln} B_{z, \rm mid}/d{\rm ln}R$) in RIAF ($f=1$). The red circles and blue triangles denote the results of the 2D and 1D models, respectively. The red and blue solid lines indicate the fitting functions for the data of the 2D and 1D models, respectively. The fitting functions are shown in equations (\ref{eq:DBz_Pm_2tanh_2D}) and (\ref{eq:DBz_Pm_2tanh_1D}).
    The dashed black line represents the analytical solution for the power-law index in the 1D model \citep{Okuzumi2014}.
    }
    \label{fig:Pm_pow_RIAF}
\end{figure}

Since RIAF is a self-similar disc model, $B_{z, \rm mid}$ follows a power-law dependence on $R$. We investigate the dependence of the power-law index on $P_{\rm m}$.
Fig.~\ref{fig:Pm_pow_RIAF} illustrates the relationship between $P_{\rm m}$ and the power-law index, $d{\rm ln} B_{z, \rm mid}/d{\rm ln}R$.
From the figure, we observe that the behaviour of the power-law index is similar between the 1D and 2D models. 
Furthermore, the data points for the 1D model match the analytical solution derived by \citet{Okuzumi2014} (dashed line).
The 2D model exhibits a smaller power-law index than the 1D model due to the multi-dimensional effect.
When $P_{\rm m}=1$, the power-law indexes in the 2D model and the analytical solution by \citet{Okuzumi2014} are $D_{B_z0} \approx 0.55$ and $D_{B_z0} \approx 0.69$, respectively.
Therefore, they agree with approximately 80\% accuracy.
For $P_{\rm m} \lesssim 1$, the power-law index is proportional to $P_{\rm m}$, while in the advection-dominated regime of $P_{\rm m} \gg 1$, the power-law index approaches $-2$. In the limit of advection dominance, the magnetic field distribution asymptotically follows $B_z \propto D R^{-2}$, as analytically derived by \citet{Okuzumi2014}.
Since our RIAF model possesses $D = \rm const.$, the asymptotic behaviour is consistent with the analytical estimation.

By fitting the data in Fig.~\ref{fig:Pm_pow_RIAF}, we derive an approximate expression for the power-law index as a function of $P_{\rm m}$:
\begin{numcases}{D_{B_z0} = \frac{d{\rm ln} B_{z, \rm mid}}{d{\rm ln}R} \approx}
    -1.93 \tanh (0.279 P_{\rm m}) & \textrm{for 2D model}  \label{eq:DBz_Pm_2D}  \\
     1.85 \tanh (0.390 P_{\rm m}) & \textrm{for 1D model}  \label{eq:DBz_Pm_1D}
\end{numcases}
The slight deviation of the asymptotic value for $P_{\rm m}\rightarrow \infty$ from $-2$ arises from the limited range of data used in the fitting, which is restricted to $P_{\rm m}\lesssim 20$. The maximum errors of this fitting function are approximately 5 and 7\% for the 2D and 1D models, respectively.
Additionally, considering the analytical solution by \citet{Okuzumi2014}, we performed fitting using a function with the coefficient of $\tanh$ fixed to $-2$ as $P_{\rm m}\rightarrow \infty$ (solid lines in Fig.~\ref{fig:Pm_pow_RIAF}):
\begin{numcases}{D_{B_z0} = \frac{d{\rm ln} B_{z, \rm mid}}{d{\rm ln}R} \approx}
    -2 \tanh (0.262 P_{\rm m}) & \textrm{for 2D model}  \label{eq:DBz_Pm_2tanh_2D}  \\
    -2 \tanh (0.335 P_{\rm m}) & \textrm{for 1D model}  \label{eq:DBz_Pm_2tanh_1D}
\end{numcases}
The maximum error of this fitting function is approximately 7\% for the 2D model and 13\% for the 1D model.
By using the results for the 2D model (equation~\ref{eq:DBz_Pm_2tanh_2D}) and equation~(\ref{eq:D_eff}), we can obtain the relationship between $D_{\rm eff}$ and $P_{\rm m}$ as follows:
\begin{align} 
 D_{\rm eff}^{-1} \approx
    \left\{
    \begin{array}{ll}
    0.363 P_{\rm m} & \textrm{for}~P_{\rm m} \lesssim 1 \\
    0.643 P_{\rm m} \approx D^{-1} & \textrm{for}~P_{\rm m} \gg 1 
    \end{array}
    \right. \label{eq:D_eff_Pm}
\end{align}

The higher-order components of the magnetic field also depend on $P_{\rm m}$. Using equations (\ref{eq:Bz2_Bz0}) and (\ref{eq:DBz_Pm_2tanh_2D}), we have
\begin{align} 
 \frac{B_z^{(2)}}{B_z^{(0)}} \approx
    \left\{
    \begin{array}{ll}
    0.0518 P_{\rm m}^2 & \textrm{for}~P_{\rm m} \lesssim 1 \\
    0.344 P_{\rm m} & \textrm{for}~P_{\rm m} \gg 1 
    \end{array}
    \right. \label{eq:Bz2_Bz0_Pm}
\end{align}
When $P_{\rm m} \lesssim 1$, $B_z^{(2)}/B_z^{(0)} \propto P_{\rm m}^{2}$ and strongly depends on $P_{\rm m}$. However, since the magnitude of $B_z^{(2)}/B_z^{(0)}$ itself is much smaller than 1, the higher-order components can be neglected (see also Fig.~\ref{fig:2DBz2_Bz0}).
In the case of $P_{\rm m} \gg 1$, $B_z^{(2)}/B_z^{(0)}>1$, and the higher-order components will significantly affect the inclination of the disc field. In other words, the approximation $B_R/B_z |_{z=H}\approx B_R^{(1)}/B_z^{(0)} (= D_{\rm eff}^{-1})$ is no longer valid.

\begin{figure}
    \centering
    \includegraphics[keepaspectratio, width=0.8\columnwidth]{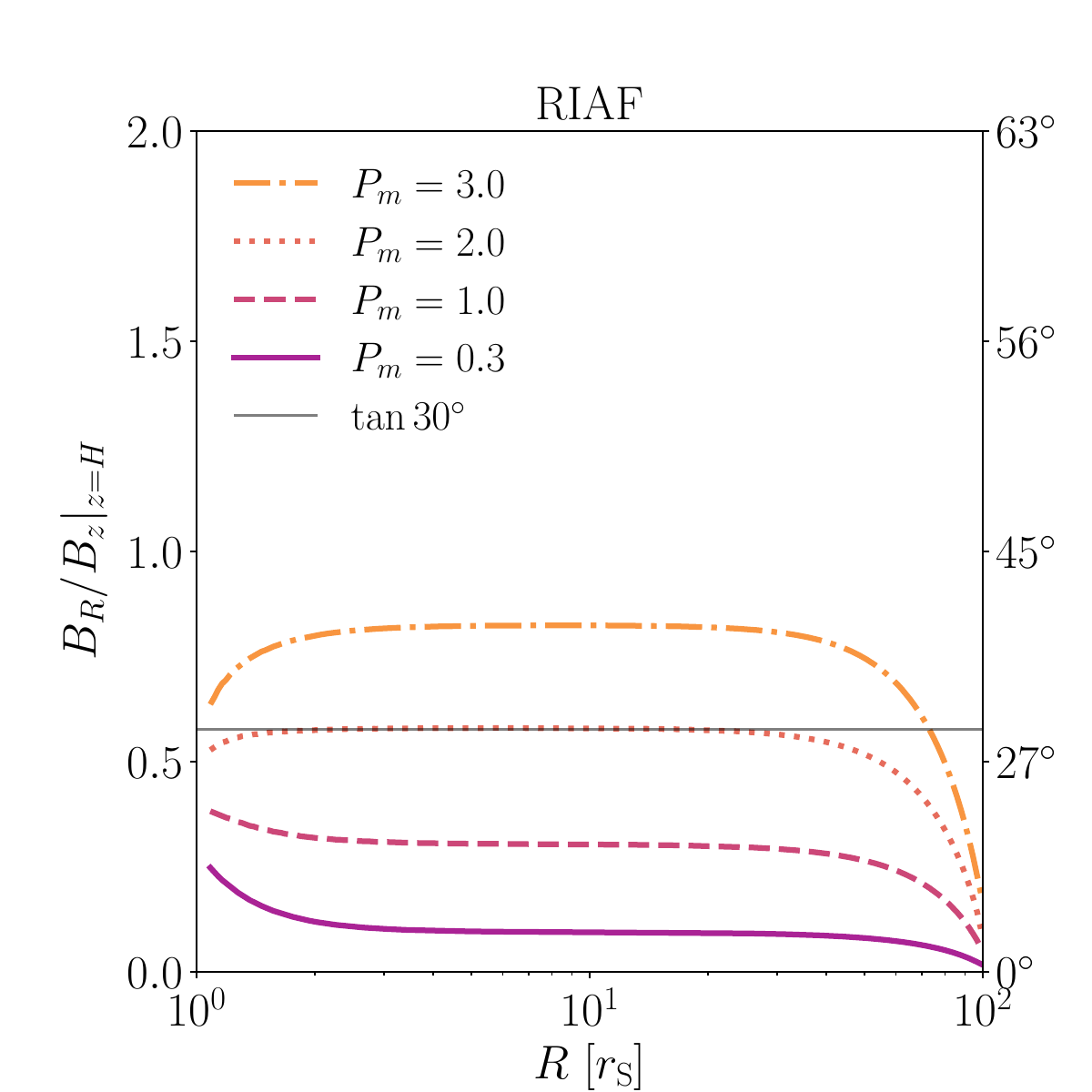}
    \includegraphics[keepaspectratio, width=0.8\columnwidth]{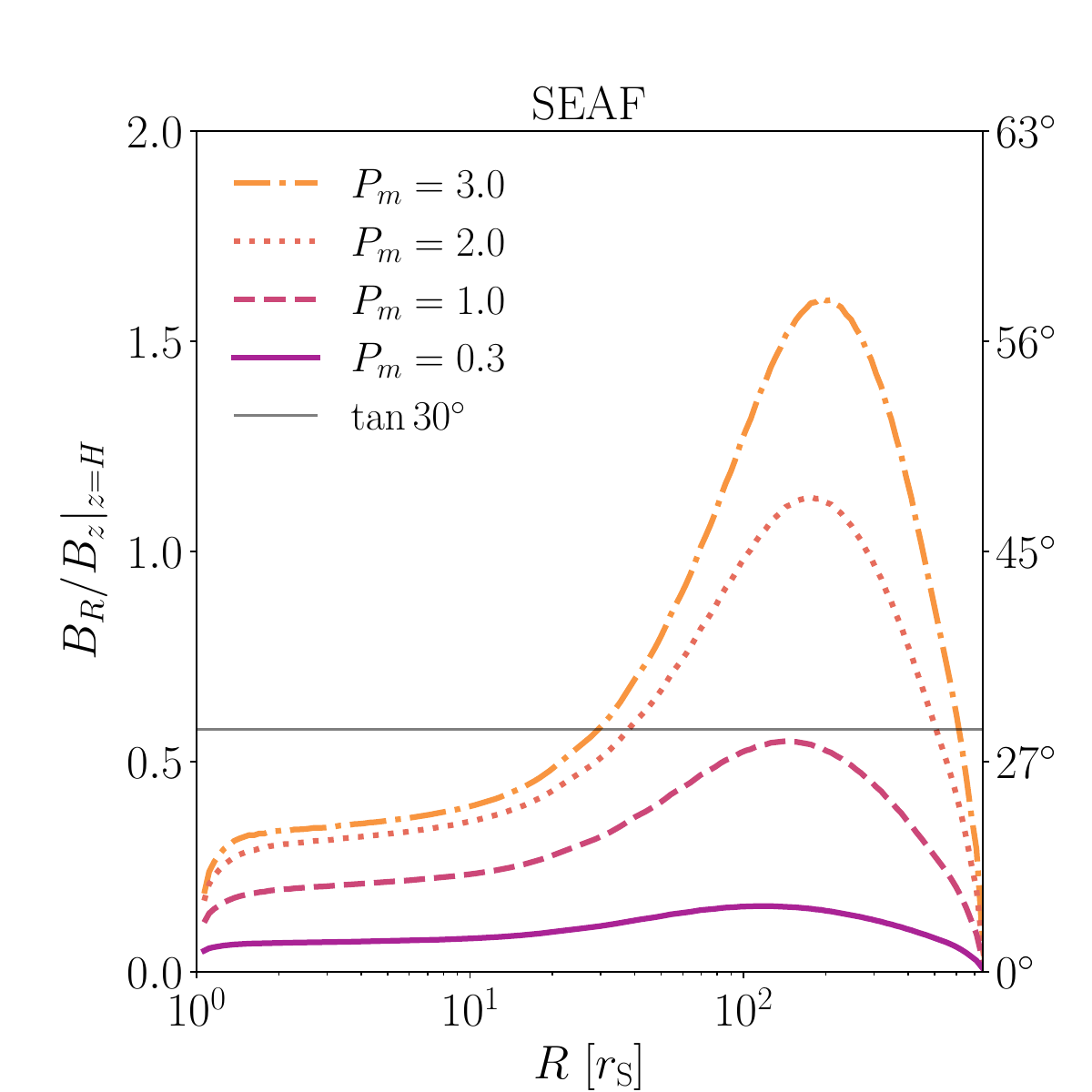}
  \caption{
  The $P_{\rm m}$ dependence of the field inclination $B_R/B_z |_{z=H}$. The grey lines denote the inclination angle of $30^\circ$.
  } \label{fig:BR_Bz_Pm}
\end{figure}

The field inclination also depends on the Prandtl number, as shown in Fig.~\ref{fig:BR_Bz_Pm}.
In our formulation, the thick RIAF ($h\approx0.5$) can drive a BP wind from almost the entire disc when $P_{\rm m} \gtrsim 2$. 
In the case of SEAF, a region outside the slim region ($R \gtrsim 85r_{\rm S}$) satisfies the conditions for driving a BP wind when $P_{\rm m} \gtrsim 1$. However, within the range of $P_{\rm m}$ investigated ($P_{\rm m} \lesssim 3$), the slim region ($R \lesssim 85r_{\rm S}$) never satisfy the condition.

\section{Discussion}

\subsection{Comparison with previous studies about the vertical structure}
\citet{Guilet2012} investigated the vertical structure of magnetic fields in geometrically thin discs analytically and numerically by using the non-ideal MHD equations. Their model, which combines the force-free magnetic field model outside the disc (determined by the Lorentz force) and the passive magnetic field model (neglecting feedback from the Lorentz force inside the disc), resembles our kinematic mean-field model.
In their analysis, the gradient of the background magnetic field ($D_{B_{z0}}$) was an input parameter and remained undetermined. However, we directly obtained $D_{B_{z0}}$ from numerical calculations using our 2D models based on the same equations as those of our analytic formulation, which allows us to examine the validity of our analytic analysis in a self-consistent way. As a result, we showed that the new parameter $D_{\rm eff}$, an extended version of $D$, is a useful indicator of magnetic flux transport efficiency (equation~\ref{eq:D_eff}).

We have shown in Section~\ref{subsec:multi-dim-effects} that $D_{\rm eff}$ is a good measure of the field inclination when the high-order terms are unimportant.
We compare our result with the result of \citet{Guilet2012}. They derived the following relation for the inclination (see their equation 129):
\begin{align} \label{eq:inclination_from_GO}
    \left. \frac{B_R}{B_z} \right|_{z = H} 
    = hP_{\rm m} \zeta_B \left[-\frac{1}{2} + \frac{2}{4\alpha^2+1}\left(4\alpha^2 + \frac{\zeta_B^2}{3}\right)\right],
\end{align}
Here, $\zeta_B$ is defined as $\zeta_B \equiv [\ln\beta_{z, \rm mid}^2]^{1/2}$ ($\beta_{\rm mid}$ is the plasma beta at the equatorial plane of the disc given by $\beta_{\rm mid} \equiv 8\pi p_{\rm mid}/B^2_{z, \rm mid}$). The accretion flow due to MRI turbulence exists in the region $-\zeta_B \leq \zeta \leq \zeta_B$. $\zeta_B$ corresponds to approximately 3.7 and 5.3 for $\beta_{\rm mid}=10^3$ and $10^6$, respectively. For comparison, giving the typical parameters of RIAF in this study ($h=0.5$, $\alpha=0.01$), the inclination of the magnetic field $B_R/B_z |_{z=H}$ for $\beta_{\rm mid}=10^3$ and $10^6$ becomes approximately $16P_{\rm m}$ and $47P_{\rm m}$, respectively.
However, our results indicate that $B_z/B_R|_{z=H} \approx 0.36P_{\rm m}$ when $P_{\rm m} \lesssim 1$ (equation~\ref{eq:D_eff_Pm}). Therefore, equation~(\ref{eq:inclination_from_GO}) predicts larger values by approximately an order of magnitude than ours.

We consider that the differences between our study and \citet{Guilet2012} originate from the following two points. The first point is that they approximate the disc thickness as negligible. However, as we have shown in Section~\ref{subsec:multi-dim-effects}, the multi-dimensional effects due to the disc thickness enhance the effective magnetic diffusivity. The second point is the behaviour of the accretion flow. In their model, the vertical range of the accretion flow increases as the magnetic field weakens. Therefore, a weaker magnetic field results in an effectively thicker disc ($h$ is larger, and $D$ is smaller). On the other hand, our model assumes that the accretion flow exists only within the vertical range of $|z|<H$. Future detailed investigations using 3D MHD simulations of accretion flows will enable us to model a realistic velocity structure.

\citet{Lovelace2009} argued that $P_{\rm m}\gtrsim 2.7$ is required for the disc to blow the BP wind. Although their constraint is similar to ours ($P_{\rm m}\gtrsim 2$ from Fig.~\ref{fig:BR_Bz_Pm}), their argument is based on the thin disc approximation. In fact, the field inclination (their equation~13) is essentially the same as equation~(\ref{eq:BRs_Bz}) which ignores the disc thickness effect. On the other hand, they solved not only the induction equation but also the momentum equation to take into account the outflow motion. Including the back-reaction of the outflow is our future work.

We have ignored the vertical structure of the accretion speed in this study just for simplicity. However, near the disc surfaces, efficient angular momentum loss due to the magnetic field can lead to a very fast accretion, as seen in MHD simulations \citep[e.g.,][]{Matsumoto1996, Beckwith2009, Zhu2018, Takasao2018, Takasao2019, Mishra2020, JacqueminIde2021}.
Such a coronal accretion will enhance the efficiency of the field transport toward the centre.
\citet{Li2021} extended the method of \citet{Lubow1994}'s kinematic mean-field model to two dimensions and demonstrated the importance of coronal accretion.

\begin{figure} 
    \centering
    \includegraphics[width=0.495\columnwidth]{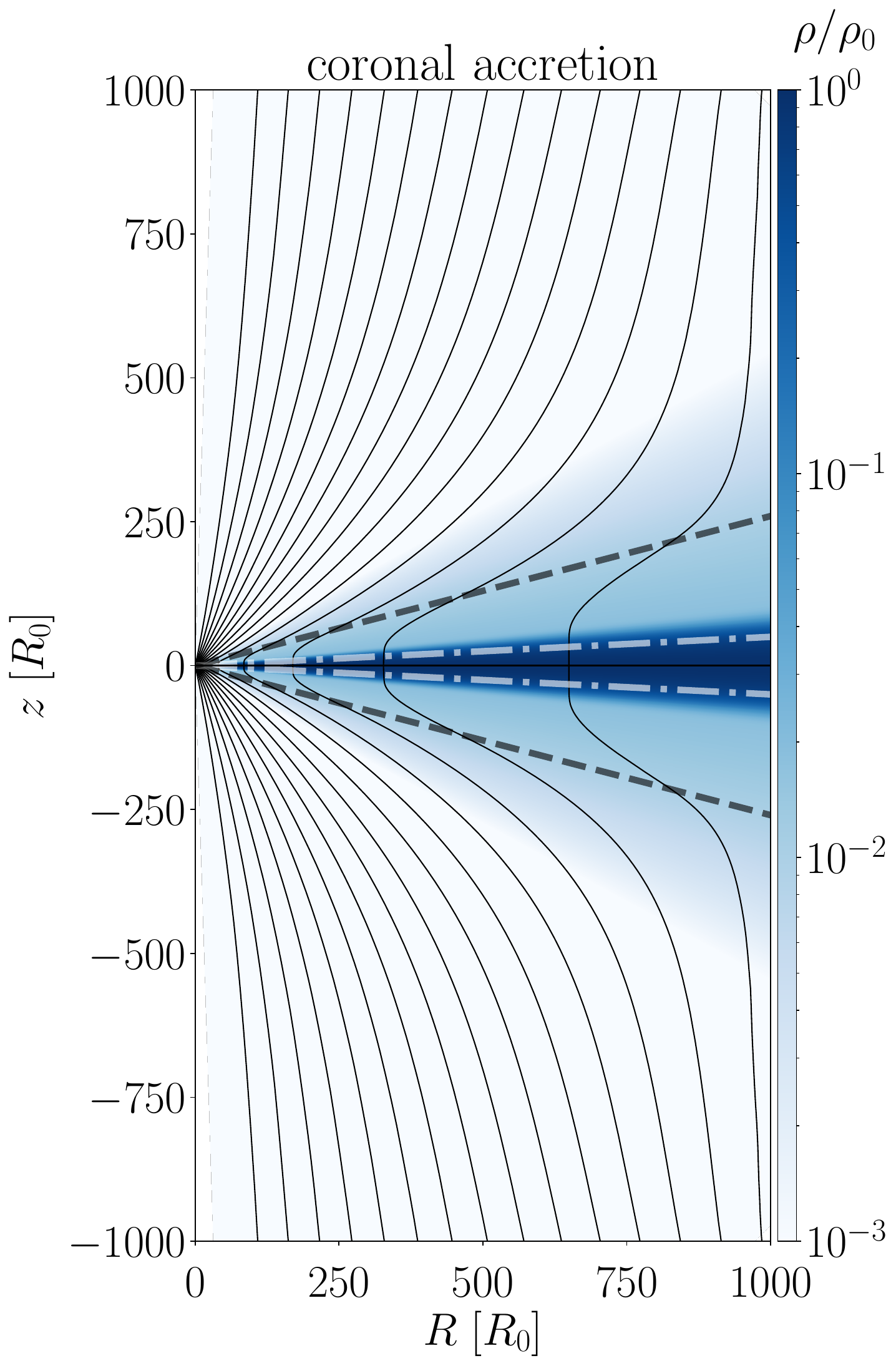}
    \includegraphics[width=0.495\columnwidth]{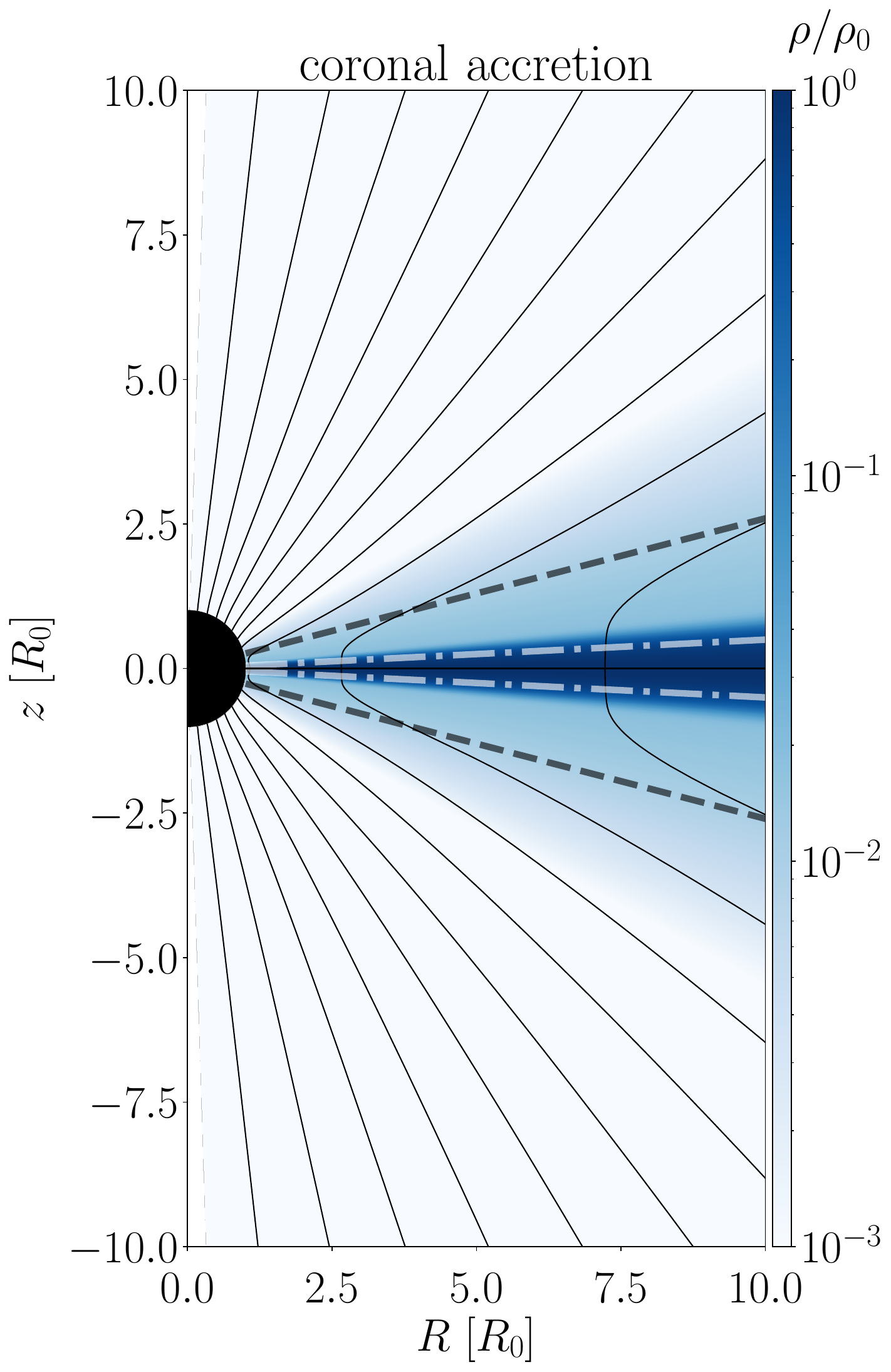}
  \caption{
  The poloidal field structure for RIAF with coronal accretion. The right panel is a zoom-in image of the left panel. The velocity field and the magnetic diffusivity are adopted from \citet{Li2021}. The dashed-dotted lines indicate the disc surfaces ($z = 0.05R$), and the dashed lines denote the coronal surfaces ($z=0.2598R$). The colour shows the density.
  }
  \label{fig:field_line_LiCao}
\end{figure}

We briefly compare our 2D model with the model of \citet{Li2021}.
Their model is based on steady-state equations, but our model can solve the time evolution of the system. Therefore, our model enables us to study how the disc magnetic field builds up to drive BP winds, for example.
Our 2D model can accept more complicated distributions of the physical quantities and can also handle coronal accretion.
To demonstrate this, we perform the flux transport calculation of a model of \citet{Li2021} using our 2D code. Fig. \ref{fig:field_line_LiCao} shows the result for the model with a scale height of $H = 0.05R$ and a height of the corona surface at $z_h = 0.2598 R$ (see the left figure of Fig. 7 in their paper). In this model, the accretion speed drastically changes with height, and the advection speed in the corona (the region between the dotted and dashed lines in Fig.~\ref{fig:field_line_LiCao}) reaches a maximum of approximately 35 times the equatorial plane velocity.
With the coronal accretion, magnetic flux is more strongly dragged toward the centre, as expected (compare Figs. \ref{fig:field_line} and \ref{fig:field_line_LiCao}).
The magnetic field distribution we obtained is generally consistent with their results.

\subsection{Implication for onset condition of MAD in RIAF}
\label{sec:MADcondition}

Considering the results in Section~\ref{sec:Results}, we discuss the required disc magnetic field strength for RIAFs to become MAD.
Previous investigations of general relativistic magnetohydrodynamic (GRMHD) simulations have found that the disc becomes MAD when the following MAD parameter exceeds $\sim 40$ \citep[e.g.,][]{Tchekhovskoy2011, Narayan2012, White2019}:
\begin{align} \label{eq:phi_BH}
    \phi_{\rm BH} = \Phi_{\rm in}\left(\dot{M}r_{\rm g}^2c\right)^{-1/2}.
\end{align}
$\Phi_{\rm in}$ represents the magnetic flux threading through the black hole's horizon, given by $\Phi_{\rm in}=\Phi(r=r_{\rm S})$, where
\begin{align} \label{eq:phi_in}
    \Phi_{\rm in} =  2\pi r_{\rm H}^2 
    \int^{\pi/2}_0 \left|B_r(r=r_{\rm H}, \theta)\right|\sin\theta d\theta, 
\end{align}
$r_{\rm H}$ is the radius of the black hole's horizon, and for a non-rotating black hole, $r_{\rm H} = r_{\rm S}$.
$\Phi_{\rm in}$ is related to the flux function $\psi_{\rm in, mid} = \psi (r=r_{\rm S}, \theta = \pi/2)$ at the innermost edge of the equatorial plane as follows:
\begin{align} \label{eq:phi_in_psi_mid}
    \Phi_{\rm in} = 2\pi \psi_{\rm in, mid},
\end{align}
where $\psi_{\rm in, mid} \equiv \psi(r=r_{\rm S}, \theta=\pi/2)$.

Another requirement for MAD is given by the plasma $\beta$. Recent simulations suggest that the plasma $\beta$ is less than or similar to 1 inside the MAD region \citep[e.g.,][]{Ressler2023}.

We first consider the constraint from the MAD parameter $\phi_{\rm BH}$.
Suppose that the disc has a poloidal field with a strength of $B_{\rm ext}$ at a radius of $R_{\rm ext}$. One may regard $R_{\rm ext}$ as the radius of the outer edge of the RIAF disc. We seek the condition for $B_{\rm ext}$ that the BH can achieve $\phi_{\rm BH}\approx 40$.
The poloidal field in RIAF can be expressed as 
\begin{align} \label{eq:Bz_pow}
B_{\rm z,mid}(R)=B_{\rm ext}\left(\frac{R}{R_{\rm ext}}\right)^{D_{B_{z0}}},
\end{align}
(equation~\ref{eq:DBz_Pm_2tanh_2D}), and therefore $\psi \propto R^{D_{Bz0}+2}$ (equation~\ref{eq:1DBz}). Combining them, we obtain
\begin{align} \label{eq:psi_mid_in}
    \psi_{\rm mid, in} = \frac{1}{2} B_{\rm ext} R_{\rm ext}^2 \left(\frac{r_S}{R_{\rm ext}}\right)^{D_{Bz0}+2}.
\end{align}
Equations~(\ref{eq:phi_in_psi_mid}) and (\ref{eq:psi_mid_in}) lead to the following relation:
\begin{align} \label{eq:phi_in_psi}
    \Phi_{\rm in} = \pi B_{\rm ext} R_{\rm ext}^2  \left(\frac{r_S}{R_{\rm ext}}\right)^{D_{Bz0}+2}.
\end{align}
By using equations~(\ref{eq:phi_BH}) and (\ref{eq:phi_in_psi}), we finally obtain the expression for $B_{\rm ext}$ in terms of the MAD parameter $\phi_{\rm BH}$:
\begin{align} \label{eq:B_ext}
    B_{\rm ext} = \frac{(\dot{M}r_{\rm g}^2c)^{1/2}}{\pi r_{\rm S}^2} \left(\frac{R_{\rm ext}}{r_S}\right)^{D_{Bz0}} \phi_{\rm BH}.
\end{align}
As seen in Section~\ref{subsec:Pm_dependences},
the power exponent $D_{Bz0}$ depends only on the nondimensional parameter $P_{\rm m}$.

We apply equation~(\ref{eq:B_ext}) to XRBs and discuss the condition for MAD.
It is possible that the accretion disc consists of the inner RIAF part plus the outer standard disc part. 
Even in such a case, observations estimate that the radius of the outer boundary of the RIAF part is of the order of $10^1-10^2r_{\rm g}$ \citep[e.g.,][]{Petrucci2010, Marino2021, Mercel2022, Barnier2022}.
Considering this, we adopt $R_{\rm ext} = 100r_{\rm g}$ as a fiducial value. The field strength required to realise MAD is estimated as follows:
\begin{align} \label{eq:Bext_MAD_condition_phiBH}
    &B_{{\rm ext, 1}} 
    \approx 1.9\times 10^6 ~ {\rm G}
    \left(\frac{R_{\rm ext}}{100r_{\rm g}}\right)^{D_{B_{z0}}(P_{\rm m}=1) } \nonumber \\
    &~~~~~~~~~~~~~~~~~~~~~~~~~~~~~~~~~\times\left(\frac{M}{10M_\odot}\right)^{-\frac{1}{2}}
    \left(\frac{\dot{M}}{10^{-3}\dot{M}_{\rm Edd}}\right)^{\frac{1}{2}}
    \left(\frac{\phi_{\rm BH}}{40}\right).
\end{align}
In the above equation, we have used equation (\ref{eq:DBz_Pm_2tanh_2D}) to calculate $D_{B_z0}$ ($D_{B_z0}(P_{\rm m}=1)\approx -0.51$).

The constraint from the plasma $\beta$ is considered.
The pressure distribution in the equatorial plane of RIAF is given by
\begin{align}
    p_{\rm gas} &=  \frac{\Pi}{\sqrt{2\pi}H} \nonumber \\
    &\approx 4.9\times10^{11}~{\rm erg/cm} \nonumber \\
    &\left(\frac{R_{\rm MAD}}{10r_{\rm g}}\right)^{-5/2} 
    \left(\frac{M}{10M_{\odot}}\right)^{-1}
    \left(\frac{\dot{M}}{10^{-3}\dot{M}_{\rm Edd}}\right)^{\frac{1}{2}}
    \left(\frac{\alpha}{0.01}\right)^{-1},
\end{align}
where $R_{\rm MAD}$ is the MAD radius, which is undetermined. We assume $R_{\rm MAD}\approx 10r_{\rm g}$ in this estimate.
Combining this with equation~(\ref{eq:Bz_pow}), we can derive the expression of the plasma $\beta$. The field strength corresponding to $\beta=1$ at $R=R_{\rm MAD}$ is given as
\begin{align} \label{eq:Bext_MAD_condition_beta}
    B_{\rm ext,2} &\approx 1.1\times10^{6} ~{\rm G} \nonumber \\
    &\left(\frac{R_{\rm MAD}}{10r_{\rm g}}\right)^{-5/4 - D_{B_{z0}}(P_{\rm m}=1)}
    \left(\frac{R_{\rm ext}}{100r_{\rm g}}\right)^{D_{B_{z0}}(P_{\rm m}=1)} \nonumber\\
    &\left(\frac{M}{10M_{\odot}}\right)^{-1/2}
    \left(\frac{\dot{M}}{10^{-3}\dot{M}_{\rm Edd}}\right)^{\frac{1}{4}}
    \left(\frac{\alpha}{0.01}\right)^{-1/2}.
\end{align}

The plasma $\beta$ gives a constraint on the field distribution, $D_{B_{z0}}$.
The plasma $\beta$ has the radial dependence of
\begin{align}
    \beta \propto R^{-5/2 - 2D_{B_{z0}}}.
\end{align}
If $-5/2 - 2D_{B_{z0}} < 0$, the low $\beta$ region must appear in the outer disc, and the stably rotating disc will not form there. Therefore, we focus on RIAF with $-5/2 - 2D_{B_{z0}} \ge 0$.
In addition, $D_{B_{z0}} \geq -2$, as mentioned in Section~\ref{subsec:Pm_dependences}. As a result, the range of $D_{B_{z0}}$ is written as
\begin{align} \label{eq:beta_pow_condition}
    -2 \leq D_{B_{z0}} \leq - \frac{5}{4}
\end{align}
The fitting function of equation~(\ref{eq:DBz_Pm_2tanh_2D}) shows that $D_{B_{z0}}$ satisfies the conditions of equation (\ref{eq:beta_pow_condition}) when $P_{\rm m} \gtrsim 2.8$.

\begin{figure}
    \centering
    \includegraphics[keepaspectratio, width=0.8\columnwidth]{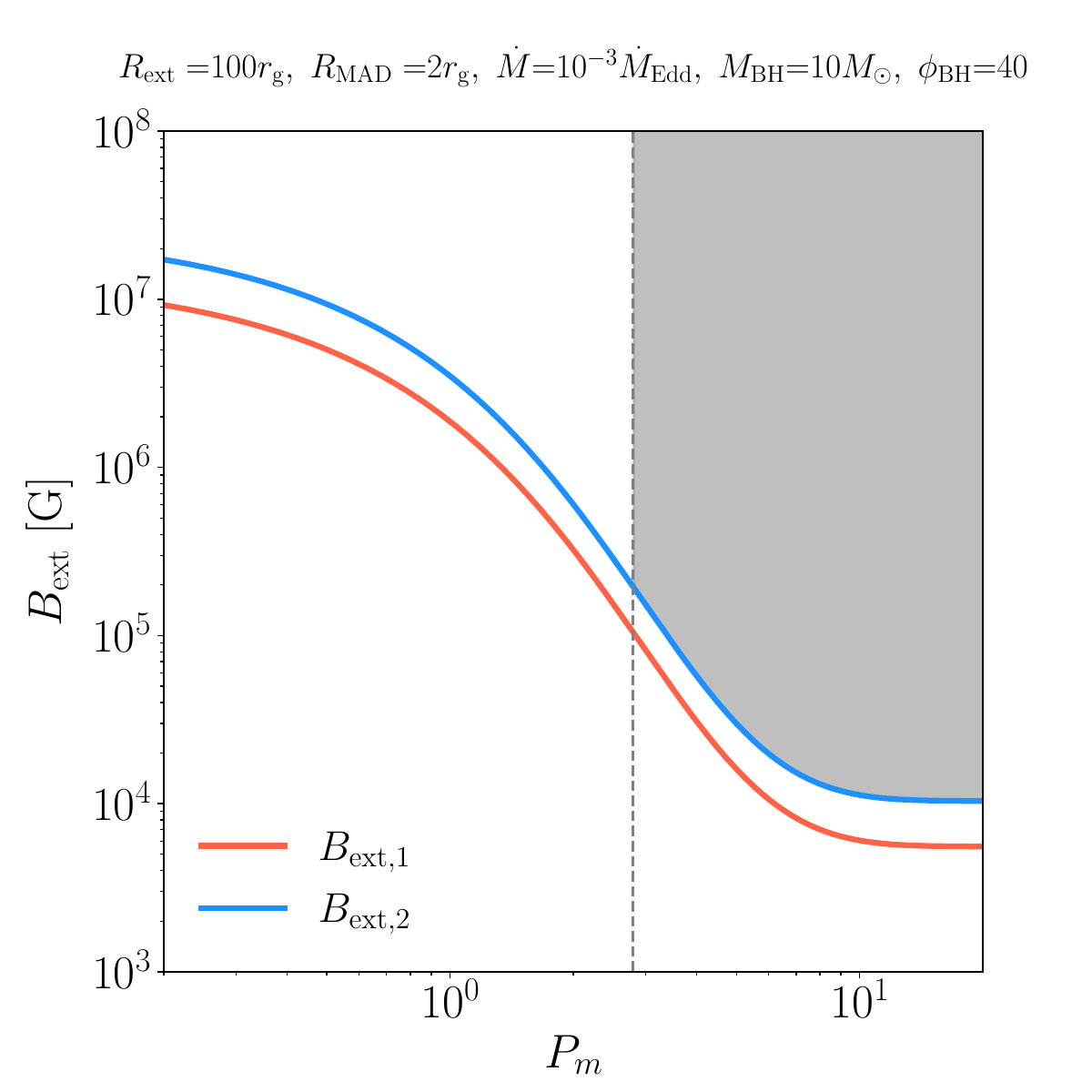}
    \includegraphics[keepaspectratio, width=0.8\columnwidth]{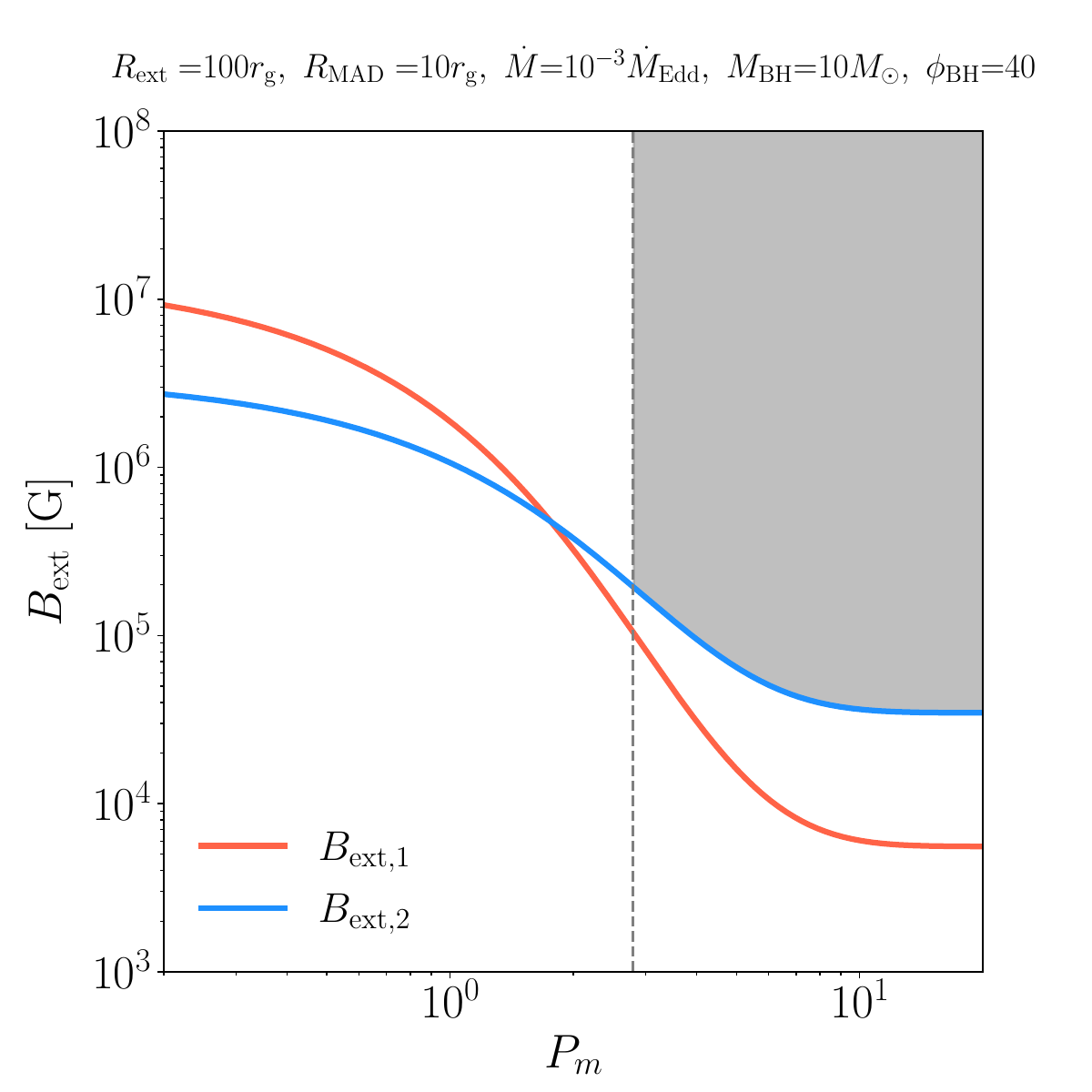}
    \caption{
    The range of $B_{\rm ext}$ that can achieve MAD in XRBs as a function of $P_{\rm m}$. The parameters are shown in the panels. The top panel shows the case in which the MAD radius $R_{\rm MAD}$ is just $2r_{\rm g}$, while the bottom panel corresponds to the case with a larger MAD radius, $R_{\rm MAD}=10r_{\rm g}$. The red and blue lines denote the lower limits given by the constraints of the MAD parameter $\phi_{\rm BH}$ and the plasma $\beta$, respectively (see equations \ref{eq:Bext_MAD_condition_phiBH} and \ref{eq:Bext_MAD_condition_beta}). The dashed vertical lines indicate the lower limit for $P_{\rm m}$ (see the description about equation \ref{eq:beta_pow_condition}).
    The grey-shaded areas correspond to the parameter spaces that satisfy all three conditions.
    }
    \label{fig:Bext_MAD_condition}
\end{figure}

Fig.~\ref{fig:Bext_MAD_condition} illustrates the parameter space of $B_{\rm ext}$ that can achieve MAD in XRBs as a function of $P_{\rm m}$.
The grey dashed line denotes the condition for $P_{\rm m}$ given by equation (\ref{eq:beta_pow_condition}). Namely, $P_{\rm m} \gtrsim 2.8$ for the given parameter set.
The grey region indicates the range of $B_{\rm ext}$ that can satisfy the possible three MAD conditions ($B_{\rm ext}\ge B_{\rm ext,1}$, $B_{\rm ext}\ge B_{\rm ext,2}$, and equation \ref{eq:beta_pow_condition}).
The top panel of Fig.~\ref{fig:Bext_MAD_condition} shows the result for a small MAD radius ($R_{\rm MAD}=2r_{\rm g}$). This case is a limiting case of MAD in the sense that the MAD size is just above the BH size. We find that the constraint from the plasma $\beta$ is slightly stronger than but is similar to the constraint from the MAD parameter ($B_{\rm ext,1}\sim B_{\rm ext,2}$). Therefore, $\phi_{\rm BH}\sim 40$ would be a good criterion for MAD formation.
The bottom panel of Fig.~\ref{fig:Bext_MAD_condition} displays the case with $R_{\rm MAD}=10r_{\rm g}$. In this case, $B_{\rm ext,1} < B_{\rm ext,2}$ in the allowed range of $P_{\rm m}$. 
Therefore, when the MAD size is much larger than the BH size, the constraint from the plasma $\beta$ seems to be more important than that from the MAD parameter.

We note that the present calculation does not consider coronal accretion. If coronal accretion exists, the effective Prandtl number will increase. Therefore, we expect that the disc can reach the MAD state with a weaker external magnetic field.

\section{Summary}

To understand the mechanism determining the magnetic field distribution in geometrically thick accretion discs around black holes, we analytically and numerically investigated magnetic flux transport in RIAF and SEAF. For the numerical study, we developed a time-dependent magnetic flux transport model in the axisymmetric spherical coordinate system (referred to as the 2D model in the text).
Below, we summarize our findings.

It has been revealed that the magnetic field behaves diffusively due to the multi-dimensional effects arising from the thickness of the disc. The multi-dimensional effects of the thickness can be evaluated by the dimensionless parameter $D_{\rm eff}$ (Sections~\ref{subsec:high_order_ana} and \ref{subsec:multi-dim-effects}). In the conventional 1D models, these multi-dimensional effects cannot be accounted for, which may lead to an overestimation of the magnetic field strength and inclination (Section~\ref{subsec:multi-dim-effects}). However, when the disc is sufficiently thin ($h\lesssim0.1$), the multi-dimensional effects are small, and the conventional 1D models are adequate for modelling (Section~\ref{subsec:aspect_dependences}).

We demonstrated that the magnetic field distribution in a thick disc strongly depends on the magnetic Prandtl number $P_{\rm m}$ (Section~\ref{subsec:Pm_dependences}). This result points out the importance of local physical quantities in determining the global distribution of the magnetic field.

To investigate the driving condition for BP winds, we studied the dependence of the inclination of the magnetic field on the disc aspect ratio ($h$) and magnetic Prandtl number ($P_{\rm m}$) (Section~\ref{subsec:aspect_dependences} and \ref{subsec:Pm_dependences}). As thicker discs can drag the poloidal field more efficiently toward the centre, they form a more inclined field. In addition, discs with a higher $P_{\rm m}$ can also do so. However, our 2D model revealed that the classical 1D model tends to overestimate the field inclination because of the lack of multi-dimensional effects. Therefore, in reality, the radial extent of the outflow driving regions could be narrower than predicted by the 1D model.

We investigated the condition for a RIAF to be MAD (Section~\ref{sec:MADcondition}). We derived the lower limit for the poloidal field strength at the radius of the outer edge of RIAF for a given $P_{\rm m}$. The diagram would be helpful for interpreting the 3D GRMHD simulations and observations.

\section*{Acknowledgements}
We thank Drs. Satoshi Okuzumi, Yusuke Tsukamoto, Chris White, Yoshiyuki Inoue, and Samuel Barnier for their fruitful comments. S.T. was supported by JSPS KAKENHI grant Nos. JP22K14074, JP22KK0043, and JP21H04487.
This work was supported by JST SPRING, Grant Number JPMJSP2138 (R.Y.).
Numerical computations were in part carried out on PC cluster at Centre for Computational Astrophysics, National Astronomical Observatory of Japan.

\section*{Data Availability}
The data underlying this article will be shared on reasonable request to corresponding authors.



\bibliographystyle{mnras}
\bibliography{ref} 




\appendix

\section{Dependence of the result on the inner boundary condition}
\label{app:2D_Inner_Boundary}
Here we explain the motivation for the inner boundary condition adopted in the main text.
The inner boundary condition for the magnetic field is nontrivial but can affect the field distribution near the boundary. We investigate the dependence of the field distribution on the inner boundary condition. We impose the boundary condition for $B_r$ such that $B_r r^a={\rm const}$ in the ghost cells of the inner boundary. If $a=2$, the magnetic flux of the radial component $B_r$ conserves in the ghost cells. In terms of $\psi$, the condition $a=2$ corresponds to the outflow boundary ($\partial \psi / \partial r = 0$). From equations~(\ref{eq:Bz_cyl}) and (\ref{eq:2DBtheta}), it is clear that $B_{z, \text{mid}} = - B_{\theta, \text{mid}} = 0$ in this case.
This condition would work best for the split monopole-type geometry and will not be appropriate for the study of the flux transport in discs. If $a<2$, a fraction of the magnetic field touching the inner boundary cannot go deeper into the ghost cells. As a result, a poloidal field with a finite $z$ component $B_z$ will form just around the boundary.

\begin{figure}
    \centering
    \includegraphics[keepaspectratio, width=0.8\columnwidth]{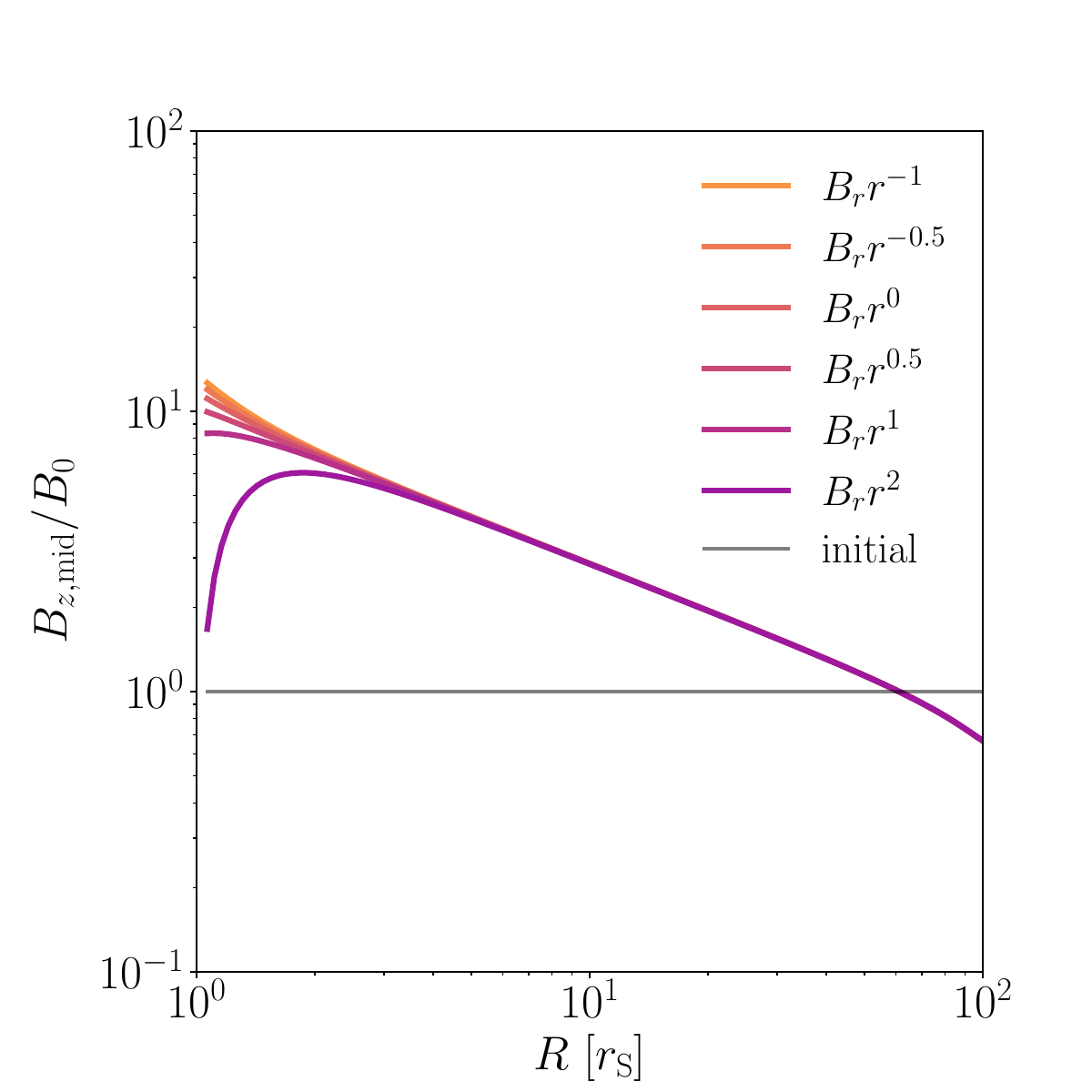}
    \includegraphics[keepaspectratio, width=0.8\columnwidth]{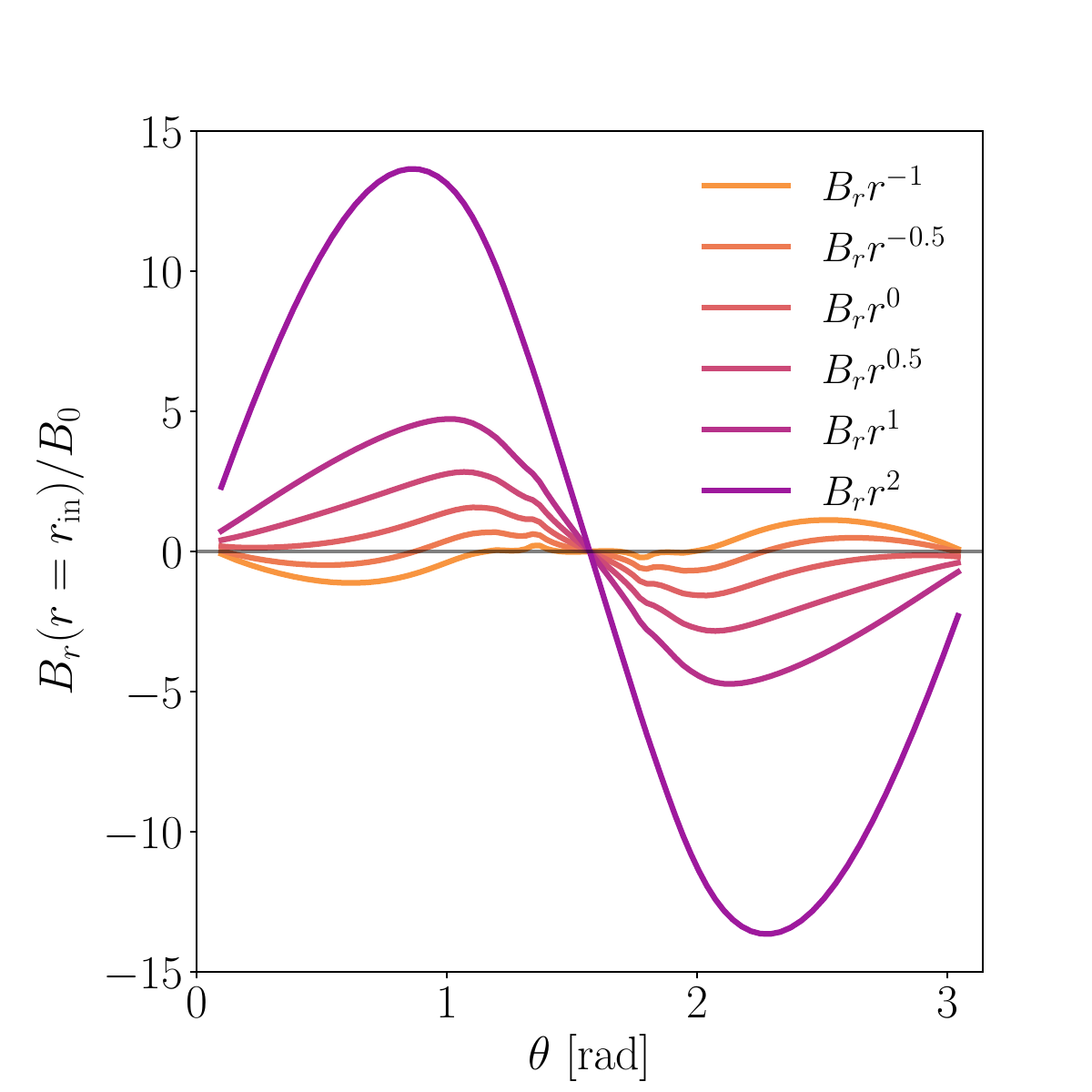}
  \caption{
  The field structures for different inner boundary conditions. We change $a$ of the constraint $B_r r^a = \rm const.$ The top panel shows the $B_{z,\rm mid}$ profiles, and the bottom panel displays the latitudinal distributions of $B_r$ at the inner boundary at a radius of $r_{\rm in}$.
  } \label{fig:Bz_Br_InnerBC}
\end{figure}

Fig.~\ref{fig:Bz_Br_InnerBC} shows the RIAF results for different values of $a$. The models with $a=-1,-0.5,0, 0.5, 1$, and $2$ are investigated.
Looking at the $B_{z,\rm mid}$ distribution (upper panel), we find no significant differences among the results for $a < 2$. However, the result for $a = 2$ shows a sharp decrease near the boundary, as expected.
The lower panel shows the $B_r$ profile in the latitudinal direction at the inner boundary. 
As we impose a positive $B_z$ in the disc, the dragged field tends to form a positive $B_r$ in the northern hemisphere ($0\le \theta \le \pi/2$), and vice versa.
However, we notice that the sign of $B_r$ flips even in a single hemisphere when $a<0$ and a complicated field geometry is formed.

From the survey, we find that adopting $0 < a < 2$ will result in a reasonable magnetic field structure. Therefore, we adopt $a=1$ as a fiducial value in the main text.

\section{Convergence check}
\label{app:convergence_check}

\begin{figure}
    \centering
    \includegraphics[keepaspectratio, width=0.8\columnwidth]{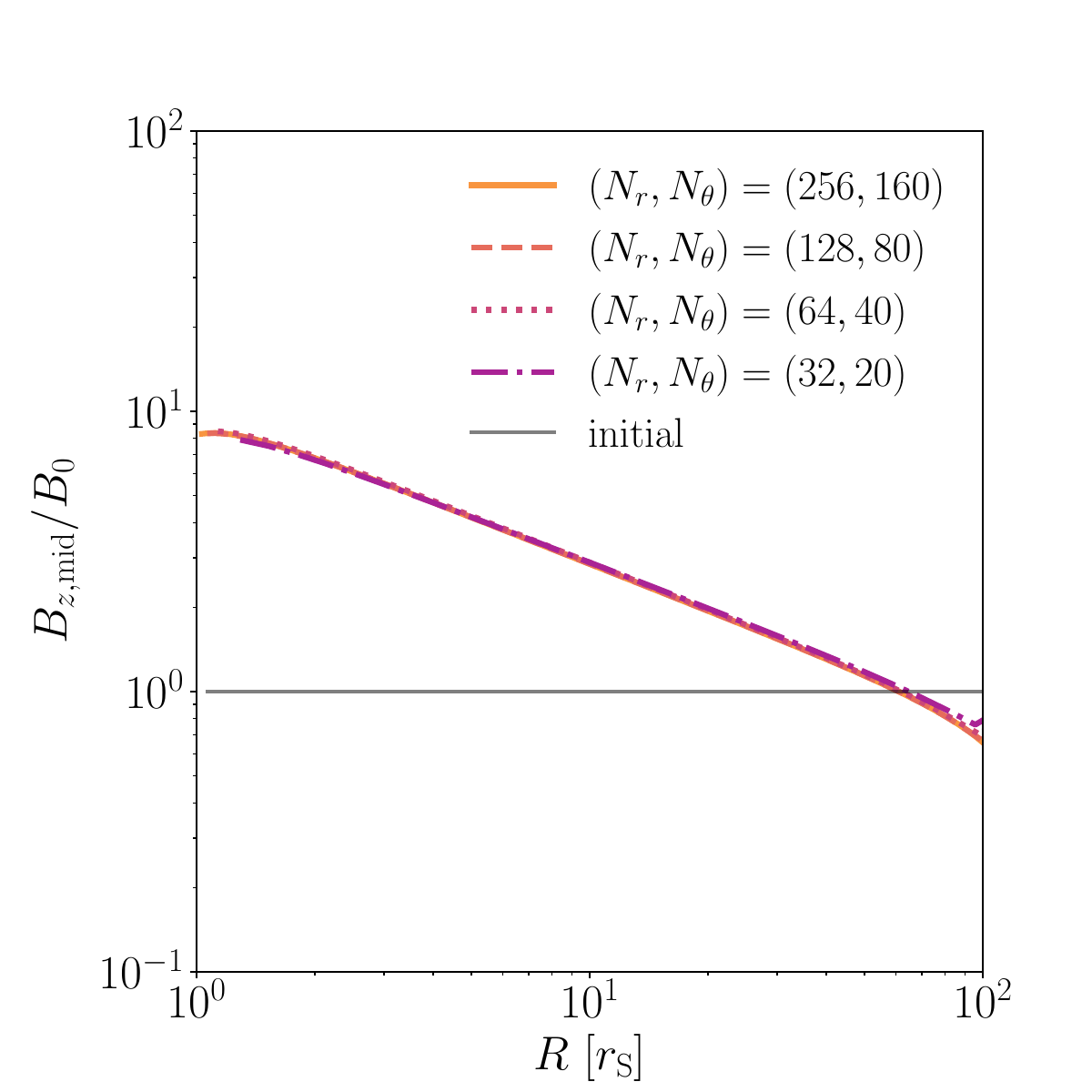}
    \includegraphics[keepaspectratio, width=0.8\columnwidth]{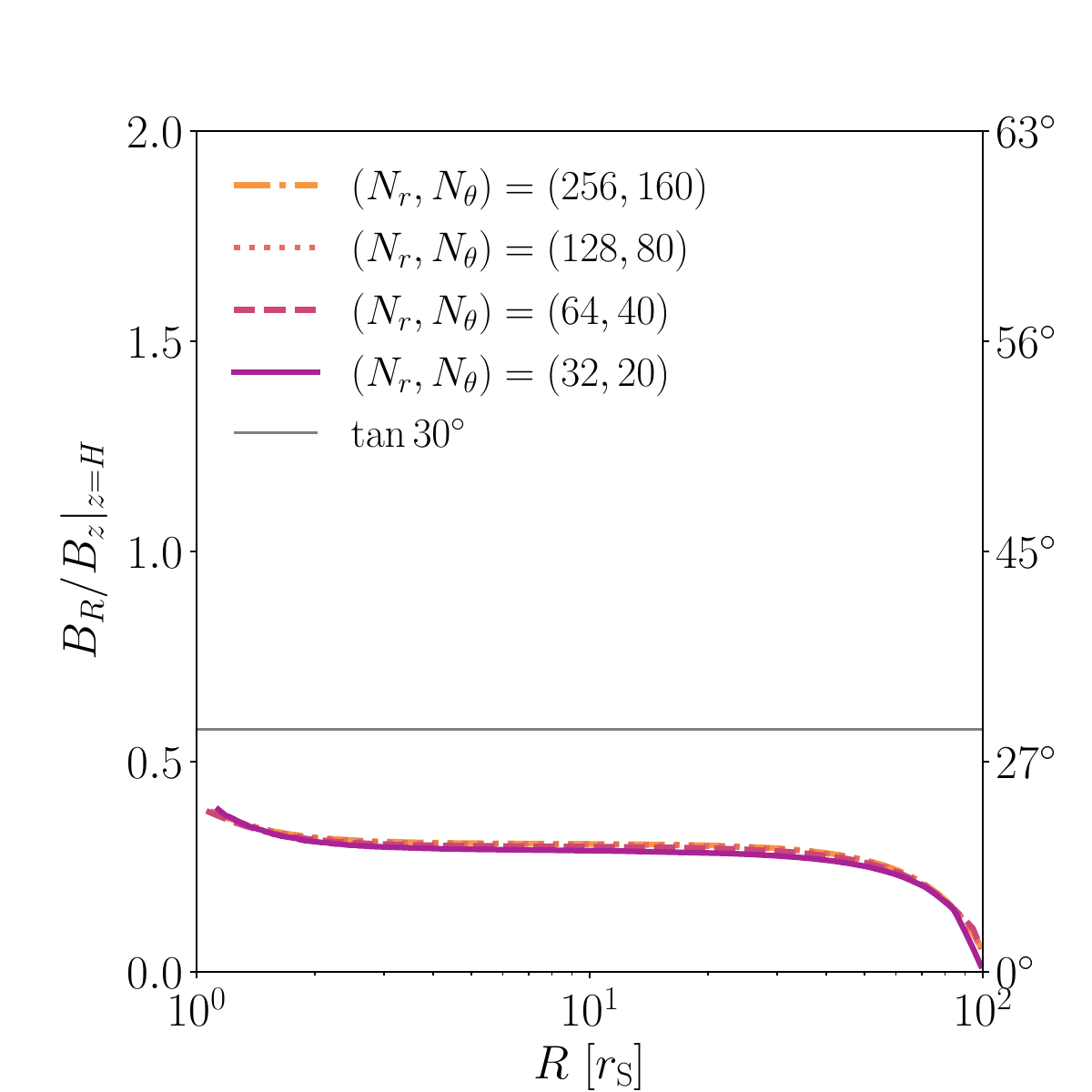}
  \caption{
  The field structures for different resolutions for RIAF. The top panel shows the $B_{z,\rm mid}$ profiles, and the bottom panel displays the inclination angle at the disc surface $B_z/B_R|_{z=H}$.
  } \label{fig:RIAF_convergence}
\end{figure}

We test the numerical convergence of the results.
Fig. \ref{fig:RIAF_convergence} summarises the results for RIAF. The figure displays the magnetic field strength at the equatorial plane of the  (upper panel) and the inclination of the magnetic field at the disc surface (lower panel) for different resolutions. The figure confirms the numerical convergence. We have also confirmed the convergence for SEAF.

\section{Caveat to 1D models}
\label{app:multidimensional_effects}

The field inclination, especially near the centre, is significantly affected by the multidimensional effects. As seen in Fig.~\ref{fig:1DBR_with_hDBz}, the 1D models based on equation~(\ref{eq:1DBRs_hDBz}) produce negative $B_{R,\rm surf}$, which should be unphysical. Fig.~\ref{fig:field_line_slim_1D2D} compares the poloidal field structures between the 1D and 2D models for SEAF. One can see that the field near the centre is inclined inward. That is unrealistic because the disk field must surround the magnetic flux accumulated around the pole, as demonstrated by the 2D model.

We explain why the 1D models produce the negative $B_{\rm R,\rm surf}$. Combining equations (\ref{eq:1DBRs_hDBz}) and (\ref{eq:Bz_Kphi}) to eliminate $K_\phi$, we obtain
\begin{align}
    B_{R, \rm surf} = h(D_*^{-1}+D_{B_z0})B_{z, \rm mid}.\label{eq:c1}
\end{align}
This equation is equivalent to equation (\ref{eq:BR1_Bz0}) in the case that $B_R^{(1)}=B_{R, \rm surf}$. As equation (\ref{eq:c1}) shows, the sign of $B_{R, \rm surf}$ depends on that of $D_*^{-1}+D_{B_z0}$.
When $P_{\rm m} = 1$,  $D_*^{-1} = 1.5$ in SEAF. $D_{B_z0}$ in the slim region ($R=20 r_S$) is approximately $-1.58$. Therefore, $D_*^{-1}+D_{B_z0}=-0.08<0$ and $B_{R, \rm surf}<0$. We note that the 2D model exhibits a smaller negative $D_{B_z0}$ ($-1.21$ at $R=20 r_S$) because the multi-dimensional effects increase effective magnetic diffusivity (see, e.g., section~\ref{subsec:multi-dim-effects}). The 1D models produce an unphysical magnetic structure because of the lack of multi-dimensional effects.


\bsp	
\label{lastpage}
\end{document}